\documentclass[12pt]{article}
\usepackage{graphicx,amsmath,setspace,amssymb}
\usepackage{units}
\usepackage{color}
\usepackage{cite}
\usepackage{hyperref}
\usepackage{multirow}
\usepackage{pbox}
\usepackage{array}
\usepackage{float}
\usepackage{lineno}
\usepackage{hyperref}
\usepackage[caption=true]{subfig}
\usepackage[font=small,labelfont=bf]{caption}

\newlength{\dinwidth}
\newlength{\dinmargin}
\def\lapproxeq{\lower .7ex\hbox{$\;\stackrel{\textstyle                                                    
<}{\sim}\;$}}                                                    
\def\gapproxeq{\lower .7ex\hbox{$\;\stackrel{\textstyle                                                    
>}{\sim}\;$}}                                                    
\def\be{\begin{equation}}                                                    
\def\ee{\end{equation}}                                                    
\def\bea{\begin{eqnarray}}                      
\def\eea{\end{eqnarray}}

\def\bb{\vec{b}'}

\def\bb{b\bar{b}}
\def\cc{c\bar{c}}

\def\GeV{\rm GeV}
\def\TeV{\rm TeV}

\def\slash#1{#1 \hskip-0.55em /}

\def\sh{\hat s}
\def\sh2{{\hat s}^2}

\def\MS{\overline{\rm MS}}

\def\a{\alpha_S(M_Z^2)}

\begin{document}



\begin{center}
{\Large \bf An investigation of the $\alpha_S$ and heavy quark\\ \vspace*{0.3cm} mass dependence in the MSHT20 global PDF analysis}

\vspace*{1cm}
T. Cridge$^a$, L. A. Harland-Lang$^{b}$, A. D. Martin$^c$, 
and R.S. Thorne$^a$\\                                               
\vspace*{0.5cm}                                                    

$^a$ Department of Physics and Astronomy, University College London, London, WC1E 6BT, UK \\           
$^b$ Rudolf Peierls Centre, Beecroft Building, Parks Road, Oxford, OX1 3PU   \\  
   
$^c$ Institute for Particle Physics Phenomenology, Durham University, Durham, DH1 3LE, UK                   \\                                 
                                                    

\begin{abstract} 
\noindent We investigate the MSHT20 global PDF sets, demonstrating the effects of varying 
the strong 
coupling $\a$ and the masses of the charm and bottom quarks. 
We determine the preferred value, and accompanying uncertainties, when we 
allow $\a$ to be a free parameter in the MSHT20 global
analyses of  deep-inelastic and related hard scattering data, at both NLO and NNLO in QCD 
perturbation theory. We also study the constraints on $\a$ which come from the
individual data sets in the global fit by repeating the NNLO and NLO global analyses at
various fixed values of $\a$,
spanning the range $\a=0.108$ to $0.130$ in units of $0.001$. We make 
all resulting PDFs sets available. 
We find that the best fit values are $\a=0.1203\pm 0.0015$ and $0.1174\pm 0.0013$
at NLO and NNLO respectively.
We investigate the relationship between the variations in $\a$ and the
uncertainties on the PDFs, and illustrate this by calculating the cross sections
for key processes at the LHC. We also perform fits where we
allow the heavy quark masses $m_c$ and $m_b$ to vary away from their 
default values and make PDF sets available in steps of 
$\Delta m_c =0.05~\GeV$ and $\Delta m_b =0.25~\GeV$, using 
the pole mass definition of the quark masses. As for varying $\a$ values, we present the 
variation in the PDFs and in the predictions. We examine the   
comparison to data, particularly the HERA data on charm and bottom cross sections 
and note that our default values are very largely compatible with best fits to data.
We provide PDF sets with 3 and 4 active quark flavours, as well as the 
standard value of 5 flavours.  
\end{abstract}                                                        
\vspace*{0.5cm}

\end{center}

\begin{spacing}{0.8}
\clearpage

\tableofcontents
\clearpage
\end{spacing}

\section{Introduction  \label{sec:1}} 

In recent years there has been a significant improvement in both the precision and in the variety of
the available data for deep--inelastic and related hard--scattering processes which
can be used in determinations of the parton distribution functions (PDFs). This has been matched by the increasing precision of the theoretical calculations for the accompanying cross sections.  
These have both contributed to the recent PDF update, known as the MSHT20 PDFs~\cite{MSHT20}, 
which supersede the previous MMHT2014 PDFs~\cite{MMHT14} obtained using the same 
general framework.  
Particular additions to the data used in the global fit have been the final
HERA combined H1 and ZEUS data on the total and the heavy flavour cross sections, some
final precision Tevatron asymmetry data and new Drell Yan, top quark pair, jet and
$Z \,\,p_T$ data sets obtained at the LHC. For some of the LHC data this is the first of
our PDF determinations for which the full NNLO calculations have been available.  
Additionally, the procedures used in the global PDF analyses have
been improved, particularly the parameterisation, 
allowing the partonic structure of the proton to be determined
with improved accuracy and reliability. Indeed, a new result is that the NNLO PDF set is found to be greatly favoured in comparison to the NLO PDF set \cite{MSHT20}. However, in our default PDF determination 
we have presented PDFs with fixed, pre-determined values of the strong coupling constant
$\a$, and only made passing reference to the preferred values.  
Here we extend the recent MSHT20 global PDF analysis to study in turn the
preferred value and uncertainty on $\a$, the constraints from individual data sets in the fit, and the implications 
for predictions for processes at the LHC.
Moreover, we make available NLO and NNLO PDF sets for various fixed values of $\a$,
spanning the range $\a=0.108$ to $0.130$ in units of $0.001$.

Similarly, our default fit uses fixed pole masses of the charm and bottom quarks, $m_c=1.4~\GeV$ 
and $m_b=4.75~\GeV$.  
Here, we extend the MSHT20 global PDF analysis~\cite{MSHT20} to study the 
dependence of the PDFs, and the quality of the comparison to data, under 
variations of these masses away from their default values. We  investigate the resulting 
predictions for processes at the LHC. We make 
available central PDF sets for
$m_c=1.2-1.6~\GeV$ in 
steps of $0.05~\GeV$ and $m_b=4.00-5.50~\GeV$ in 
steps of $0.25~\GeV$, and also make available the standard MSHT20 PDFs, as well as the 
sets with alternate masses, in the 3 and 4 flavour number schemes.  

\section{The strong coupling $\alpha_S(M_Z^2)$}

\begin{figure} 
\begin{center}
\includegraphics[scale=0.22]{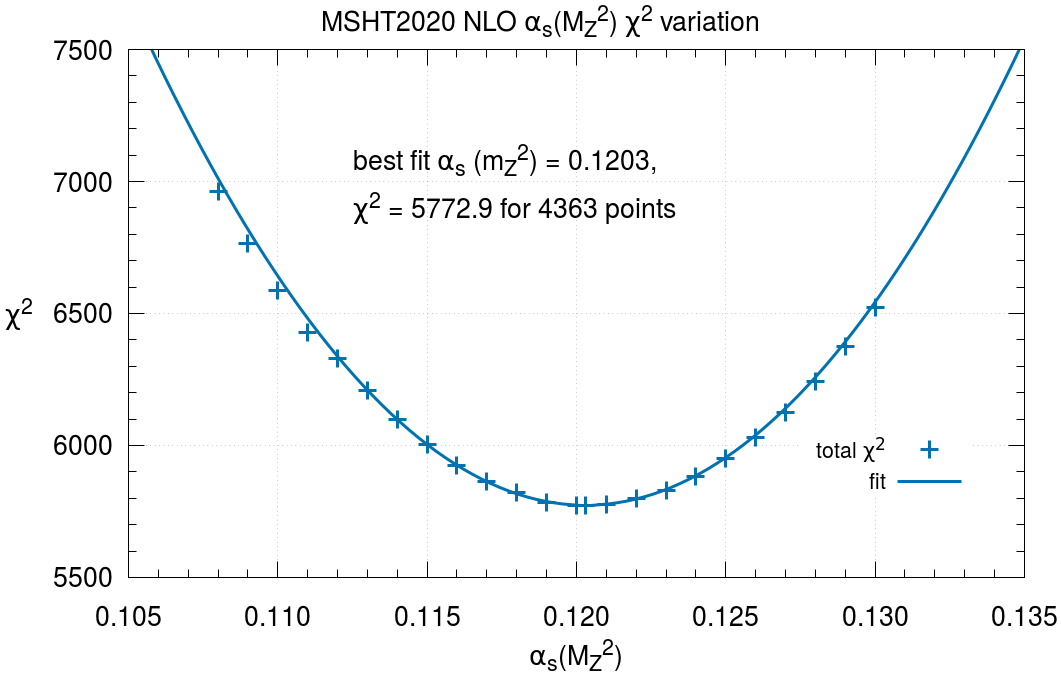}
\includegraphics[scale=0.22]{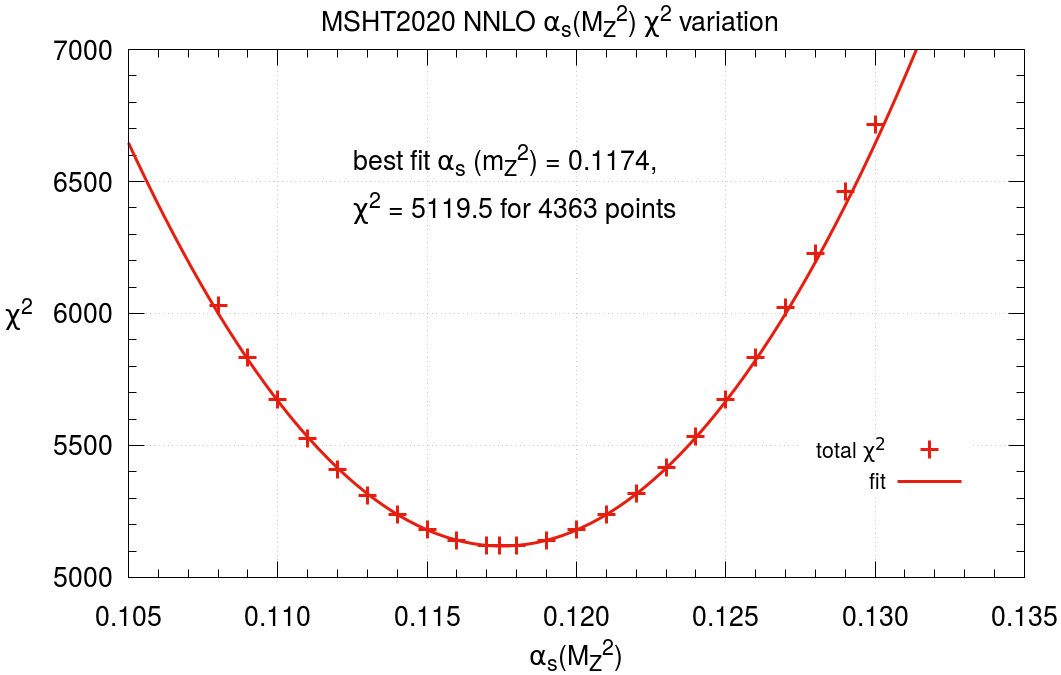}
\caption{\sf The left and right plots show total $\chi^2$ as a function of the value of the 
parameter $\a$ for the NLO (left) and NNLO (right) MSHT20 fits, respectively.}
\label{fig:total}
\end{center}
\end{figure}

In our default PDF study we fix the value $\a=0.118$ at NNLO, in order to be consistent with the 
world average value~\cite{PDG2020}. At NLO we consider the same value $\a=0.118$, and also use 
$\a=0.120$ since the best-fit 
value of $\a$ in NLO PDF studies consistently lies $\sim 0.002$ above that at NNLO \cite{MMHTas}.
In \cite{MSHT20} we noted, however, that as for the MMHT2014 PDFs the best fit value of 
$\a$ at NNLO is just a little below the default of $\a=0.118$ while at NLO it is again close to   
$\a=0.120$. Here we present the variation with $\a$ in more detail. 
At both NLO and NNLO we allow the value of $\a$ to vary as a free parameter in the global fit. 
The best values are found to be
\bea
\alpha_{S,{\rm NLO}}(M_Z^2) & = & 0.1203 \label{eq:optNLO}\\
\alpha_{S,{\rm NNLO}}(M_Z^2) & = & 0.1174\;. \label{eq:optNNLO}
\eea
The corresponding total
$\chi^2$ profiles versus $\a$ are shown in Fig.~\ref{fig:total}. The points indicate the fits performed with different fixed $\alpha_S(M_Z^2)$ values whilst the line represents a quadratic fit. These plots
indeed clearly show the reduction in the optimum value of $\a$ as we go from the NLO to
the NNLO analysis. It is also clear that the global $\chi^2$ shows a very good quadratic behaviour as a function of $\alpha_S(M_Z^2)$, even for the extreme $\alpha_S(M_Z^2)$ values taken well away from the best fits. We also provide in Table~\ref{tab:deltachisq_global} the $\Delta \chi^2$ values as one moves away from the best fit values of $\alpha_S(M_Z^2)$ of 0.1203 at NLO or 0.1174 at NNLO. In the next section we show how the individual data sets
contribute to produce this $\chi^2$ profile versus $\a$, and also determine the uncertainty on $\a$.

\begin{table}
\begin{center}
\renewcommand\arraystretch{1.25}

\begin{tabular}{|l|l|l|l|}
\hline
   $\alpha_S(M_Z^2)$       &  $\Delta \chi^2_{\rm global}(\rm NLO)$ &   $\Delta\chi^2_{\rm global}(\rm NNLO)$  
        \\ \hline
0.108 & 1188.6 & 909.6 \\ 
0.109 & 991.0 & 715.0 \\ 
0.110 & 813.6 & 553.1 \\ 
0.111 & 654.8 & 405.4 \\ 
0.112 & 556.5 & 290.0 \\ 
0.113 & 434.4 & 192.6 \\ 
0.114 & 324.5 & 118.2 \\ 
0.115 & 230.2 & 61.8 \\ 
0.116 & 151.7 & 21.8 \\ 
0.117 & 91.3 & 2.6 \\ 
0.1174 & - & 0 \\ 
0.118 & 50.3 & 2.7 \\ 
0.119 & 10.7 & 22.1 \\ 
0.120 & 1.1 & 61.1 \\ 
0.1203 & 0 & - \\ 
0.121 & 3.3 & 119.3 \\ 
0.122 & 27.1 & 197.9 \\ 
0.123 & 56.1 & 296.1 \\ 
0.124 & 110.8 & 414.4 \\ 
0.125 & 177.5 & 553.8 \\ 
0.126 & 257.8 & 715.0 \\ 
0.127 & 351.2 & 902.0 \\ 
0.128 & 469.0 & 1107.8 \\ 
0.129 & 602.0 & 1344.6 \\ 
0.130 & 748.6 & 1596.7 \\ 
\hline
    \end{tabular}
\end{center}

\caption{\sf The quality of the global fit versus $\alpha_S(M_Z^2)$ at NLO and NNLO relative to the best fits  at $\alpha_S(M_Z^2)=0.1203, 0.1174$ respectively. The number of data points in the global fit is 4363.}
\label{tab:deltachisq_global}   
\end{table}

It is a matter of debate whether one should actually extract the value of
$\alpha_S(M_Z^2)$ from PDF global fits or simply use a fixed value, i.e. 
the world average value~\cite{PDG2020}.
Our opinion has always been that a very accurate and precise value of the coupling can 
be obtained from PDF fits, and hence we have traditionally performed fits in order to determine 
this parameter and its uncertainty. Indeed the extracted value of $\alpha_S(M_Z^2)$ in the NNLO
MSHT analyses continues the trend of our extraction of being close to the world average of
$\alpha_S(M_Z^2)=0.1179\pm 0.001$~\cite{PDG2020}, and as in previous studies our NLO value is 
 a little higher, a result frequently seen in extractions of $\a$. Hence,
the result from our PDF fit is entirely consistent with the independent determinations of
the coupling. We also note that the quality of the global fit to the data increases by only 2.7
units in $\chi^2$ at NNLO when we move away from our absolute best-fit value to the default of 
$\a=0.118$, and so our default PDFs give an extremely good representation of our PDFs at the best
fit value of $\a$.
Hence, since for the use of PDF sets by external users
it is preferable to present PDFs at common (and hence `rounded')
values of $\alpha_S(M_Z^2)$ in order to compare and combine with PDF
sets from other groups, for example as in \cite{PDF4LHC1, PDF4LHC2, bench1, bench2,PDF4LHC15},
we continue to define those extracted for the choice $\a=0.118$ as the default. 
At NLO we also make a set available with the same 
$\alpha_S(M_Z^2)=0.118$ following the same reasoning, but in this case the $\chi^2$ increases by 
50 units from the best fit value. Therefore the default PDFs also contain the set at the round value of 
$\a=0.120$, extremely close to the best-fit value - differing by only 1.1 units of $\chi^2$. In \cite{MSHT20} we provided PDF sets
corresponding to the best fit for $\alpha_S(M_Z^2)$ values $\pm 0.001$ and $\pm 0.002$ relative
to the default values, in order for users to determine the
$\alpha_S(M_Z^2)$ uncertainty in predictions if so desired. Here we will extend the range of $\alpha_S(M_Z^2)$ values provided, and return to the issue of PDF+$\alpha_S(M_Z^2)$ uncertainty later.

\subsection{Description of data sets as a function of $\alpha_S(M_Z^2)$} \label{sec:desc}

The MSHT20 global analysis \cite{MSHT20} presented 
PDF sets at LO, NLO and NNLO in $\alpha_S$, where for NNLO we use the splitting functions calculated in~\cite{Moch:2004pa,Vogt:2004mw} and for structure function data, the massless coefficient functions calculated in~\cite{vanNeerven:1991nn,Zijlstra:1991qc,Zijlstra:1992kj,Zijlstra:1992qd,Moch:2004xu,Vermaseren:2005qc}. The fit is based on a fit to 61
different sets of data on deep--inelastic and  hard scattering
processes\footnote{Full data references can be found in \cite{MSHT20}.}. 
These comprise: 10 structure functions data sets from the
fixed--target charged lepton--nucleon experiments of the SLAC, BCDMS, NMC and
E665 collaborations; 6  neutrino data sets on $F_2,~xF_3$ and dimuon
production from the NuTeV, CHORUS and CCFR collaborations; 2 fixed-target Drell--Yan data
sets from E886/NuSea; eight  data sets from HERA involving the combined
H1 and ZEUS structure function data and heavy flavour structure function data;
 8 data sets from the Tevatron, namely
 measurements of inclusive jet, $W$ and $Z$ production by the CDF and D{\O}
collaborations; and finally, in a dramatic increase from the MMHT2014 study, 27 data sets from 
the ATLAS, CMS and LHCb collaborations at the LHC. 
The goodness--of--fit, $\chi^2_n$, for each of the data sets is given for the NLO and 
NNLO global fits in the
Tables 6 and 7 of~\cite{MSHT20}, and the $\chi^2$ definition is explained in Section
2.4 of the same article.  The references to all of the data that are fitted are also given 
in~\cite{MSHT20}.

\begin{figure} 
\begin{center}
\includegraphics[scale=0.22]{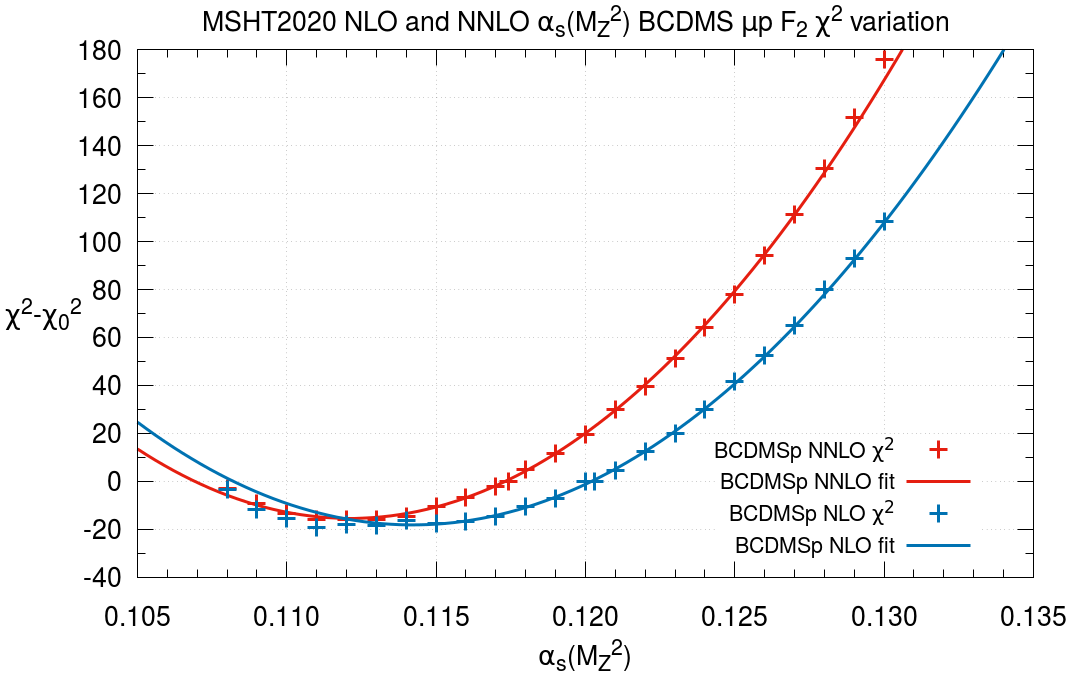}
\includegraphics[scale=0.22]{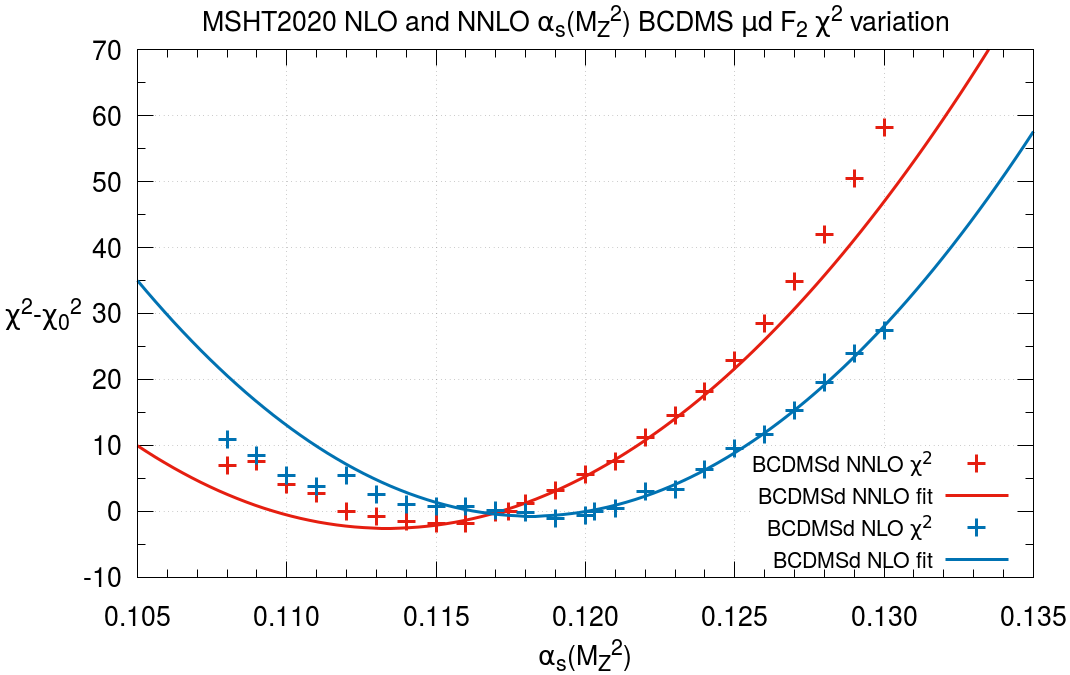}
\includegraphics[scale=0.22]{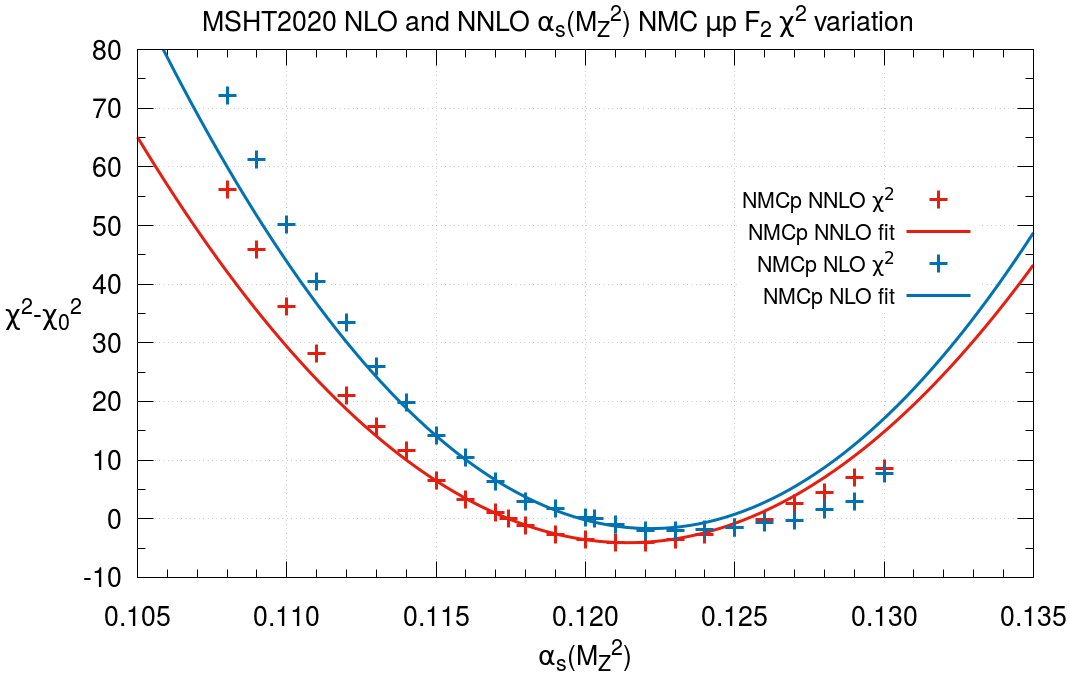}
\includegraphics[scale=0.22]{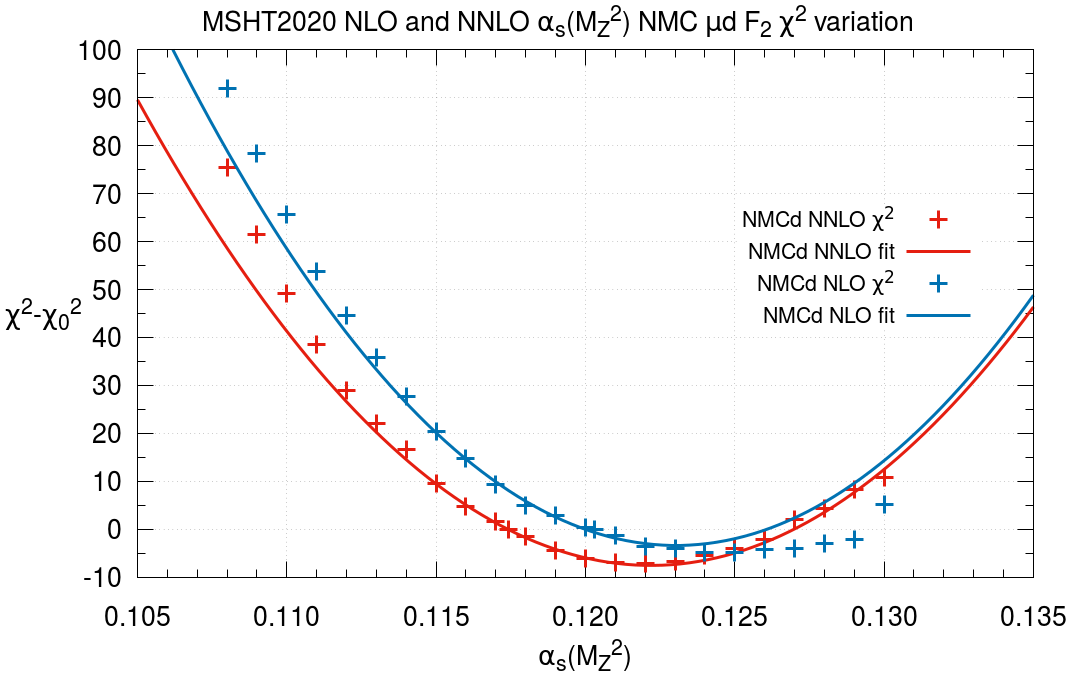}
\includegraphics[scale=0.22]{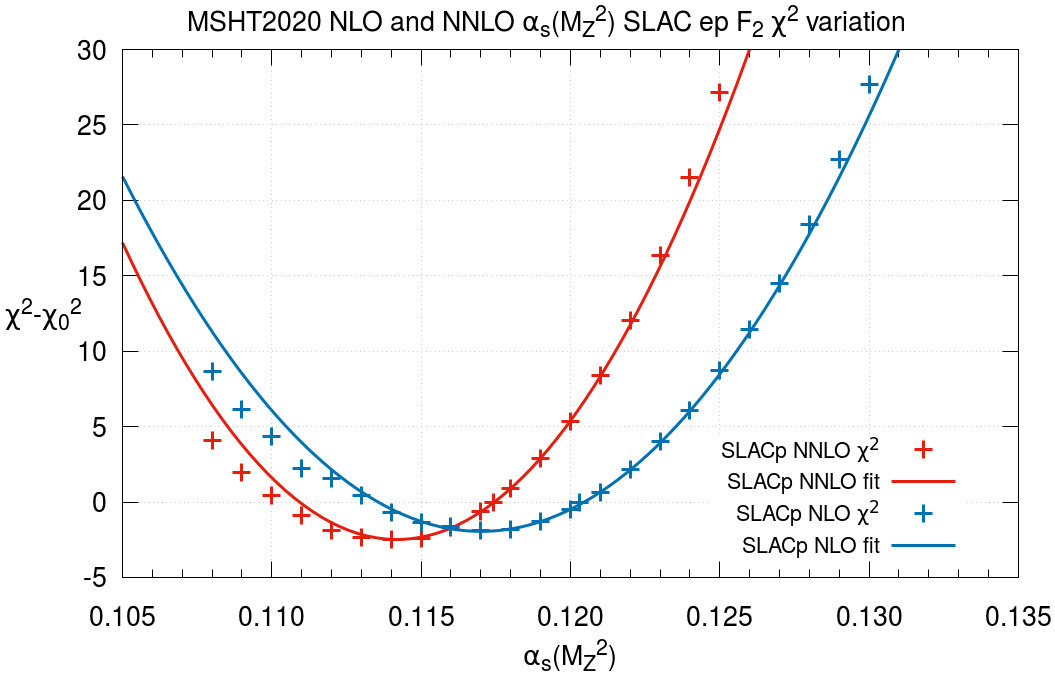}
\includegraphics[scale=0.22]{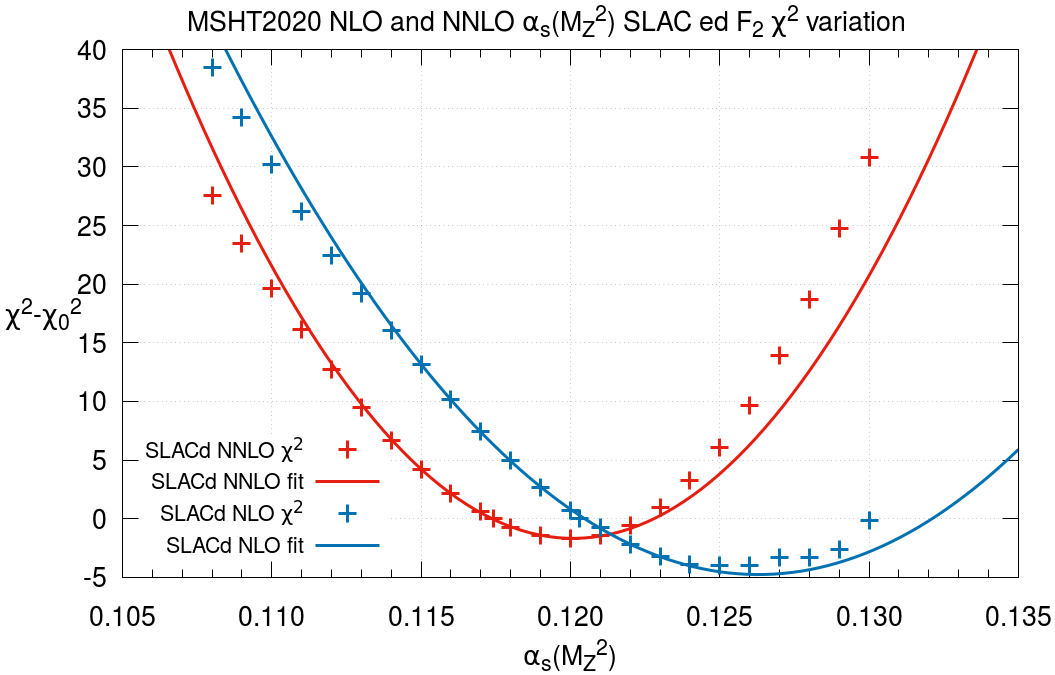}
\includegraphics[scale=0.22]{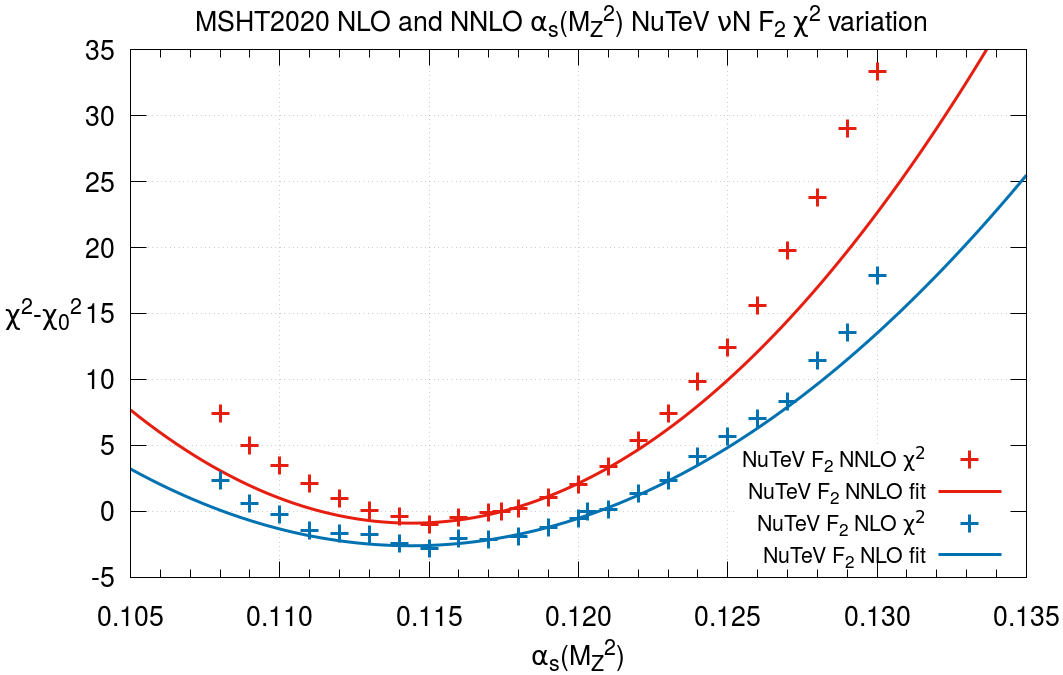}
\includegraphics[scale=0.22]{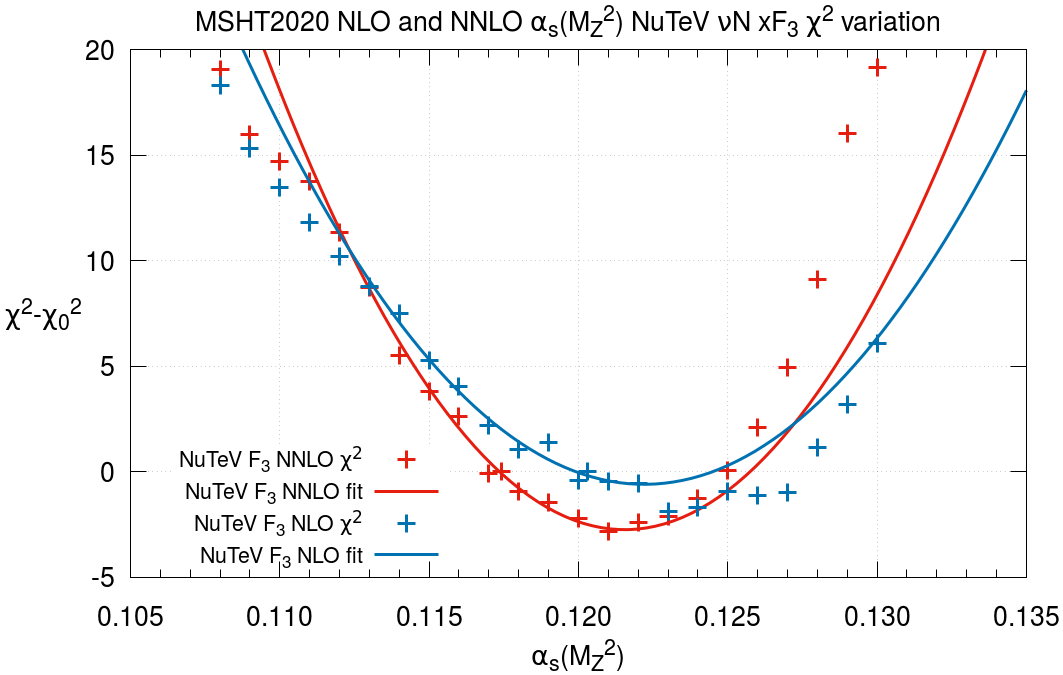}
\caption{\sf Difference in the $\chi^2$ relative to the $\chi_0^2$ obtained at the global best fit $\alpha_S(M_Z^2)$, as a function of the value of $\a$ for the NLO (blue) and NNLO (red) MSHT20 fits, respectively. The points are the results of the fits at a variety of fixed $\a$ values, whilst the curves are quadratic fits made to these in the vicinity of the central values. Here the most notable fixed target data sets are shown.} 
\label{fig:datasets1_alphas}
\end{center}
\end{figure}

\begin{figure} 
\begin{center}
\includegraphics[scale=0.22]{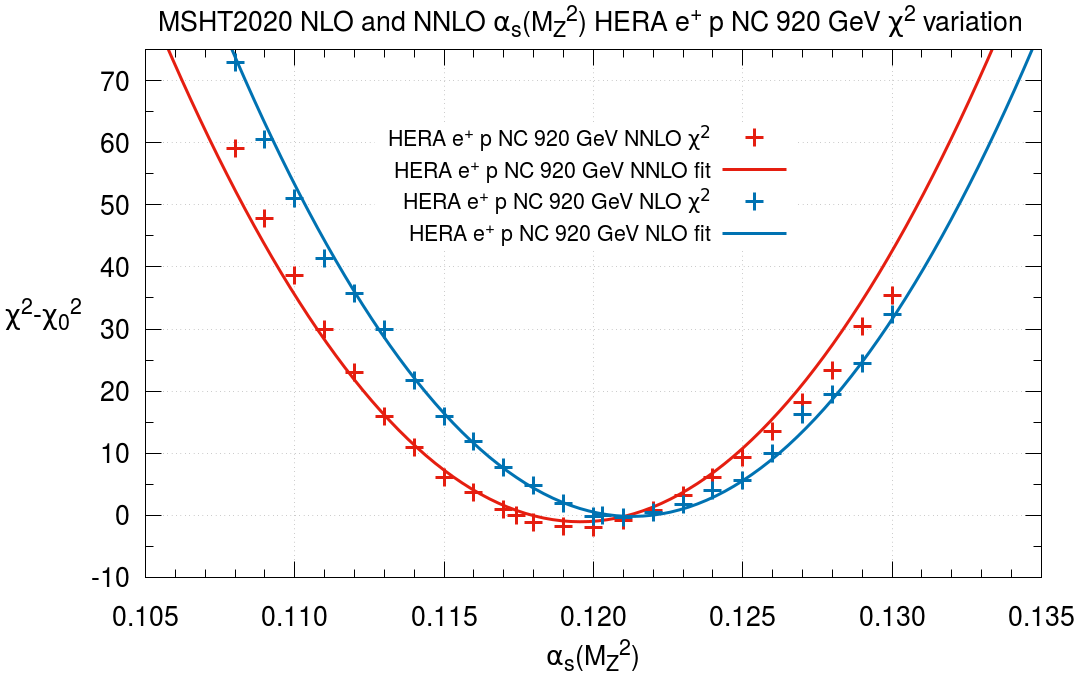}
\includegraphics[scale=0.22]{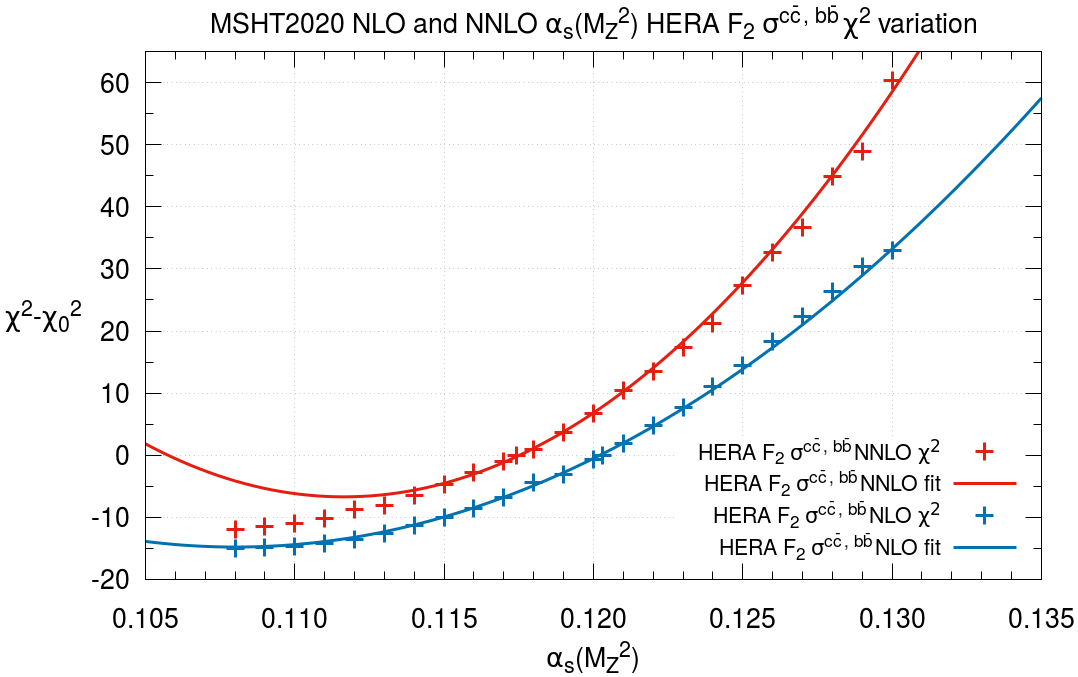}
\includegraphics[scale=0.22]{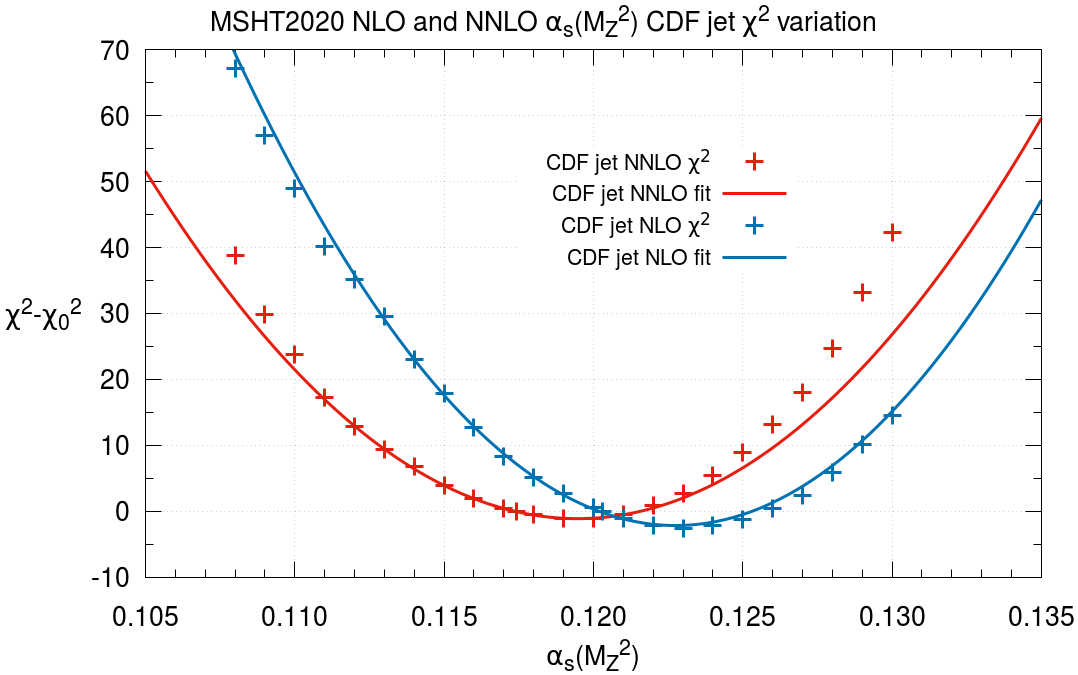}
\includegraphics[scale=0.22]{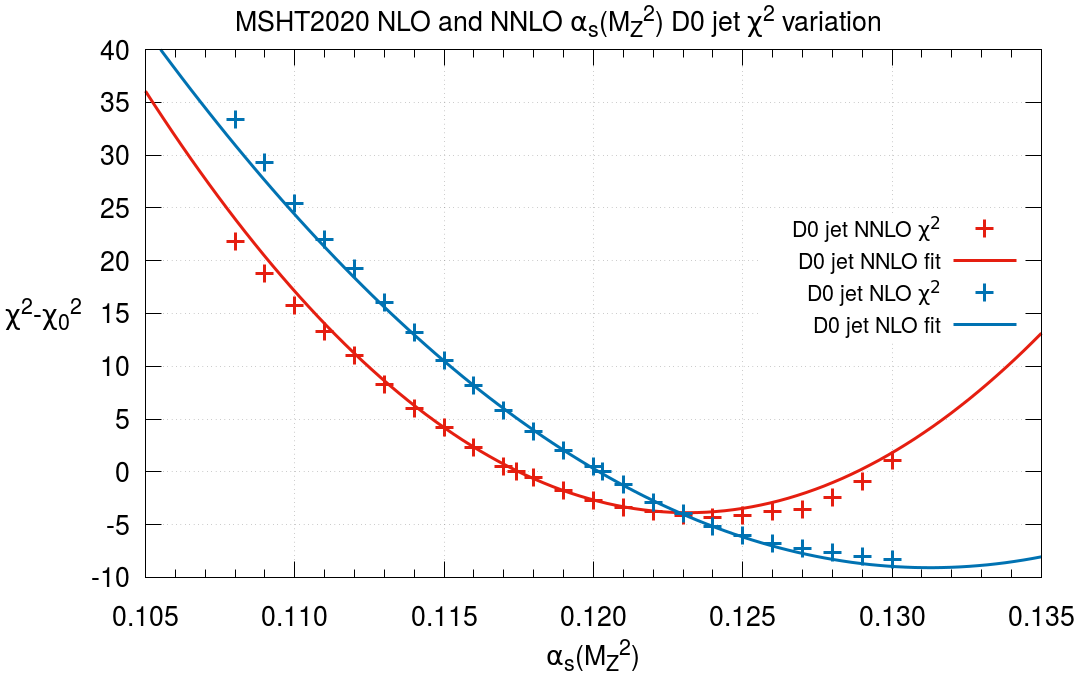}
\includegraphics[scale=0.22]{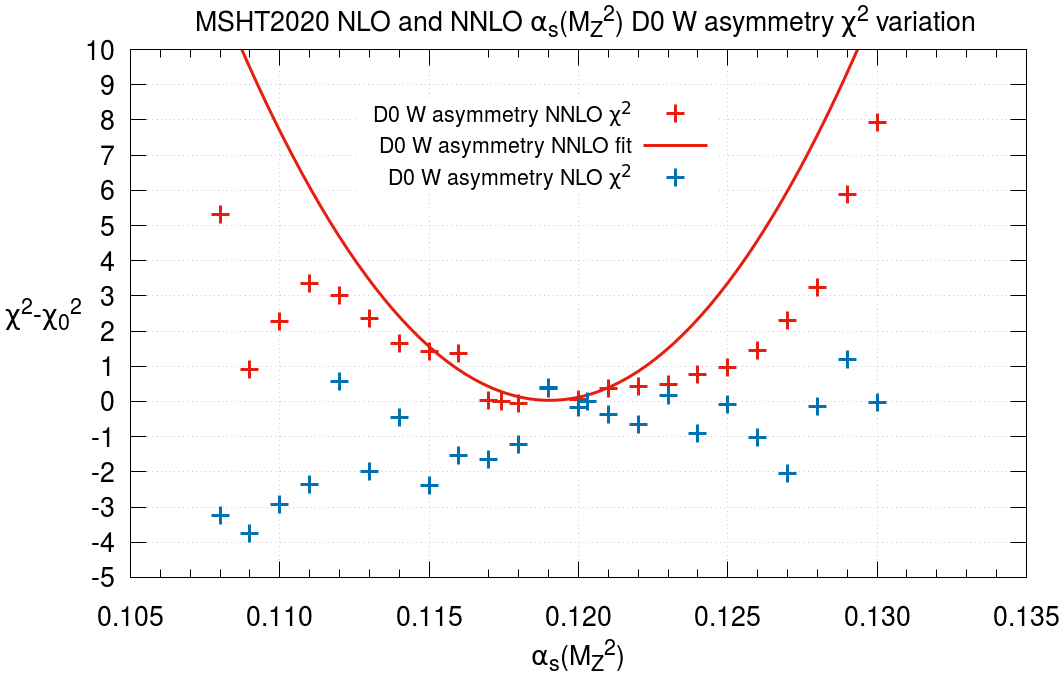}
\includegraphics[scale=0.22]{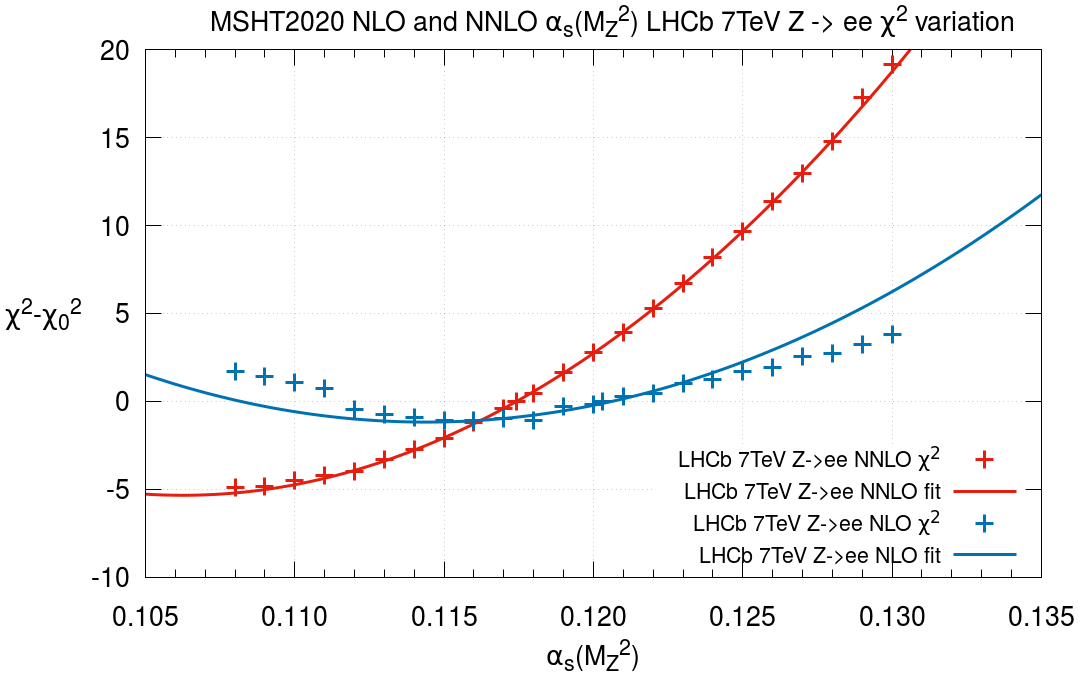}
\includegraphics[scale=0.22]{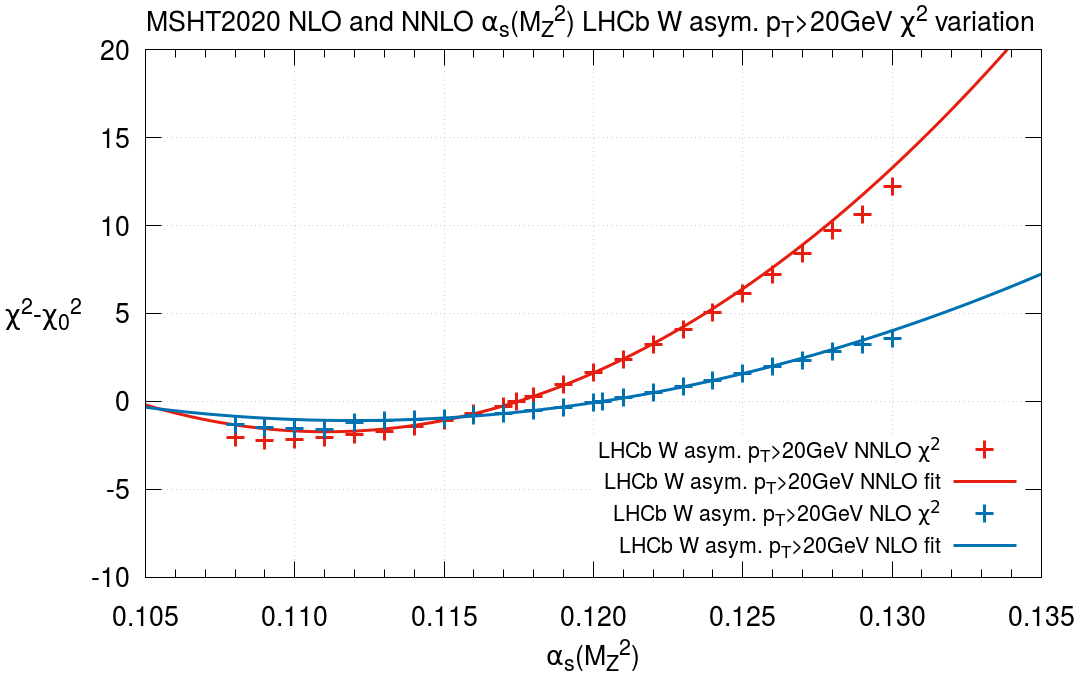}
\includegraphics[scale=0.22]{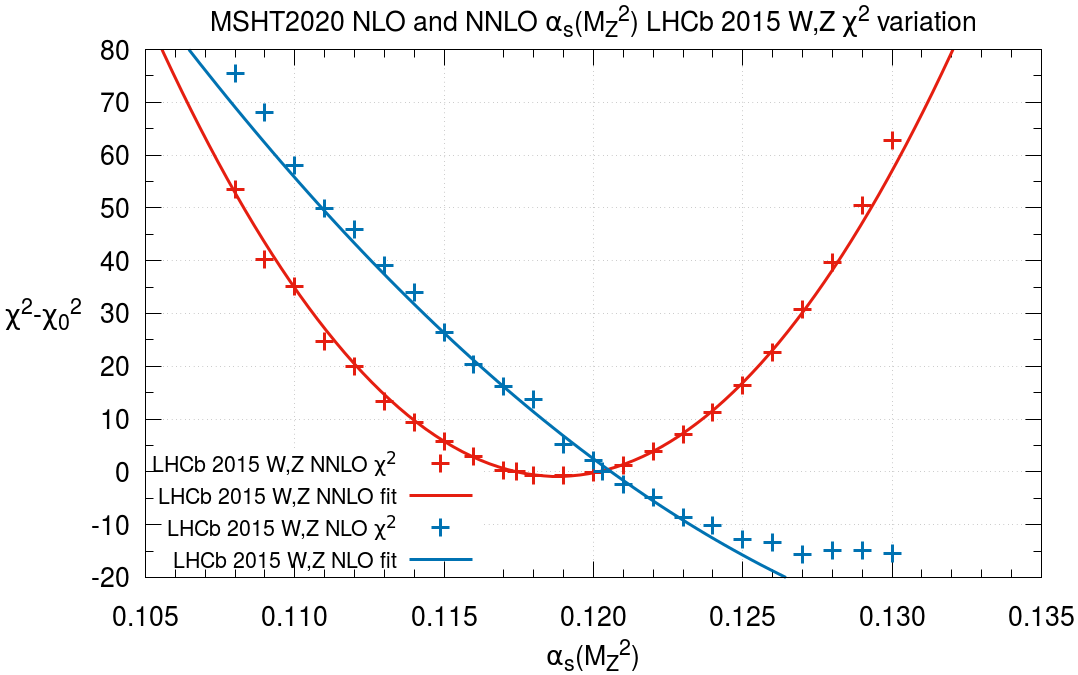}
\caption{\sf Difference in the $\chi^2$ relative to the $\chi_0^2$ obtained at the global best fit $\alpha_S(M_Z^2)$, as a function of the value of  $\a$ for the NLO (blue) and NNLO (red) MSHT20 fits, respectively. The points are the results of the fits at a variety of fixed $\a$ values, whilst the curves are quadratic fits made to these in the vicinity of the central values. Here the most notable HERA (first row), Tevatron (second row and third row left) and LHCb (third row right and bottom row) data sets are shown. }
\label{fig:datasets2_alphas}
\end{center}
\end{figure}

\begin{figure} 
\begin{center}
\includegraphics[scale=0.22]{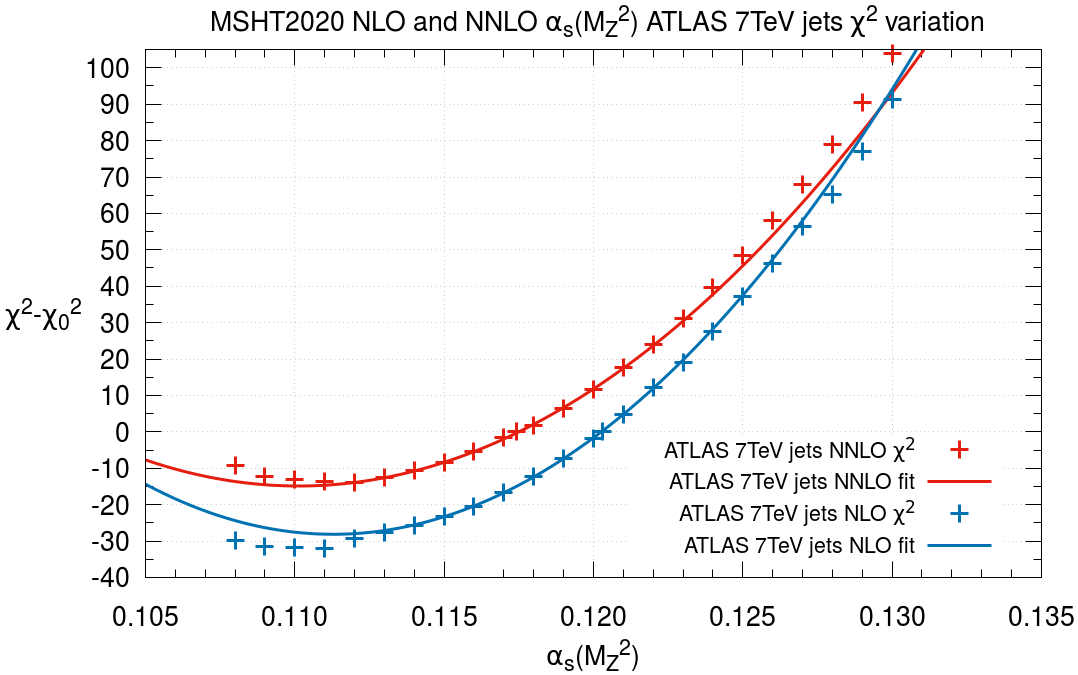}
\includegraphics[scale=0.22]{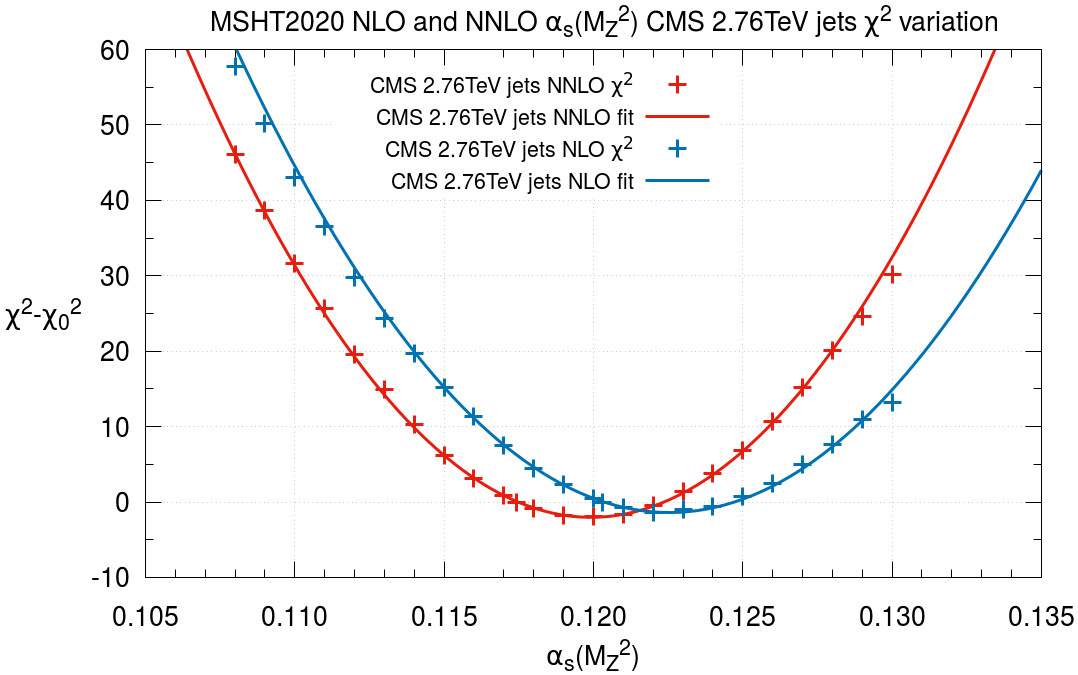}
\includegraphics[scale=0.22]{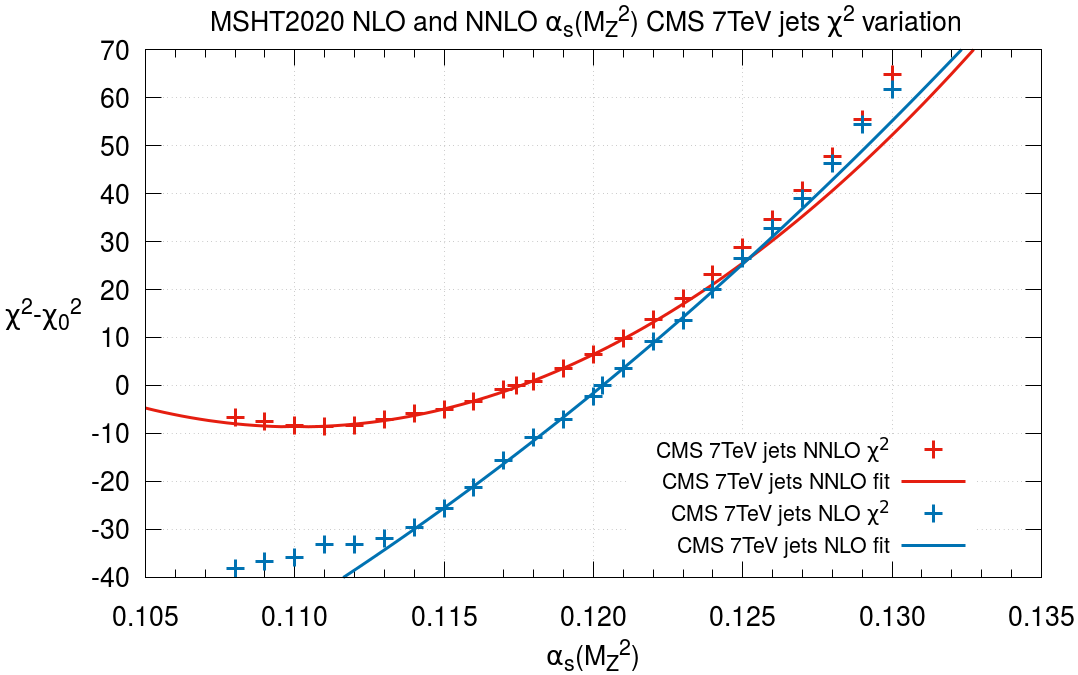}
\includegraphics[scale=0.22]{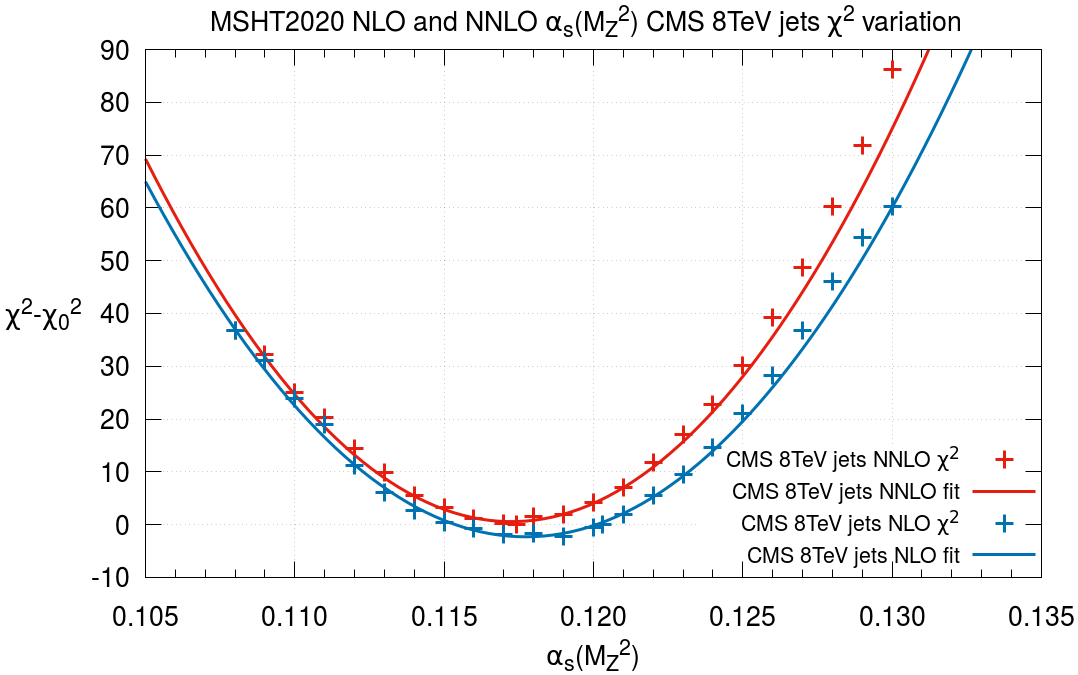}
\includegraphics[scale=0.22]{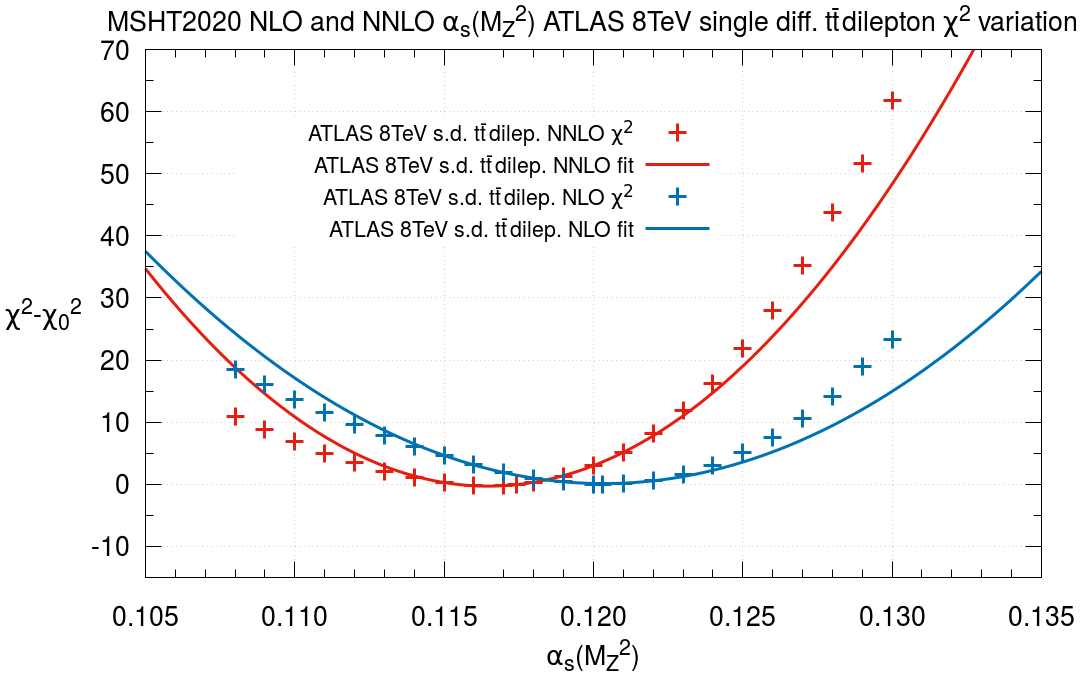}
\includegraphics[scale=0.22]{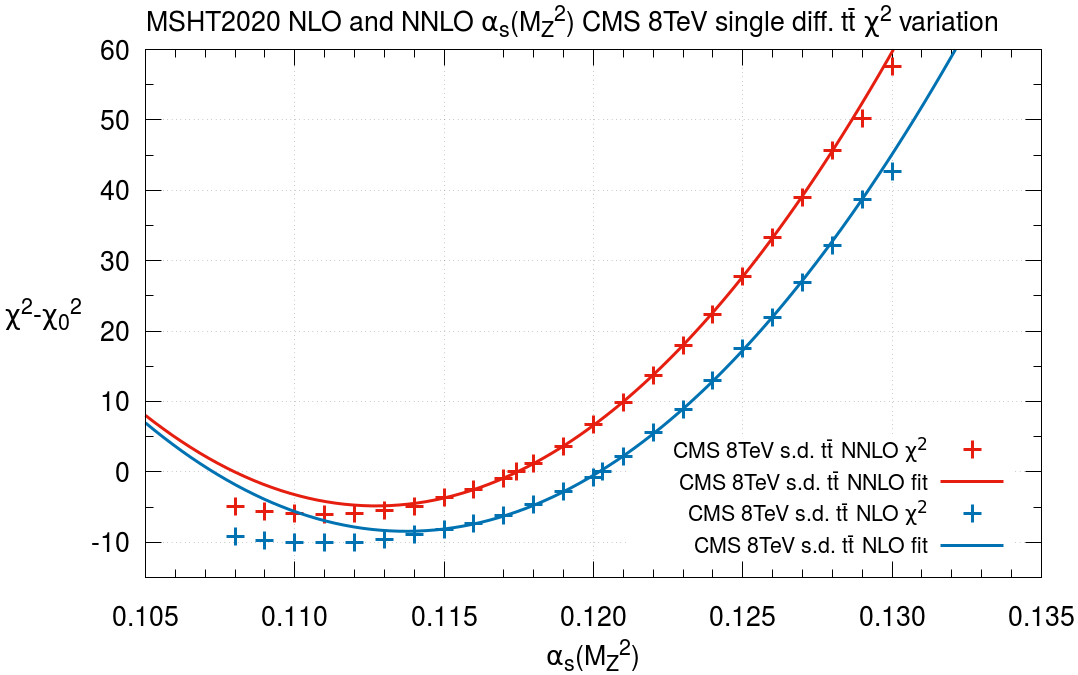}
\includegraphics[scale=0.22]{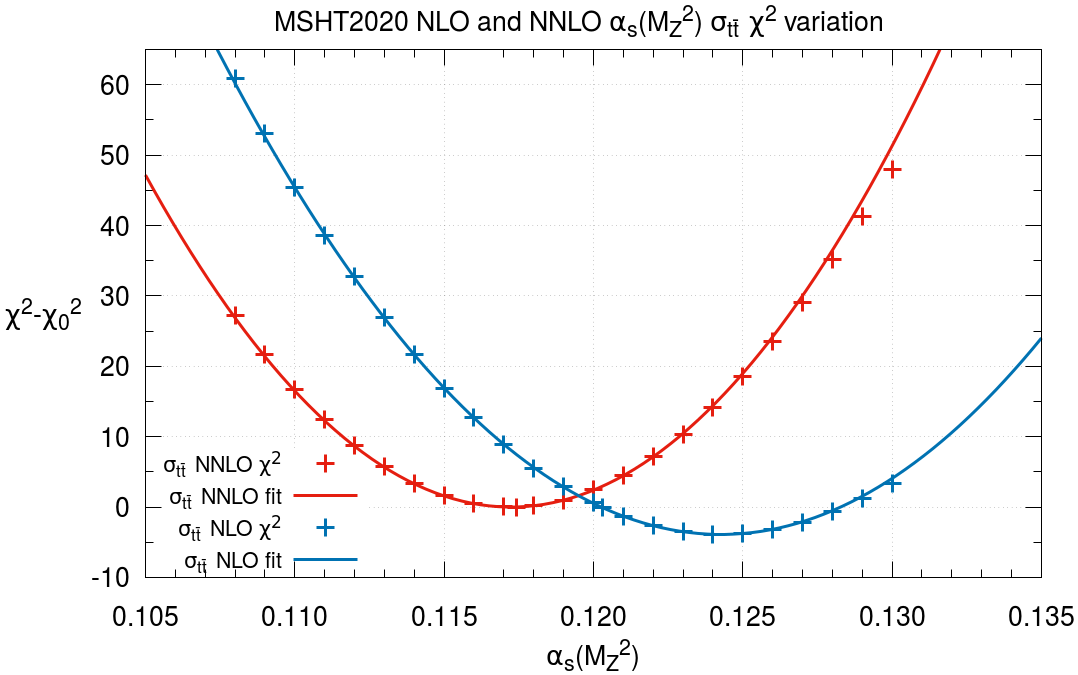}
\includegraphics[scale=0.22]{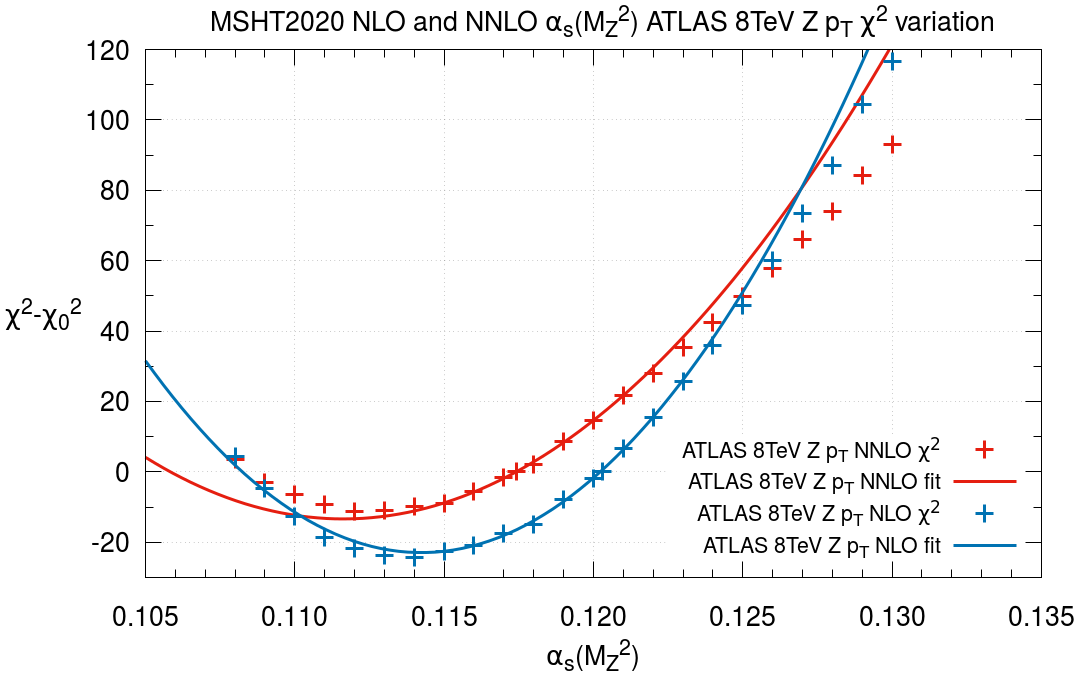}
\caption{\sf Difference in the $\chi^2$ relative to the $\chi_0^2$ obtained at the global best fit $\alpha_S(M_Z^2)$, as a function of the value of  $\a$ for the NLO (blue) and NNLO (red) MSHT20 fits, respectively. The points are the results of the fits at a variety of fixed $\a$ values, whilst the curves are quadratic fits made to these in the vicinity of the central values. Here the LHC data sets with direct sensitivity to $\a$ are shown, i.e. jet, $t\bar{t}$ and $Z$ $p_T$ datasets.}
\label{fig:datasets3_alphas}
\end{center}
\end{figure}

\begin{figure} 
\begin{center}
\includegraphics[scale=0.22]{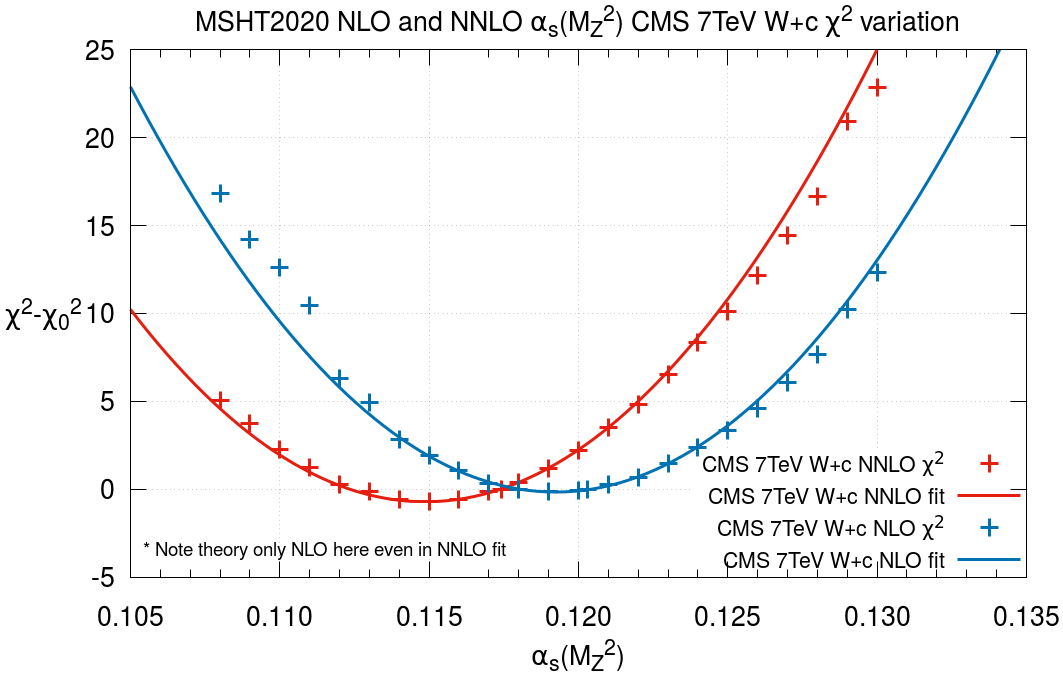}
\includegraphics[scale=0.22]{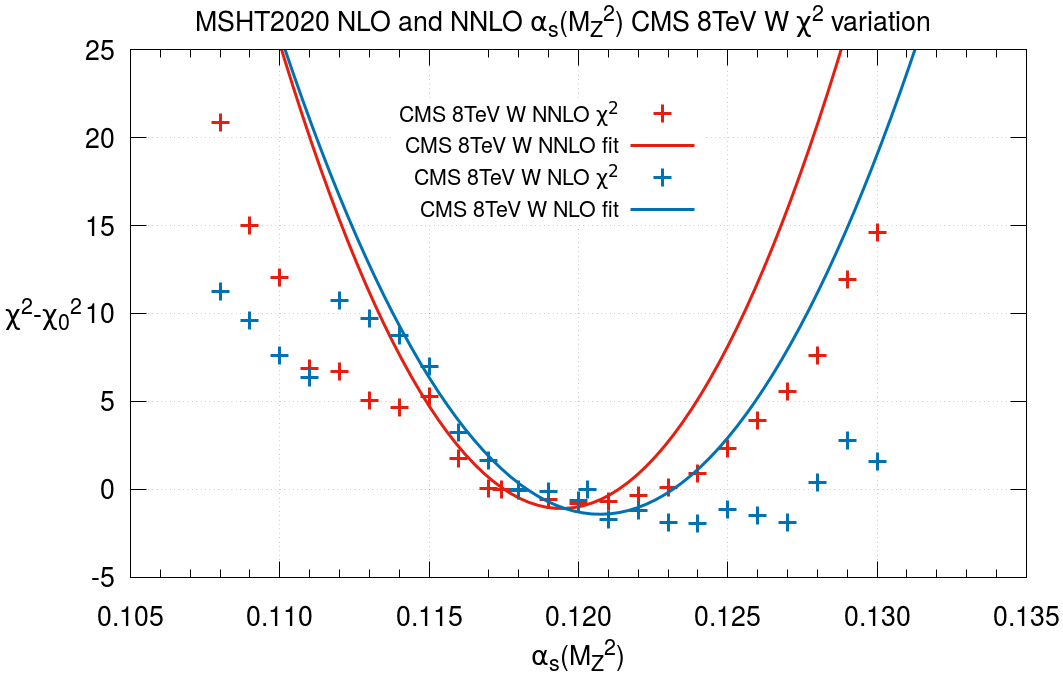}
\includegraphics[scale=0.22]{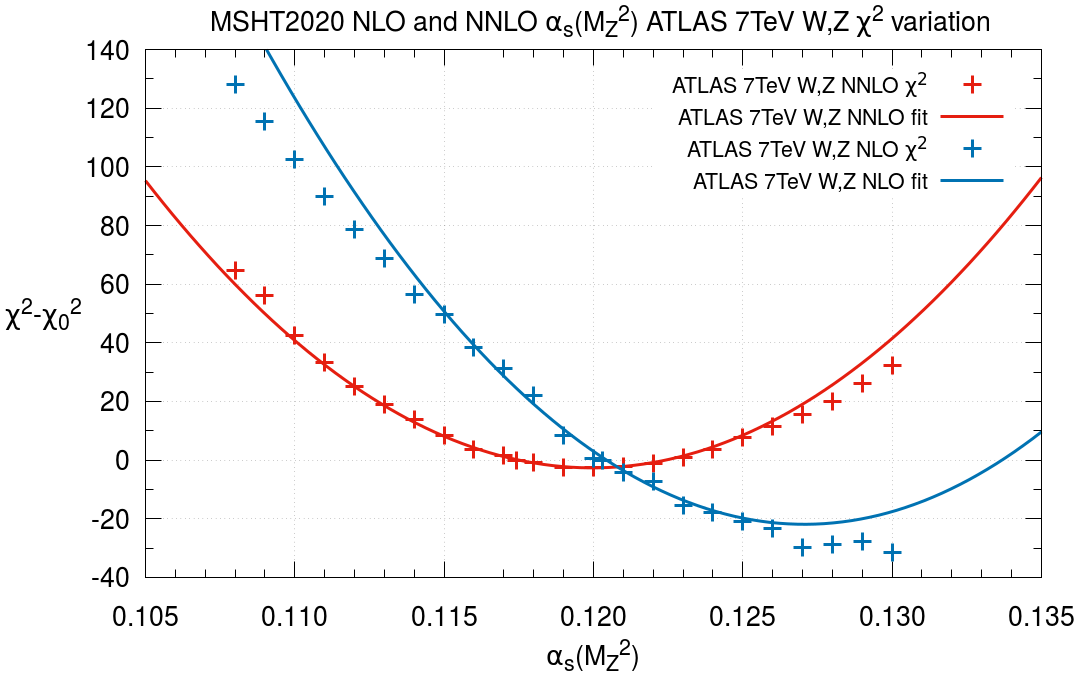}
\includegraphics[scale=0.22]{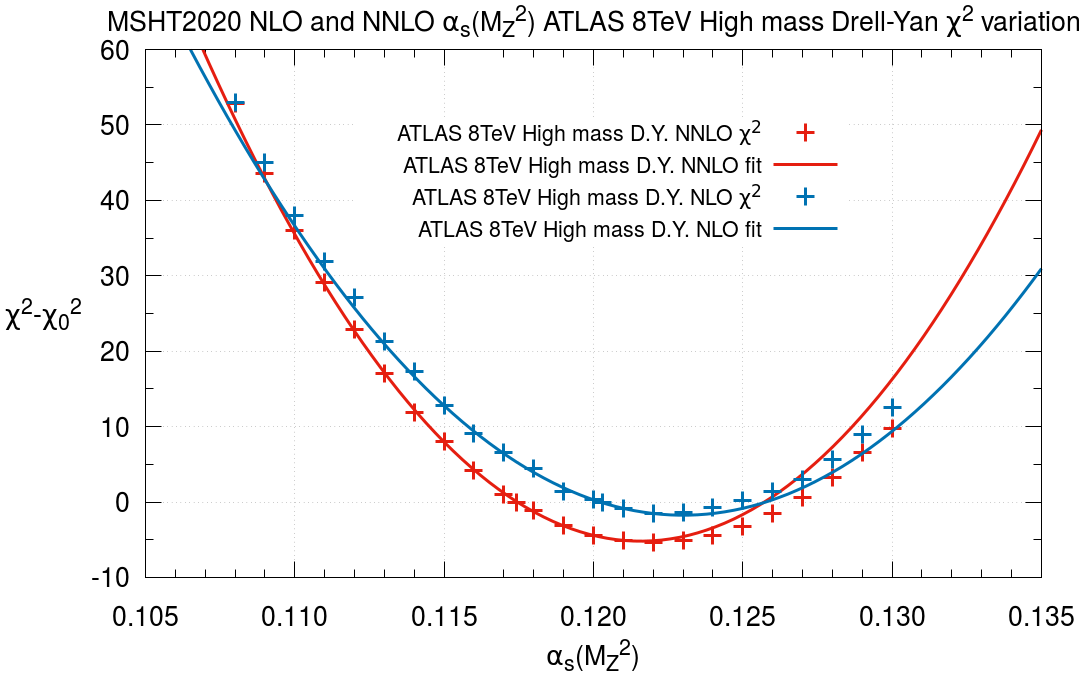}
\includegraphics[scale=0.22]{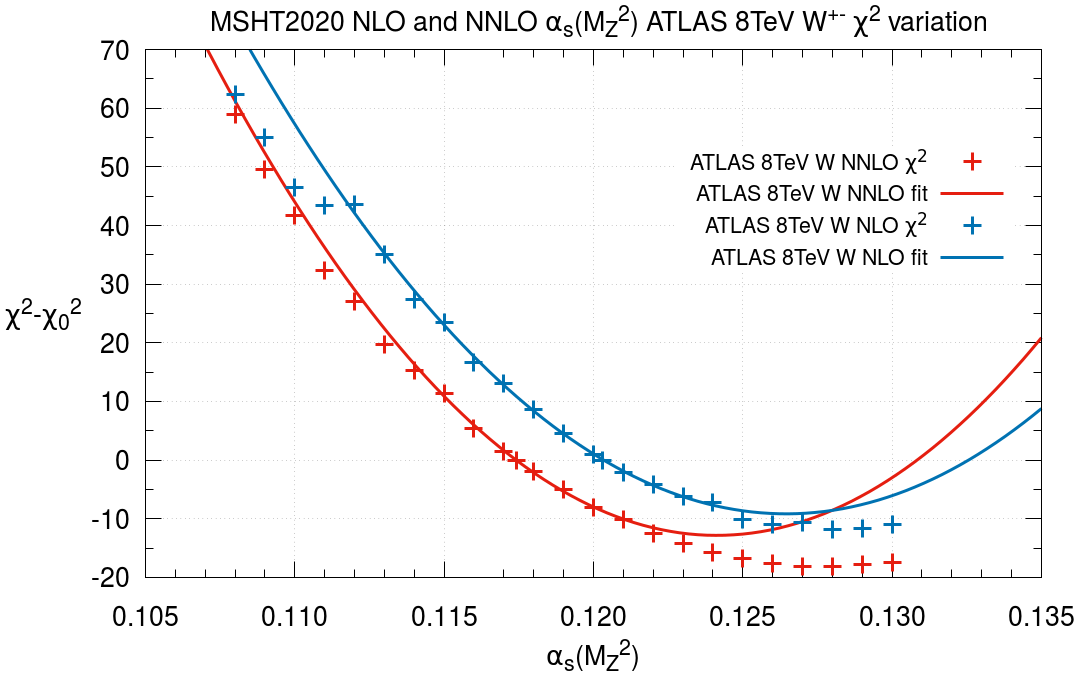}
\includegraphics[scale=0.22]{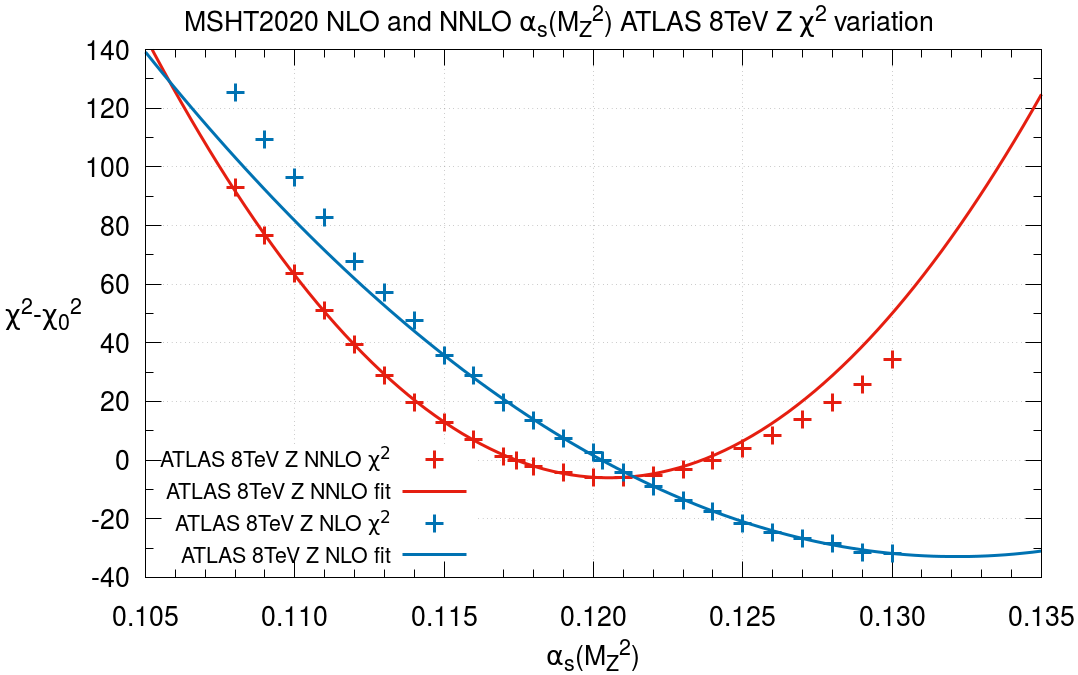}
\caption{\sf Difference in the $\chi^2$ relative to the $\chi_0^2$ obtained at the global best fit $\alpha_S(M_Z^2)$, as a function of the value of  $\a$ for the NLO (blue) and NNLO (red) MSHT20 fits, respectively. The points are the results of the fits at a variety of fixed $\a$ values, whilst the curves are quadratic fits made to these in the vicinity of the central values. Here the LHC data sets with indirect sensitivity (through their precision) to the value of $\a$ are shown, that is the 7 and 8~TeV $W,Z$ data sets. }
\label{fig:datasets4_alphas}
\end{center}
\end{figure}

For the NNLO global fit of~\cite{MSHT20}, we denote the contribution to the 
total $\chi^2$ from data set $n$ by $\chi^2_{n}$ and 
we investigate the $\chi^2_n$ profiles as a function of $\alpha_S(M_Z^2)$ by 
repeating the global fit for different fixed values of $\alpha_S(M_Z^2)$ in the 
neighbourhood of $\a=0.118$. The results 
for data sets which show significant dependence on $\a$ are shown in Figs.~\ref{fig:datasets1_alphas}~-~\ref{fig:datasets4_alphas}, where
we plot the $\chi^2_n$ profiles when varying 
$\alpha_S(M_Z^2)$ for data set $n$ as the difference from 
the value at the global minimum, $\chi^2_{n,0}$. Unlike in \cite{MMHTas} we do not show all 
data sets, largely because the number has now expanded significantly. 
The points (\textbf{$+$}) in Figs.~\ref{fig:datasets1_alphas}~-~\ref{fig:datasets4_alphas} are generated for fixed 
values of $\alpha_S(M_Z^2)$ between 0.108 and 0.130 in steps of 0.001.  These 
 are then fitted to a quadratic function of $\alpha_S(M_Z^2)$ over the central region of the $\a$ variation, shown by 
the continuous curves, and included as a guide to the eye.  The profiles satisfy 
$(\chi^2_n-\chi^2_{n,0})=0$ at $\alpha_S(M_Z^2)=0.1174$, corresponding to the 
value of $\alpha_S(M_Z^2)$ at the NNLO global minimum.  If all data sets behaved in the same manner 
with respect to $\a$  then each would show a quadratic minimum about this point.  Of course, 
in practice, the various data sets pull in varying degrees to smaller or larger values 
of $\alpha_S(M_Z^2)$. There is also some point--to--point fluctuation
for the values of $(\chi^2_n-\chi^2_{n,0})$, even near the minimum, but this is generally 
small. A small number of data sets show some non-quadratic behaviour, but these do not 
include the sets with the most significant dependence on $\a$. Note that the fact that the minimum of 
$\chi^2$ for an individual set within the global fit may be very different from the value of $\a$ 
preferred by the global fit highlights the issues in obtaining a value of $\a$ by comparison with 
a single data set using global fit PDFs, as discussed in \cite{Forte:2020pyp}.

We comment first in detail on the NNLO profiles, then we repeat this exercise at NLO, where the $(\chi^2_n-\chi^2_{n,0})$ profiles and fits are shown on the same figures (Figs.~\ref{fig:datasets1_alphas}~-~\ref{fig:datasets4_alphas}) as the NNLO. The profiles in this case satisfy  
$(\chi^2_n-\chi^2_{n,0})=0$ at $\alpha_S(M_Z^2)=0.1203$. We make fewer comments on the NLO profiles as it was clearly shown in \cite{MSHT20} that the NNLO fit is now preferred by the data, with NNLO needed to adequately describe many of the newer LHC data sets which became available for the MSHT20
analyses.

The fixed--target structure function data play an important role in 
constraining the value of $\alpha_S(M_Z^2)$,
and there is some tension between these data sets. These are shown in Fig.~\ref{fig:datasets1_alphas}. At NNLO the BCDMS data prefer values of 
$\alpha_S(M_Z^2)$ around 0.111 and 0.115 respectively for the $p$ and $d$ data, with both showing a clear preference for low $\alpha_S(M_Z^2)$ values. On the other hand, 
the NMC data prefer values around 0.121, whilst the SLAC  $F_2^{p,d}$ data prefer 
$\alpha_S(M_Z^2)$ values around 0.114 and 0.120 respectively. The BCDMS and SLAC $F_2^{d}$ data
come from similar regions of $x$ and $Q^2$ and are both subject to the deuteron corrections 
described in Section 2.2 of \cite{MSHT20}. These are determined from a function with 4 free parameters
which are allowed to vary, and have uncertainties, as $\a$ varies in the fits. However, the correction depends on $x$ only, 
so there is perhaps some indication from this observation of $\a$ difference between proton and 
deuteron that the deuteron correction may prefer also some $Q^2$ dependence.  
The neutrino 
$F_2$ and $xF_3$ data prefer $\alpha_S(M_Z^2) \sim 0.115$ and 0.121 respectively. Neutrino 
dimuon production (not shown) has little dependence on $\alpha_S(M_Z^2)$, since the 
extra $B(D\to \mu)$ branching ratio parameter, which we allow to vary with a 10\% uncertainty,  
can partially compensate for the changes in $\alpha_S(M_Z^2)$.

The combined H1 and ZEUS structure function data from HERA, shown in the top row of Fig.~\ref{fig:datasets2_alphas}, do not provide a strong constraint on $\a$,
though this is largely because the data play such a central role in the fit that the gluon distribution
varies with $\a$ in a manner very much aligned with keeping the fit to these data at its 
optimum. The most sensitive inclusive data, i.e. the 920 GeV $e^+ p$ data, mildly (given the very large number of points in this data set - 402) prefer a 
value of $\alpha_S(M_Z^2)$ of about 0.120 at NNLO, while the heavy flavour structure function data 
prefer a quite low value of $\alpha_S(M_Z^2)$ at NNLO. 
These data, particularly at low $Q^2$, are sensitive to the value of the charm 
mass $m_c$, and there is some correlation between its value and
$\a$, as discussed later in Section~\ref{Sec:Heavy_quark_mass_dependence}.

The Tevatron data sets with the most interesting effects and largest constraints on the value of $\alpha_S(M_Z^2)$ are shown in the second row and third row (left) of Fig.~\ref{fig:datasets2_alphas}. The Tevatron data consist largely of $Z$-rapidity data and charge--lepton asymmetry measurements 
arising from $W^{\pm}$ production, which are a ratio of cross sections, 
and therefore generally have little dependence on the value of $\alpha_S(M_Z^2)$.  
However, the latest D{\O} $W$-asymmetry, which is the most precise of these measurements, does have some limited sensitivity at NNLO, at least in the vicinity of the global best fit $\a$ values, with a preferred value very close to the global best-fit value. Nonetheless, given its small number of data-points and small $\chi^2$, the profiles are subject to significant noise, whilst the quadratic behaviour clearly reduces significantly away from these central $\a$ variations. The most constraining data sets from the Tevatron are therefore, predictably, the jet data sets. The 
CDF jet data prefer a slightly higher value, near $\a \sim 0.119$, at NNLO and disfavour low values, while the D{\O} jet data have a clear preference for higher values near $\a \sim 0.124$. Tevatron data on the total top pair production cross-section are also a part (along with LHC data) of the $\sigma_{t\bar{t}}$ dataset, whose $\chi^2$ variation is given in the bottom left of Fig.~\ref{fig:datasets3_alphas}, which will be commented on later.

The big change since the MMHT2014 analysis in terms of data is the number of high precision 
LHC data sets and the variety of NNLO cross sections which can be used for these in the PDF 
analysis. Some of these display direct sensitivity to $\a$, e.g. inclusive jet cross sections, data
on top-pair production and the $p_T$ distribution of the $Z$ boson (all shown in Fig.~\ref{fig:datasets3_alphas}), while others, e.g. 
determinations of the rapidity dependence of $W$ and $Z$ boson production have relatively indirect
dependence on $\a$, but provide constraints due to their extreme precision. These are shown largely in Fig.~\ref{fig:datasets4_alphas}, with the LHCb data in the bottom two rows of Fig.~\ref{fig:datasets2_alphas}. Again, we only show those LHC data sets with 
clear sensitivity to $\a$. Note, however, that in some cases there is little apparent sensitivity since, similarly to the HERA data, the best fit PDFs change with $\a$ in such a manner 
as to compensate the direct $\a$ dependence.  
  
In general there are two contrasting trends in the LHC data. Overall, those data sets with direct 
dependence on $\a$ in their cross sections tend to more frequently prefer lower values of $\a$, as is clear in Fig.~\ref{fig:datasets3_alphas}. 
This includes the ATLAS and CMS 7~TeV inclusive jet data,  the ATLAS 8~TeV $Z$ $p_T$ data, 
the ATLAS $t \bar t$ single differential dilepton data and CMS $t \bar t$ single differential data
and CMS $W+c\,\,$ jet data. Note, however, that the differential $t \bar t$ data (third row of Fig.~\ref{fig:datasets3_alphas}) is calculated with a 
fixed value of $m_t=172.5~\GeV$, whereas the total cross section (bottom row left of Fig.~\ref{fig:datasets3_alphas}) allows the mass to vary (with a penalty), and the best fit value varies as $\alpha_S(M_z^2)$ does.  In addition, the  $W+c\,\,$ jet 
cross section calculation (shown in Fig.~\ref{fig:datasets4_alphas} top left) is at NLO. 
In detail, the ATLAS 7~TeV jets data, CMS 7~TeV jets data, CMS 8~TeV single differential $t\bar{t}$ data, and ATLAS 8~TeV $Z$ $p_T$ data at NNL0 all prefer $\a$ values in the $\sim 0.111-0.112$ range.
Meanwhile some of the other such data sets, including perhaps the most constraining jet data set
in the fit, the 8~TeV CMS inclusive jet data, prefers an $\a$ value near the best fit value. In that case the $\chi^2$ in the vicinity of the best fit depends
relatively weakly on $\a$ compared to the large number of points, perhaps suggesting the high-$x$ gluon varies in such a way as to moderate the
$\a$ dependence for this data set. The 2.76~TeV CMS inclusive jet data, ATLAS 8~TeV single differential $t\bar{t}$ dilepton data, Tevatron and LHC total $t\bar{t}$ cross-section data also prefer a moderate $\a$ value. Finally, the CMS 7~TeV $W+c\,\,$ jet data shown in the top left of Fig.~\ref{fig:datasets4_alphas} also support a slightly lowered value of $\a \sim 0.115$.

In contrast, the precision $W,Z$ data from ATLAS and CMS, shown in the remainder of Fig.~\ref{fig:datasets4_alphas}, tend to 
prefer higher values of $\a$. At NNLO these tend to be only slightly raised relative to the best fit, but the trend at NLO is both clearer and more noticeable. 
In particular the CMS 8~TeV $W$ data,  ATLAS high-precision 7~TeV $W,Z$ data,  ATLAS 8~TeV High-mass Drell-Yan and ATLAS 8~TeV $Z$ data $\chi^2$ profiles minimise in the $\a \sim 0.120$ region, while for the ATLAS 8~TeV $W$ data this occurs in the $\a \sim 0.128$ region, though  in some cases the profiles are rather flat in the vicinity of the minima.

Given the relatively weak dependence of the cross section to $\a$ of these data sets the effect is most likely due to the 
manner in which $\a$ affects the evolution of the quark and anti-quark PDFs. Some of the high-rapidity LHCb data prefers lower values, for example the LHCb 7~TeV $Z \rightarrow ee$ and the LHCb $W$ asymmetry $p_T>20$ GeV data shown in the bottom two rows (third row right and bottom row left) of Fig.~\ref{fig:datasets2_alphas} prefer values of $\a \sim 0.108, 0.109$ at NNLO. On the other hand the LHCb 2015 $W,Z$ data prefer $\a \sim 0.119$ (bottom row right of Fig.~\ref{fig:datasets2_alphas}) and the LHCb 8~TeV $Z \rightarrow ee$ (not shown here) have a slight preference for the best fit $\a$ value. In any case these LHCb data sets have less constraining power. 
Hence, as with other general types of data, the LHC data do not 
provide a consistent trend for either low or high $\a$ compared to the overall best fit, as different 
individual data sets can pull in different directions. 

At NLO similar conclusions for the $\a$ pulls of the different data set types and individual data sets can be drawn. For the structure function data the NNLO corrections to the structure functions are positive and speed up the evolution. In order to compensate for these missing corrections,  the optimum values of $\alpha_S(M_Z^2)$ is generally larger than  at NNLO. The difference $\alpha_{S,{\rm NNLO}}<\alpha_{S,{\rm NLO}}$ is clearly evident in the majority of the corresponding plots in Fig.~\ref{fig:datasets1_alphas}. The behaviour of the HERA data sets (Fig.~\ref{fig:datasets2_alphas} top row) is similar to that at NNLO, with the heavy flavour structure function data again preferring a low value of $\alpha_S(M_Z^2)$. As for the Tevatron jet data, both the CDF and D{\O} jet data prefer a higher value of $\a$ at NLO than NNLO, in order to compensate for the missing positive NNLO corrections. The D{\O} $W$ asymmetry data on the other hand at NLO are completely dominated by fluctuations due to the small number of points and weak dependence on $\a$. As a result we plot no quadratic fit in this case. For this dataset, whilst a preference for $\a$ around the best fit value could be seen for NNLO, this is no longer the case at NLO, with no clear trend obvious. The LHC data sets directly sensitive to $\a$, such as the jets, $t\bar{t}$ and $Z$ $p_T$ (Fig.~\ref{fig:datasets3_alphas}) generally show a similar behaviour at NLO to NNLO, mainly preferring lower values of $\a$ and doing so for the same data sets as reported at NNLO. As was the case for the Tevatron jets, for these LHC data sets there is perhaps a slight tendency for slightly increased values of $\a$ being preferred relative to NNLO. One of the most notable examples of this is the total $t\bar{t}$ cross-section data (from Tevatron and LHC) at NLO preferring a value around 0.124, compared to its preference for the best fit value at NNLO, however in this specific case this is the result of a poor fit at NLO, reflecting large NNLO corrections missing at that order. Finally, we comment on the precise LHC Drell-Yan data sets of Fig.~\ref{fig:datasets4_alphas} at NLO. We have seen in \cite{MSHT20} that several of these, particularly the ATLAS 7 and 8~TeV precision $W$, $Z$ data sets, are poorly fit at NLO, clearly preferring NNLO, therefore few conclusions can be drawn at NLO as a result. This is evidenced by the significantly different behaviour of the $\chi^2$ profiles with $\a$ for these data sets at NLO relative to NNLO. Several data sets can also be observed to show less quadratic behaviour at NLO than NNLO, for example not just the D{\O} $W$ asymmetry but also the LHCb 7~TeV $Z \rightarrow ee$, LHCb 2015 $W,Z$, CMS 7~TeV jets, ATLAS 7~TeV $W,Z$ and ATLAS 8~TeV $Z$ data sets.

\subsection{The best fit values and uncertainty on $\alpha_S(M_Z^2)$ \label{sec:uncsq}}

\begin{figure} 
\begin{center}
\includegraphics[scale=0.4]{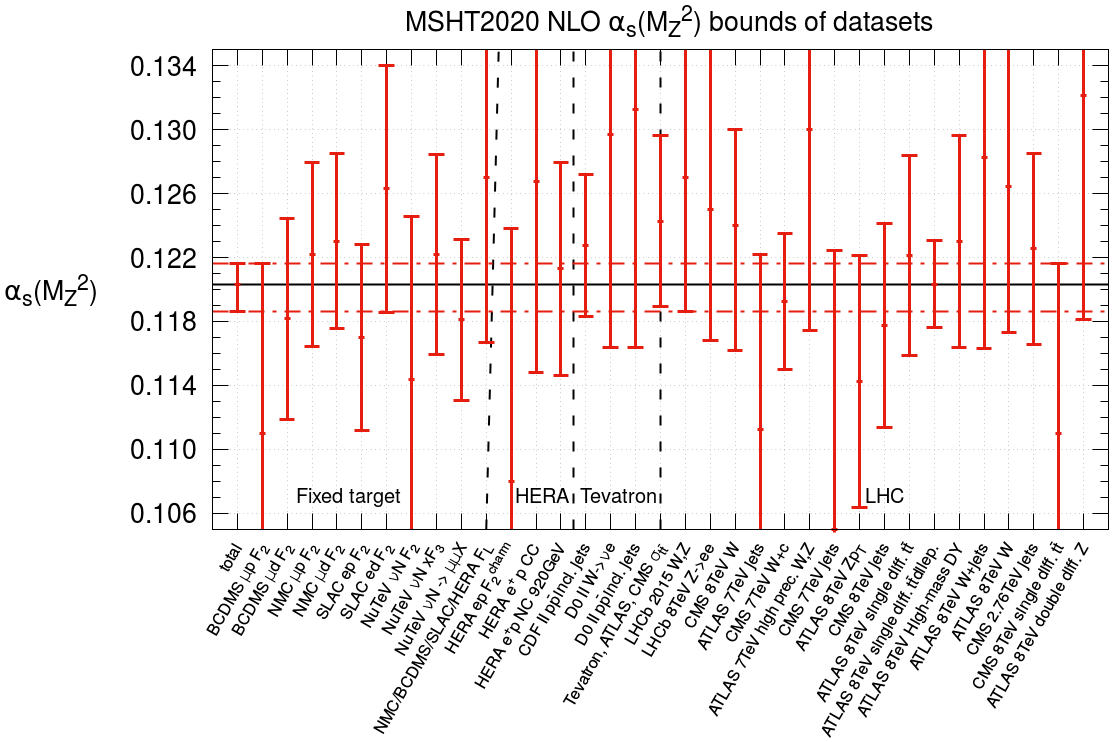}
\includegraphics[scale=0.4]{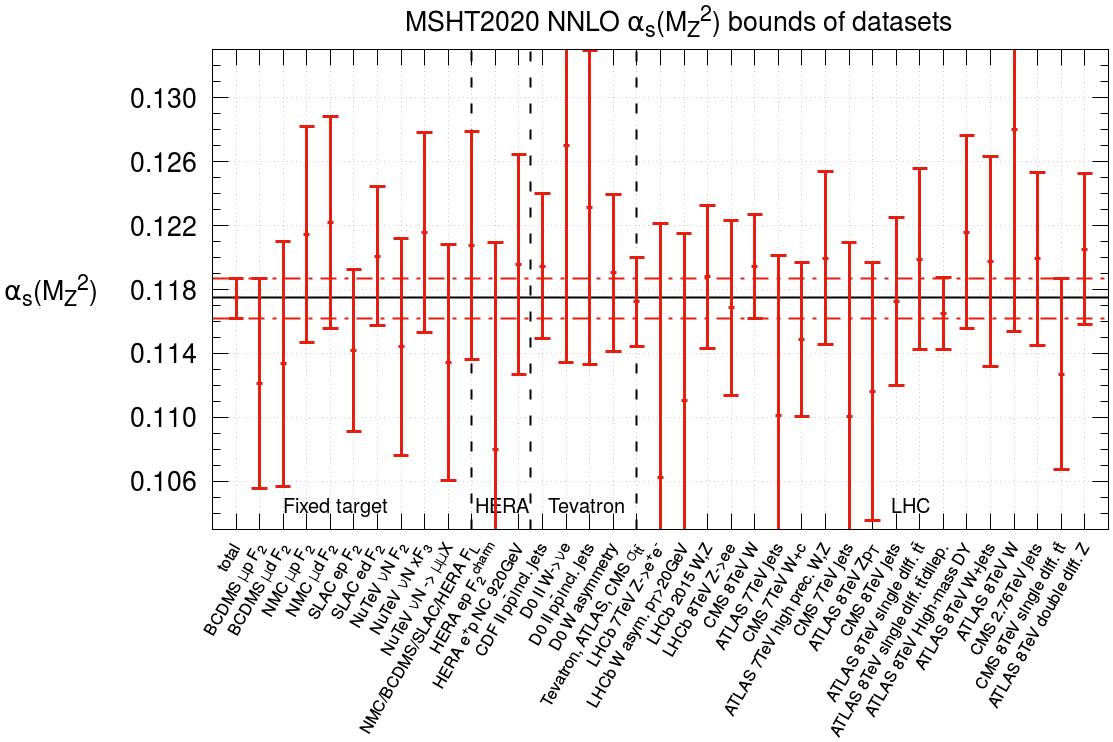}
\caption{\sf The upper and lower plots show the value of $\alpha_S(M_Z^2)$ corresponding to the best fit, together with the upper and lower 1$\sigma$ constraints on $\alpha_S(M_Z^2)$ from the more constraining data sets at NLO and NNLO respectively. The overall upper and lower bounds taken are given by the horizontal dashed red lines.}
\label{fig:datasets_alphasbounds}
\end{center}
\end{figure}

Ever since the MSTW08 analysis of PDFs \cite{MSTW} we have determined the uncertainties of the PDFs 
using the Hessian approach with a dynamical tolerance procedure. That is, we obtain PDF `error' 
eigenvector sets, each corresponding to 68$\%$ confidence level uncertainty, where the eigenvectors 
are orthogonal to each other and span the PDF parameter space. As in the MSTW and MMHT studies, we  
again determine the uncertainty on $\alpha_S(M_Z^2)$ at NLO and NNLO 
by using the same technique as for the PDF eigenvector uncertainties, i.e. 
we apply the tolerance procedure to determine the uncertainty in each direction away from
the value at the best fit when one data set goes beyond its $68\%$ confidence
level uncertainty. The values at which each data set does reach its 
$68\%$ confidence level uncertainty, and the value of $\alpha_S(M_Z^2)$ 
for which each data set has its best fit (within the context of a global fit)
are shown at NLO and NNLO in Fig.~\ref{fig:datasets_alphasbounds}, where we include only the most constraining data sets. 

The dominant constraint on $\a$ in the downwards direction at 
NNLO is the CMS 8~TeV data on the $W$ rapidity distribution, which gives 
$\Delta \a = -0.0013$, though this is closely followed by the 
ATLAS 8~TeV $Z$ data, the SLAC deuteron data and ATLAS 8~TeV high-mass Drell Yan data, which give $\Delta \a = -0.0017,-0.0018,-0.0019$ respectively.   
In the upwards direction, at NNLO, the BCDMS proton structure function
data give $\Delta \a = +0.0012$. 
This is very closely followed by the CMS $t \bar t$ single differential data and ATLAS $t \bar t$ single differential dilepton 
data, though both of these data sets
have the caveat that the top mass is fixed, as discussed above. A bound of 
$\Delta \a = +0.0018$ is provided by the SLAC proton structure function data. In 
both directions there are other data sets that are not much less constraining than 
those mentioned explicitly. Hence, it is far from being a single data set which is 
overwhelmingly providing a dominant constraint on the upper or lower limit on $\a$. 

The dominant constraint at NLO in the downwards direction is from the 
top pair cross section data and, using the dynamical tolerance procedure, this gives 
an uncertainty of $\Delta \a =-0.0014$. However, even though the mass dependence is 
correctly accounted for when varying $\a$, the fit to this data set is poor
compared to that at NNLO, with $\chi^2 =17.5$ at NLO at the best fit value of $\a$, as opposed to 
$\chi^2=14.3$ at NNLO. Moreover, this is achieved for $m_t=163.7~\GeV$, an unrealistically low pole mass value, and as $\a$ decreases $m_t$ has to decrease further in order to try to maintain 
the fit quality. It was noted in \cite{MSHT20} that at NLO a number of data sets were simply fit 
poorly, and the $t \bar t$  total cross section is one of these, with large NNLO corrections 
missing from the cross section calculation. Hence, we do not use the constraint from this 
data set as our lower limit of $\a$. The next strongest constraint in the downwards direction
is $\Delta \a =-0.0017$ from LHCb 7 and 8~TeV $W,Z$ 
data, and we take this 
value as our lower limit. An almost identical constraint is provided by SLAC deuterium structure function data. In the upwards direction the strongest constraint is again from BCDMS proton structure 
function data with an uncertainty of $\Delta \a = +0.0013$. The next strongest constraint is nominally from the CMS 8~TeV
$t \bar t$ single differential data with $\Delta \a = +0.0014$.
However, given the fixed value of $m_t$ in the cross section used we would not include this constraint 
in any case; we note that there is a detailed study of the constraints on both $\a$ and 
$m_t$ in \cite{Cooper-Sarkar:2020twv}. Constraints of $\Delta \a \approx +0.0018, +0.0019,+0.0021$
are set by the following LHC data sets respectively: ATLAS 8~TeV
$Z$ $p_T$ distribution, ATLAS 7~TeV inclusive jets, and the CMS 7~TeV inclusive jets, although in the latter case the profile indicates a potential lack of quadratic behaviour. As at NNLO, the SLAC proton structure function data also provides an upper bound, this time of $\Delta \a = +0.0023$.

The uncertainties in the upwards and downwards directions are 
slightly asymmetric, but for simplicity we chose to symmetrise these. 
Hence at NLO and NNLO we average the two uncertainties 
(obtained without the $\sigma_{t \bar t}$ constraint). We obtain 
\bea
\alpha_{S,{\rm NLO}}(M_Z^2) & = & 0.1203\pm 0.0015 \label{eq:optNLOunc}\\
\alpha_{S,{\rm NNLO}}(M_Z^2) & = & 0.1174\pm 0.0013. \label{eq:optNNLOunc}
\eea
This corresponds to $\Delta^{\rm NLO} \chi^2_{\rm global}=19$ and 
$\Delta^{\rm NNLO} \chi^2_{\rm global}=17$. These are the sort of tolerance values typical of 
the PDF eigenvectors, though a little towards the higher end. 

The NNLO value of $\a$ is well within 1$\sigma$ of the world average  
of $0.1179 \pm 0.001$, while the NLO value is consistent within $2 \sigma$. 
This is not surprising as most determinations of $\a$ in the world average are obtained
at NNLO, so it is effectively a NNLO value, which is lower than an NLO value would be. 
Hence, we present the values in 
eqs. (\ref{eq:optNLOunc}) and (\ref{eq:optNNLOunc}) as independent measurements of 
$\a$, but acknowledge that at NNLO, taking both this determination and the world average 
into account then a round value of $\a=0.118$ is an appropriate one at which to present the 
PDFs. At NLO we would recommend the use of $\a=0.120$ as the preferred value for the 
PDFs, but have also made eigenvector sets available at $\a=0.118$. 

\subsection{The PDF$+\alpha_S(M_Z^2)$ uncertainty on cross sections \label{sec:PDFalpha}}

Within the Hessian approach to PDF uncertainties it has been shown that  
the correct manner in which to account for the PDF+$\a$ uncertainty on any quantity, with the correlations between the PDFs and 
$\alpha_S$ included, can be obtained by simply taking the PDFs
defined at $\a \pm \Delta \a$ and treating these two PDF sets (with their accompanying
value of $\a$) as an extra pair of eigenvectors \cite{CTEQalphas}. The full uncertainty is obtained
by adding the uncertainty from this extra eigenvector pair in quadrature with the PDF
uncertainty, i.e. 
\be
\Delta \sigma~=~\sqrt{(\Delta \sigma_{\rm PDF})^2+(\Delta\sigma_{\alpha_S})^2}\;.
\label{eq:PDFaunc}
\ee
This procedure has the benefit of  being both  simple, but also separating
out the $\a$ uncertainty on a quantity explicitly from the purely PDF
uncertainty. Strictly speaking, the method only completely holds if the central PDFs are those 
obtained from the best fit when $\a$ is left free, and if the uncertainty $\Delta\a$ on 
$\a$ that is used is the uncertainty obtained from the fit. If instead we use PDFs defined 
at $\a=0.118$ at NNLO we are still very near the best fit, and the error induced by not expanding the 
eigenvector pair about the best fit value of $\a$ will
be very small. At NLO a distinctly larger error will be induced by using the PDFs defined at 
$\a=0.118$ rather than those at $\a=0.120$. Any choice of $\Delta \a$ of $0.001-0.002$, as opposed to 
the $\a$ uncertainty in the last subsection, should
only induce a small error. Hence, we advocate using this approach
with NLO PDFs defined at $\a=0.120$ and NNLO PDFs defined at $\a=0.118$. The value of 
$\Delta \a$ is open to the choice of the user to some extent, but it is recommended 
to stay close to the values of $\Delta \a$ that we have found. A simple, and 
perfectly consistent choice might be to use $\Delta \a =0.001$, similar to that for the world average.

In Section \ref{sec:cxunc} we apply the above procedure to determine the PDF$+\a$ uncertainties on the 
predictions for the cross sections for benchmark processes at the Tevatron, the LHC and at a potential 
future circular collider (FCC). First, we examine the change in the PDF sets themselves with $\a$.

\subsection{Comparison of PDF sets with varying $\a$ \label{sec:PDF}}

\begin{figure} 
\begin{center}
\includegraphics[scale=0.23]{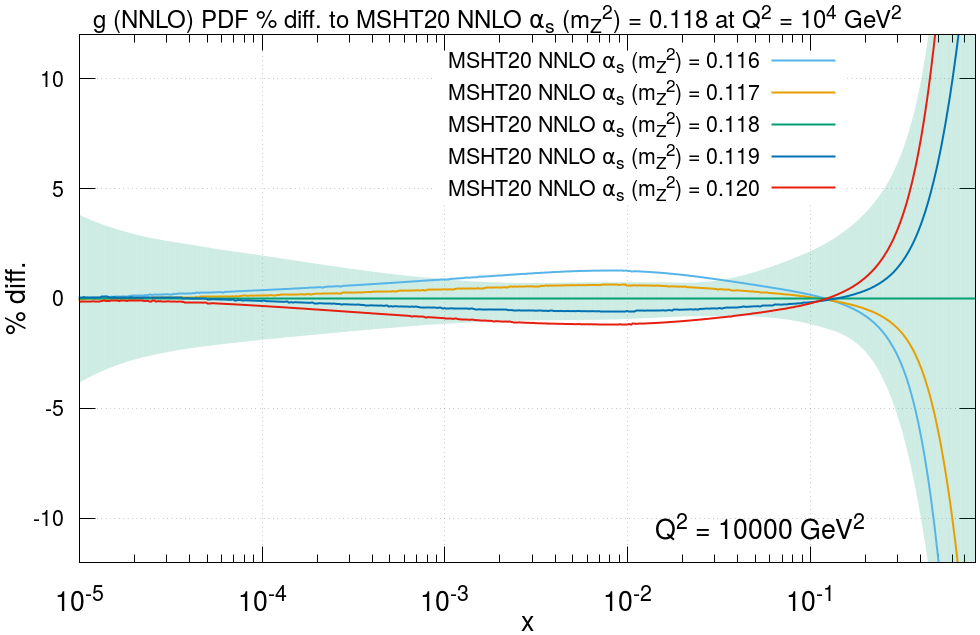}
\includegraphics[scale=0.23]{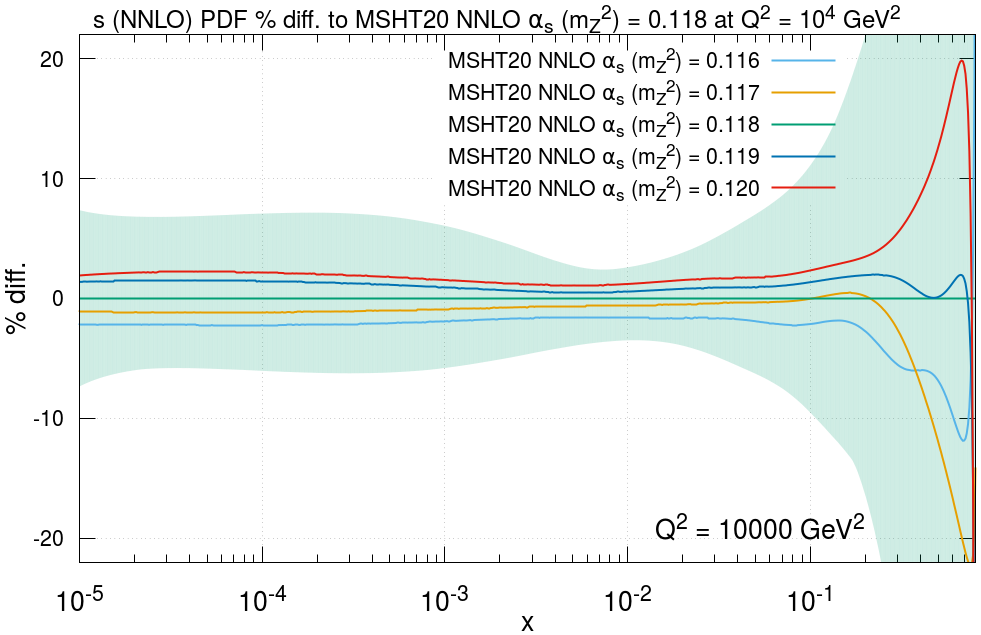}
\caption{Percentage difference in the NNLO gluon and strange quark PDFs at $Q^2=10^4$ ${\rm GeV}^2$ relative to the central ($\alpha_S(M_Z^2)=0.118$) set for fits with different values of $\alpha_S$. The percentage error bands for the central set are shown.}
\label{fig:pdfcomp1}
\end{center}
\end{figure}

\begin{figure} 
\begin{center}
\includegraphics[scale=0.23]{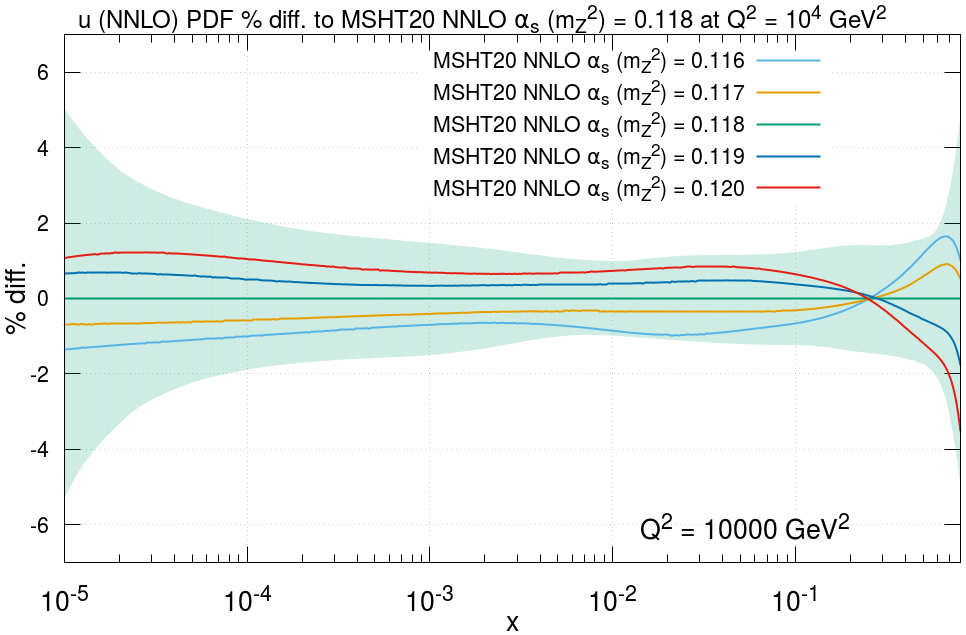}
\includegraphics[scale=0.23]{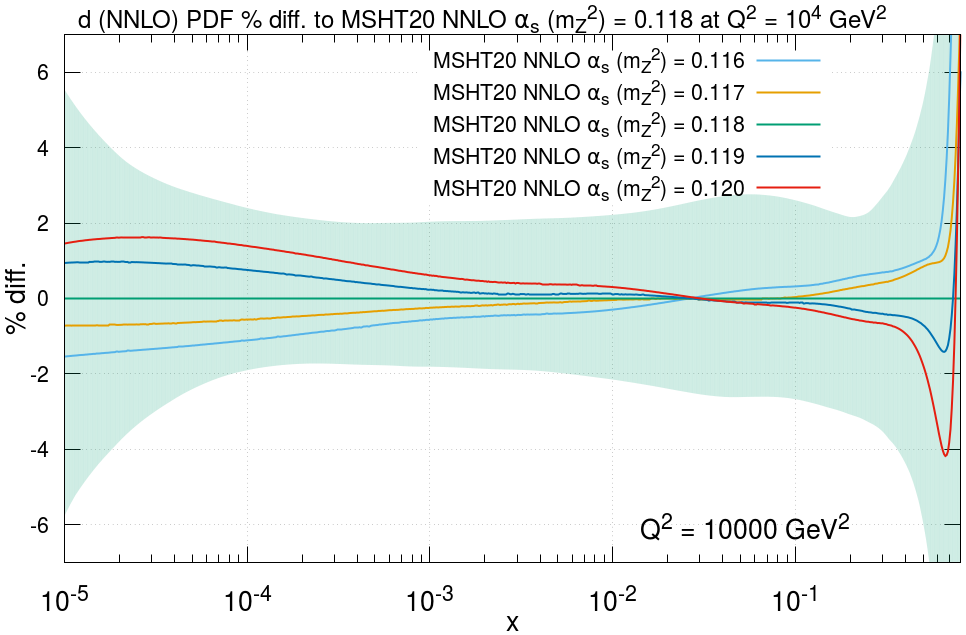}
\caption{Percentage difference in the NNLO up and down quark PDFs at $Q^2=10^4$ ${\rm GeV}^2$ relative to the central ($\alpha_S(M_Z^2)=0.118$) set for fits with different values of $\alpha_S$. The percentage error bands for the central set are shown.}
\label{fig:pdfcomp2}
\end{center}
\end{figure}

\begin{figure} 
\begin{center}
\includegraphics[scale=0.23]{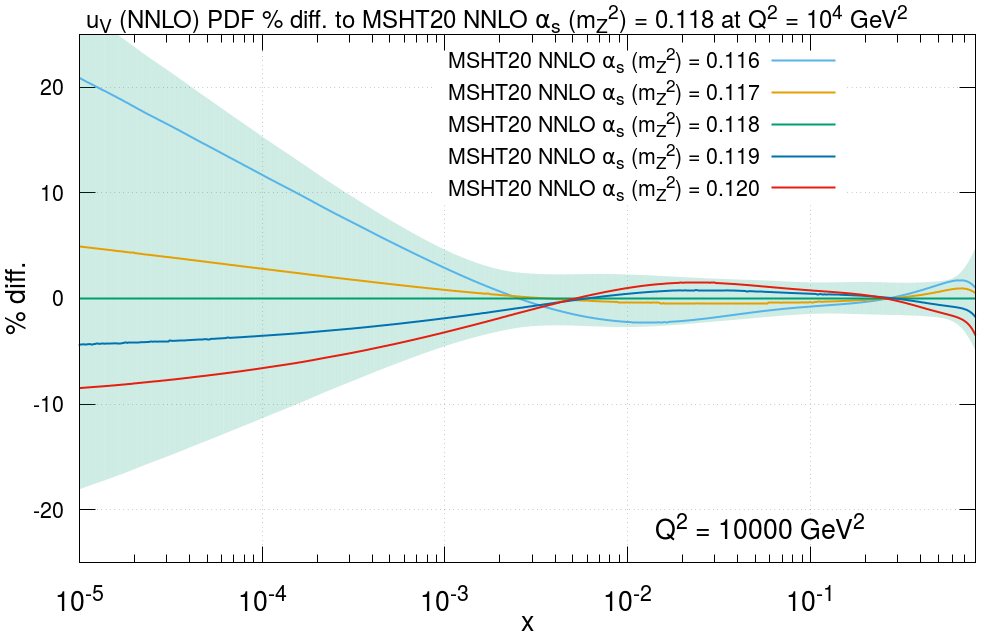}
\includegraphics[scale=0.23]{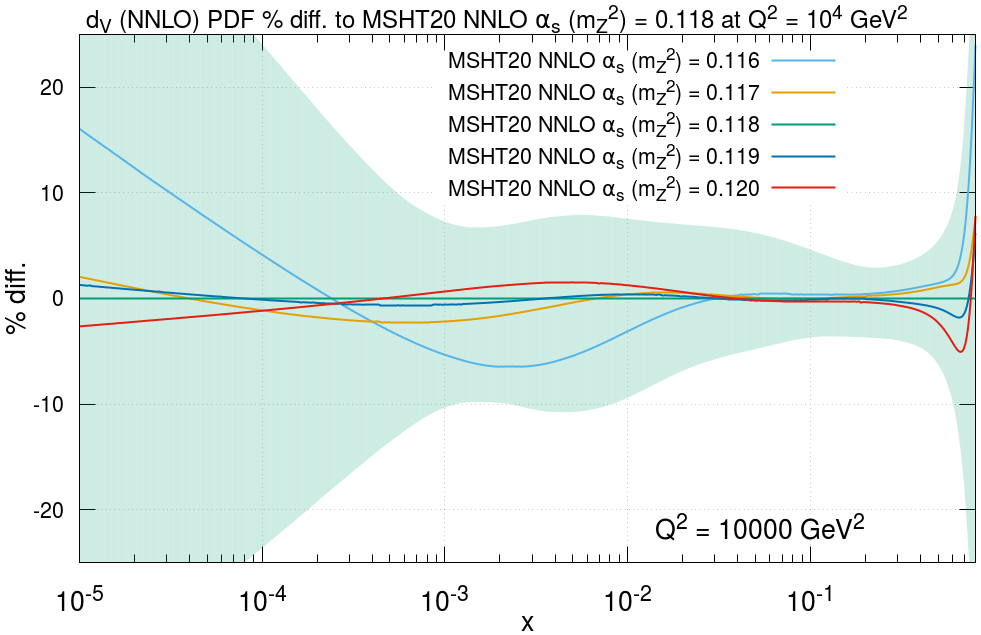}
\caption{Percentage difference in the NNLO up valence and down valence quark PDFs at $Q^2=10^4$ ${\rm GeV}^2$  relative to the central ($\alpha_S(M_Z^2)=0.118$) set for fits with different values of $\alpha_S$. The percentage error bands for the central set are shown.}
\label{fig:pdfcomp3}
\end{center}
\end{figure}

It is informative to see the changes in the PDFs obtained in global fits 
for fixed values of $\alpha_S(M_Z^2)$ relative to those obtained for the 
central value. We only consider the NNLO case here, but note that the NLO PDFs behave in a very 
similar way.
These are shown in Figs.~\ref{fig:pdfcomp1}--\ref{fig:pdfcomp3} at
$Q^2=10^4$ GeV$^2$.

In general, the changes in the PDFs for the coupling varied in the 
range $0.116<\alpha_S(M_Z^2)<0.120$ are within the PDF uncertainty bounds.
In more detail, the gluon distribution for $x<0.1$ is larger for 
$\alpha_S(M_Z^2)=0.116$ and smaller for $\alpha_S(M_Z^2)=0.120$. This approximately 
preserves the product $\alpha_S g$, which largely determines the
evolution of $F_2(x,Q^2)$ with $Q^2$ at low $x$. This is the dominant 
constraint on the gluon, and then the additional constraint of the momentum sum rule 
means a smaller low $x$ gluon leads to a larger high--$x$ 
gluon (and vice versa). 
The $u$ and $d$ PDFs have 
the opposite trend as $\alpha_S(M_Z^2)$ changes. At small $x$ values this is a
marginal effect, with the change in the gluon with $\a$ maintaining the 
evolution of the small-$x$ structure function, and hence also the small-$x$ quarks. 
At high $x$ the decreasing quark distribution with increasing $\alpha_S$ is 
due to the quicker evolution of quarks. This is seen explicitly also in the 
plots for the valence distributions. 
The relative insensitivity of 
the strange quark PDF to variations of $\alpha_S(M_Z^2)$ at low $x$ is 
partly just due to the insensitivity of all low--$x$ quarks, but 
is also  explained by   
changes in $\alpha_S(M_Z^2)$ being, to some extent, compensated by 
changes in the $B(D\to \mu)$ branching ratio parameter, which we allow to be free in the fit, with a $10\%$ uncertainty. At high $x$, the strange PDF shows the opposite trend to the $u$ and $d$ quark PDFs,  increasing with $\a$. This occurs in order to compensate for the reduction of the $u$ and, in particular, the $d$ 
valence quark PDFs in this region, ensuring the charge weighted sum remains approximately constant.

\begin{figure} 
\begin{center}
\includegraphics[scale=0.23]{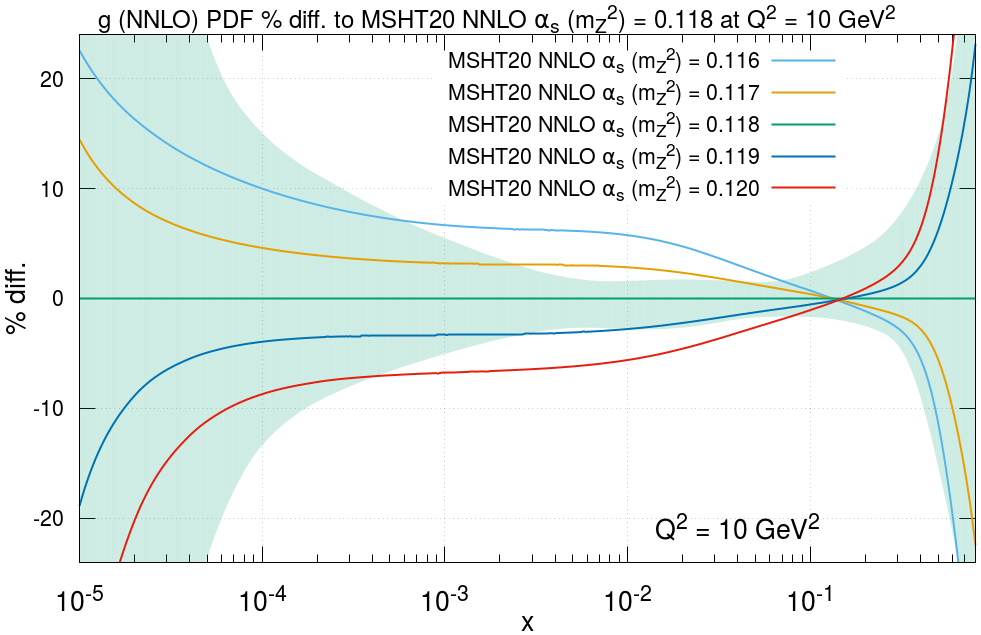}
\caption{Percentage difference in the NNLO gluon PDF at the lower scale of $Q^2=10$ ${\rm GeV}^2$  relative to the central ($\alpha_S(M_Z^2)=0.118$) set for fits with different values of $\alpha_S$. The percentage error bands for the central set are shown.}
\label{fig:pdfcomp4}
\end{center}
\end{figure}

Finally, in Fig.~\ref{fig:pdfcomp4}, we show the effect of different fixed values of $\a$ on the gluon at the lower scale of $Q^2=10$ ${\rm GeV}^2$,  much closer to the starting scale. Here, the changes in the gluon PDF now lie notably outside the uncertainty bands of the $\a = 0.118$ central fit. The reason for this difference is that at lower scales the gluon PDF is more sensitive to the larger variations in the low scale $\alpha_S(Q^2)$ value. In contrast, at the high value of $Q^2=10^4$ ${\rm GeV}^2$ relevant for LHC physics, the long evolution length means that the gluon in the data region around $x \sim 0.01$ is determined by evolution and hence a convolution over the PDFs at larger $x$, leaving it more insensitive to the $\a$ value.

\subsection{Benchmark cross sections \label{sec:cxunc}}

In this section  we show uncertainties for cross sections at the 
Tevatron, and for $8~\TeV$ and $13~\TeV$ at the LHC, as well as for a $100~\TeV$ FCC (pp). 
Uncertainties for  
$7~\TeV$ and $14~\TeV$ will be very similar to those at $8~\TeV$ and $13~\TeV$,
respectively. We calculate the cross sections for $W$ and $Z$ boson, Higgs boson via gluon--gluon 
fusion and top--quark pair production. For the $W/Z$ ratio there will be almost complete cancellation 
in the $\a$ uncertainties.

We calculate the PDF and $\a$ uncertainties for the MSHT20
PDFs \cite{MSHT20} at the default values of $\a$. 
We use a value of $\Delta \a=0.001$ as an example: we provide our PDF sets with $\a$ changes in units of $0.001$ and this is very similar to the uncertainty in the world average. 
However, for values similar to $\Delta \a=0.001$ a linear scaling
of the change in the prediction with $\a$ can be applied to a very good approximation. As explained in 
Section \ref{sec:PDFalpha}, the full PDF+$\a$ uncertainty may then be obtained by adding the 
two uncertainties in quadrature. 

To calculate the cross section at NNLO in QCD perturbation theory we use the same procedure as in
Section 9 of \cite{MSHT20}. 
As there we use LO electroweak perturbation theory, with the $qqW$ and $qqZ$ 
couplings defined by 
\begin{equation}
  g_W^2 =  G_F M_W^2 / \sqrt{2}, \qquad g_Z^2 = G_F M_Z^2 \sqrt{2}, 
\end{equation}
and other electroweak parameters as in \cite{MSTW}. We take the Higgs mass to be 
$m_H=125~\GeV$  and the top pole mass is $m_t=172.5~\GeV$.  
For the $t\overline{t}$ cross section we use \texttt{top++}~\cite{NNLOtop}.
Here our primary aim is not to present definitive predictions or to compare 
in detail to other PDF sets, as both these results are frequently provided in the literature with very 
specific choices of codes, scales and parameters which may differ from those used here.  Rather, our 
main objective is to illustrate the size of the PDF$+\a$ uncertainties.

\subsubsection{$W$ and $Z$ production}

\begin{table}
\begin{center}
\renewcommand\arraystretch{1.25}
\begin{tabular}{|l|c|c|c|}
\hline
& $\sigma$& PDF unc.& $\alpha_S$ unc.  \\
\hline
$\!\! W\,\, {\rm Tevatron}\,\,(1.96~\TeV)$   & 2.705    & ${}^{+0.054}_{-0.057}$ $\left({}^{+2.0\%}_{-2.1\%}\right)$ & ${}^{+0.018}_{-0.017}$  $\left({}^{+0.66\%}_{-0.61\%}\right)$    \\   
$\!\! Z \,\,{\rm Tevatron}\,\,(1.96~\TeV)$   & 0.2506& ${}^{+0.0045}_{-0.0046}$  $\left({}^{+1.8\%}_{-1.8\%}\right)$ &${}^{+0.0018}_{-0.0016}$ $\left({}^{+0.70\%}_{-0.62\%}\right)$ \\    
\hline 
$\!\! W^+ \,\,{\rm LHC}\,\, (8~\TeV)$        &7.075    & ${}^{+0.099}_{-0.110}$ $\left({}^{+1.4\%}_{-1.6\%}\right)$  &  ${}^{+0.064}_{-0.060}$ $\left({}^{+0.91\%}_{-0.85\%}\right)$  \\    
$\!\! W^- \,\,{\rm LHC}\,\, (8~\TeV)$        & 4.955  & ${}^{+0.071}_{-0.083}$ $\left({}^{+1.4\%}_{-1.7\%}\right)$  & ${}^{+0.044}_{-0.042}$ $\left({}^{+0.88\%}_{-0.84\%}\right)$ \\    
$\!\! Z \,\,{\rm LHC}\,\, (8~\TeV)$          & 1.122   & ${}^{+0.014}_{-0.017}$ $\left({}^{+1.3\%}_{-1.4\%}\right)$ & ${}^{+0.010}_{-0.010}$ $\left({}^{+0.90\%}_{-0.86\%}\right)$  \\ \hline    
$\!\! W^+ \,\,{\rm LHC}\,\, (13~\TeV)$       & 11.53      & ${}^{+0.16}_{-0.18}$ $\left({}^{+1.4\%}_{-1.6\%}\right)$  & ${}^{+0.12}_{-0.11}$ $\left({}^{+1.0\%}_{-0.94\%}\right)$  \\    
$\!\! W^- \,\,{\rm LHC}\,\, (13~\TeV)$       & 8.512     &${}^{+0.12}_{-0.14}$ $\left({}^{+1.4\%}_{-1.6\%}\right)$   & ${}^{+0.080}_{-0.078}$ $\left({}^{+0.94\%}_{-0.91\%}\right)$    \\    
$\!\! Z \,\,{\rm LHC}\,\, (13~\TeV)$         & 1.914  & ${}^{+0.024}_{-0.029}$ $\left({}^{+1.3\%}_{-1.5\%}\right)$  & ${}^{+0.019}_{-0.018}$ $\left({}^{+0.98\%}_{-0.94\%}\right)$   \\ 
\hline
$\!\! W^+ \,\,{\rm FCC}\,\, (100~\TeV)$       & 70.82      & ${}^{+2.46}_{-3.08}$ $\left({}^{+3.6\%}_{-4.4\%}\right)$  & ${}^{+0.94}_{-0.89}$ $\left({}^{+1.3\%}_{-1.3\%}\right)$  \\    
$\!\! W^- \,\,{\rm FCC}\,\, (100~\TeV)$       & 60.39     &${}^{+1.65}_{-2.04}$ $\left({}^{+2.9\%}_{-3.3\%}\right)$   & ${}^{+0.79}_{-0.74}$ $\left({}^{+1.3\%}_{-1.2\%}\right)$    \\    
$\!\! Z \,\,{\rm FCC}\,\, (100~\TeV)$         & 13.50  & ${}^{+0.40}_{-0.47}$ $\left({}^{+3.1\%}_{-3.4\%}\right)$  & ${}^{+0.19}_{-0.17}$ $\left({}^{+1.4\%}_{-1.3\%}\right)$   \\ 
\hline
    \end{tabular}
\end{center}
\caption{\sf Predictions for $W^\pm$ and $Z$ cross sections (in nb), including leptonic branching, obtained with the NNLO MSHT20 parton sets. The PDF and $\alpha_S$ uncertainties  are shown, where the $\alpha_S$ uncertainty corresponds to a variation of $\pm 0.001$ around its central value.  The full PDF$+\a$ uncertainty can be obtained by adding these two uncertainties in quadrature, as explained in Section~\ref{sec:PDFalpha}.}
\label{tab:sigmaWZNNLO}   
\end{table}

The predictions for the $W$ and $Z$ production cross sections  at NNLO are shown in Table \ref{tab:sigmaWZNNLO}.
In this case the cross sections contain zeroth--order contributions in $\alpha_S$, 
with positive NLO corrections of about 
$20\%$, and  much smaller NNLO contributions. Hence, an approximately $1\%$ change in $\a$ will only 
directly 
increase the cross section by a small fraction of a percent. The PDF uncertainties on the cross 
sections are about $2\%$ at the Tevatron and slightly smaller at the LHC; the lower beam energy at the
Tevatron meaning the cross sections have more contribution from higher $x$, where PDF uncertainties
increase. For these cross sections the $\alpha_S$ uncertainty is small, about $0.6\%$ at the Tevatron 
and close to $1\%$ at 
the LHC, being slightly larger at 13~TeV than at 8~TeV, and larger again at 100~TeV. 
Hence, the $\alpha_S$ uncertainty is small, but
more than the small fraction of a percent expected from the direct change in the cross section with 
$\alpha_S$. This is because the main increase in cross sections with $\alpha_S$ is due to the change in 
the PDFs with the coupling, rather than its direct effect on the cross section. 
From Fig.~\ref{fig:pdfcomp2} 
 we see that the up and down quark PDFs increase with $\alpha_S$ below 
$x\sim 0.1-0.2$, due to increased speed of evolution. 
From Fig.~\ref{fig:pdfcomp1} we see that the strange quark PDF  increases a little
with $\alpha_S$ at all $x$ values. As already mentioned, the Tevatron cross sections are more sensitive to the high--$x$ 
quarks, which decrease with increasing $\alpha_S$, so this introduces a certain amount of anti--correlation 
of the cross section with $\alpha_S$. However, even at the Tevatron the main contribution is from low enough $x$ that
the distributions increase with $\alpha_S$. The net effect is therefore an increase with $\alpha_S$, which is a little 
larger than that coming directly from the $\alpha_S$ dependence of the cross section. As the energy increases at the LHC the contributing 
quarks move to lower $x$ and the increase of the cross section with $\alpha_S$ increases. This is a smaller effect than the increase in the PDF uncertainty itself at 100~TeV due to the very small $x$ PDFs sampled. For any collider scenario the total PDF+$\alpha_S$ uncertainty 
obtained by adding the two contributions in quadrature, is only a maximum of about $20\%$ greater  
than the PDF uncertainty alone, if $\Delta \a = 0.001$ is used.     

\subsubsection{Top-quark pair production}

\begin{table}
\begin{center}
\renewcommand\arraystretch{1.25}
\vspace{0.5cm}
\begin{tabular}{|l|c|c|c|}
\hline
& $\sigma$& PDF unc.& $\alpha_S$ unc.  \\
\hline
$t\overline{t}$ $ {\rm Tevatron}\,\,(1.96~\TeV)$   & 7.24    & ${}^{+0.13}_{-0.12}$ $\left({}^{+1.8\%}_{-1.7\%}\right)$ & ${}^{+0.15}_{-0.15}$  $\left({}^{+2.1\%}_{-2.1\%}\right)$    \\   
$t\overline{t}$  ${\rm LHC}\,\, (8~\TeV)$        &243.1   & ${}^{+6.4}_{-3.9}$ $\left({}^{+2.6\%}_{-1.6\%}\right)$  &  ${}^{+4.4}_{-4.5}$ $\left({}^{+1.8\%}_{-1.9\%}\right)$  \\    
$t\overline{t}$ ${\rm LHC}\,\, (13~\TeV)$       & 796.8   &${}^{+16.0}_{-10.6}$ $\left({}^{+2.0\%}_{-1.3\%}\right)$  & ${}^{+12}_{-13}$ $\left({}^{+1.5\%}_{-1.6\%}\right)$  \\    
$t\overline{t}$ ${\rm FCC}\,\, (100~\TeV)$       & 34600   &${}^{+300}_{-400}$ $\left({}^{+0.9\%}_{-1.2\%}\right)$  & ${}^{+400}_{-400}$ $\left({}^{+1.2\%}_{-1.2\%}\right)$  \\    
\hline
    \end{tabular}
\end{center}
\caption{\sf Predictions for $t\overline{t}$ cross sections (in pb), obtained with the NNLO MSHT20 parton sets. The PDF and $\alpha_S$ uncertainties  are shown, where the $\alpha_S$ uncertainty corresponds to a variation of $\pm 0.001$ around its central value. The full PDF$+\a$ uncertainty can be obtained by adding these two uncertainties in quadrature, as explained in Section~\ref{sec:PDFalpha}.}
\label{tab:sigmatNNLO}   
\end{table}

In Table \ref{tab:sigmatNNLO} we show the analogous results for the top--quark pair  production cross 
section. 
At the Tevatron the PDFs are probed in the region $x\sim 0.2$, and the main 
production source is the $q{\bar q}$ channel.  The quark distributions are reasonably insensitive to $\a$ in this region of 
$x$, as it is in the approximate region of the transition point of the PDFs, where evolution switches from PDFs 
decreasing with scale to increasing. Hence, there is only a small change in cross section due to changes 
in the PDFs with $\alpha_S$. However, the cross section for $t{\bar t}$ production begins at order 
$\alpha_S^2$, and there is a significant positive higher--order correction at NLO, and 
still an appreciable one at NNLO. Therefore, a change in $\alpha_S$ a little lower than $1\%$ should give a direct 
change in the cross section of about $2\%$ or slightly more, which is indeed the change that is observed. This is to be compared 
with a slightly smaller PDF  uncertainty of nearly $2\%$. 

At the LHC the dominant production mechanism, due to the higher energy and proton--proton nature of the collisions is gluon--gluon fusion, with the central $x$ value probed being
$x\approx  0.05$ at 8~TeV, and $x\approx 0.03$ at 13~TeV. As seen 
from the left plot of Fig. \ref{fig:pdfcomp1} the 
gluon decreases with increasing $\a$ below $x=0.1$ and the maximum decrease is for $x\sim 0.01$. 
The $\a$ uncertainty on $\sigma_{t \bar t}$ at 8~TeV is slightly less than $2\%$, almost as large as at the 
Tevatron, with the gluon above the pivot point still contributing considerably to the cross section, such that 
the indirect $\a$ uncertainty due to PDF variation largely cancels. At 13~TeV the lower $x$ probed 
means that most contribution is below the pivot point and there is some anti--correlation between    
the direct $\alpha_S$ variation and the indirect impact via the PDFs, with a reduced $\alpha_S$ uncertainty of $1.5\%$. 
At this energy the PDF only uncertainty has also reduced to about $2\%$ due to the decreased sensitivity
to the uncertainty in high--$x$ PDFs, the gluon in this case. 
At 100~TeV we have $x\approx 0.004$, and the PDF uncertainty has approximately minimised, while the 
anti-correlation between the gluon and $\a$ has increased such that there is a reduced $\alpha_S$ 
uncertainty of $1.2\%$.  
At the 8~TeV and 13~TeV LHC the 
$\a$ uncertainty is similar to the PDF uncertainty, and the total is about 1.4 times the
PDF uncertainty alone. At the Tevatron and 100~TeV FCC the $\a$ uncertainty is slightly larger, 
such that the total uncertainty, for
 $\Delta \a = 0.001$  is about 1.6-1.7 that of the PDF uncertainty.

\subsubsection{Higgs boson production}

 \begin{table}
\begin{center}
\renewcommand\arraystretch{1.25}
\begin{tabular}{|l|c|c|c|}
\hline
& $\sigma$& PDF unc.& $\alpha_S$ unc.  \\
\hline
Higgs $ {\rm Tevatron}\,\,(1.96~\TeV)$   & 0.867    & ${}^{+0.030}_{-0.019}$ $\left({}^{+3.5\%}_{-2.2\%}\right)$ & ${}^{+0.019}_{-0.019}$  $\left({}^{+2.2\%}_{-2.2\%}\right)$    \\   
Higgs  ${\rm LHC}\,\, (8~\TeV)$        &18.44    & ${}^{+0.24}_{-0.24}$ $\left({}^{+1.3\%}_{-1.3\%}\right)$  &  ${}^{+0.29}_{-0.29}$ $\left({}^{+1.6\%}_{-1.6\%}\right)$  \\    
Higgs ${\rm LHC}\,\, (13~\TeV)$       & 42.13   &${}^{+0.47}_{-0.51}$ $\left({}^{+1.1\%}_{-1.2\%}\right)$  & ${}^{+0.64}_{-0.65}$ $\left({}^{+1.5\%}_{-1.5\%}\right)$  \\    
Higgs ${\rm FCC}\,\, (100~\TeV)$       & 708.2   &${}^{+9.5}_{-12}$ $\left({}^{+1.3\%}_{-1.7\%}\right)$  & ${}^{+12}_{-12}$ $\left({}^{+1.7\%}_{-1.7\%}\right)$  \\    
\hline
    \end{tabular}
\end{center}
\caption{\sf Predictions for the Higgs boson cross sections (in pb), obtained with the NNLO MSHT 20 parton sets. The PDF and $\alpha_S$ uncertainties are  shown, where the $\alpha_S$ uncertainty corresponds to a variation of $\pm 0.001$ around its central value. The full PDF$+\a$ uncertainty can be obtained by adding these two uncertainties in quadrature, as explained in Section~\ref{sec:PDFalpha}.}
\label{tab:sigmahNNLO}   
\end{table}

In Table \ref{tab:sigmahNNLO} we show the uncertainties in the rate of Higgs boson production from gluon--gluon fusion. As with top-pair production the cross section starts at order $\alpha^2_S$ and there are large positive NLO and NNLO contributions. Therefore,
changes in $\alpha_S$ of about $1\%$ would be expected to lead to direct changes in the cross section
of about $2-3\%$. However, even at the Tevatron the dominant $x$ range probed, i.e. $x \approx 0.06$, corresponds to a region where
the gluon distribution falls with increasing $\a$, so there is some anti-correlation. At the LHC where $x \approx 0.01-0.02$ at central rapidity the 
anti--correlation between $\a$ and the gluon distribution is near its maximum, and at the FCC where
$x \approx 0.001$, anti-correlation remains high. Hence, at the Tevatron the
total $\a$ uncertainty is a little less than the direct value, i.e. a little more than $2\%$, 
and at the LHC and FCC it is reduced to about $1.5\%$. In the former case this is slightly less than 
the PDF uncertainty of $\sim 2.8\%$, with some sensitivity to 
the relatively poorly constrained high--$x$ gluon, while at the LHC and FCC the PDF uncertainty is much 
reduced, due to the 
smaller $x$ probed, and is smaller than the $\a$ uncertainty. 
Hence for $\Delta \a = 0.001$ the Higgs boson cross 
section from gluon--gluon fusion  is about 1.6-1.7 that of the PDF uncertainty alone.   

\section{Heavy-quark masses} \label{Sec:Heavy_quark_mass_dependence}

\subsection{Choice of the range of heavy-quark masses}

In the study of heavy-quark masses that accompanied the MMHT PDFs 
\cite{MMHThq} we varied the charm and bottom quark masses, defined in the 
pole mass scheme, from $1.15~\GeV$ to $1.55~\GeV$, in steps of $0.05~\GeV$, and $m_b$ from $4.25~\GeV$
to $5.25~\GeV$ in steps of $0.25~\GeV$. 
This was an asymmetric range about our default value of $m_c=1.4~\GeV$, and 
was because in this previous study for both charm and bottom the
preferred mass values were towards the lower end of the range. In the present study, as we will show, 
there is no longer such a clear preference for lower values, so we choose for $m_c$ the 
symmetric range from $1.2~\GeV$ to $1.6~\GeV$, while for $m_b$ we expand our range slightly from 
$4~\GeV$ to $5.5~\GeV$. 

Let us consider this range compared to  
the constraint from other determinations of the quark masses. 
These are generally quoted in the $\MS$ scheme, and in \cite{PDG2020} 
are given as $m_c(m_c)=(1.27 \pm 0.02)~\GeV$ and 
$m_b(m_b)=(4.18^{+0.03}_{-0.02})~\GeV$. The transformation to the pole mass definition 
is not well-defined  due to the diverging series, i.e. there is a renormalon 
ambiguity of $\sim 0.1-0.2~\GeV$. The series is considerably less convergent for the 
charm quark, due to the lower scale in the coupling, but the renormalon 
ambiguity cancels in the difference between the charm and bottom masses. Indeed,
in this way $m_b^{\rm pole}-m_c^{\rm pole}=3.4~\GeV$ is obtained with a very small uncertainty
\cite{Bauer:2004ve,Hoang:2005zw}. Using the perturbative expression for the 
conversion of the bottom mass, and the relationship between the bottom and 
charm mass it can be determined that
\be
m_c^{\rm pole}=1.5 \pm 0.2 ~{\GeV}  ~~~~ {\rm and} ~~~~ m_b^{\rm pole}=4.9\pm 0.2~\GeV, 
\ee 
where the two uncertainties are almost completely correlated. This disfavours $m_c\leq 1.2-1.3~\GeV$ and $m_b\leq 4.6-4.7~\GeV$. 
There is some indication from PDF fits for
a slightly lower $m^{\rm pole}$ than that suggested by the simple use of the perturbative series out 
to the order at which it starts to show lack of convergence for the central pole mass value. 
As the fit
quality prefers values slightly in this direction, we allow some values a little
lower than this in our scan. 
In the upper direction the fit quality clearly deteriorates, 
so our upper values are not too far beyond the central values quoted above. 
We now consider 
the variation with $m_c$ and $m_b$ in more detail. 

\subsection{Dependence on $m_c$}

\begin{figure} [h]
\begin{center}
\vspace*{-0.0cm}
\includegraphics[scale=0.22]{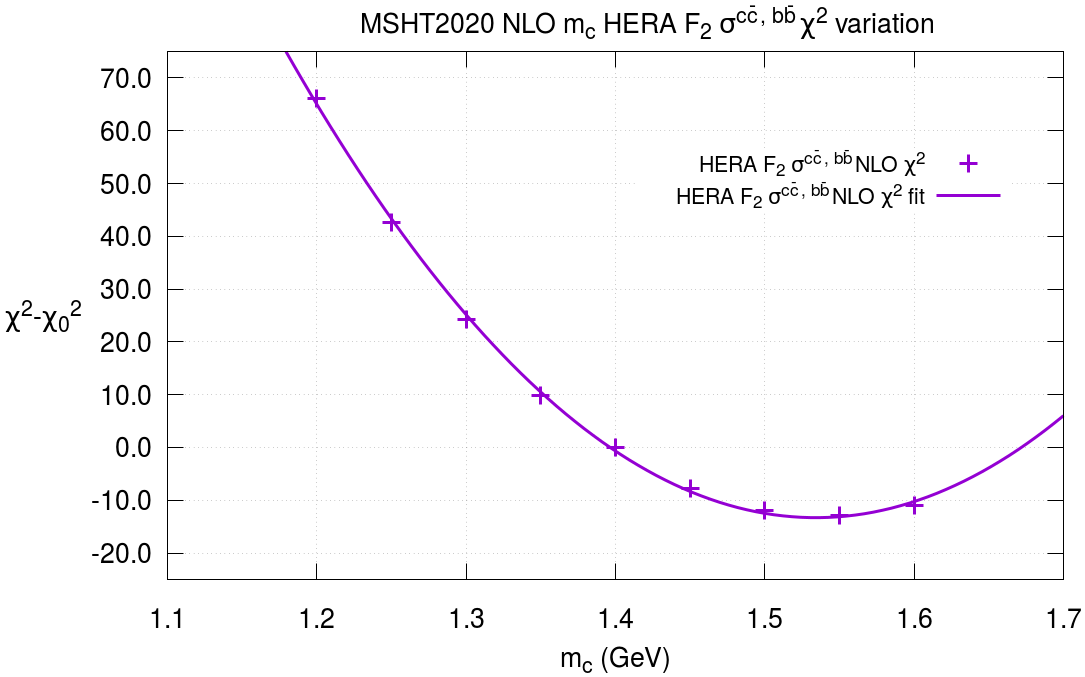}
\includegraphics[scale=0.22]{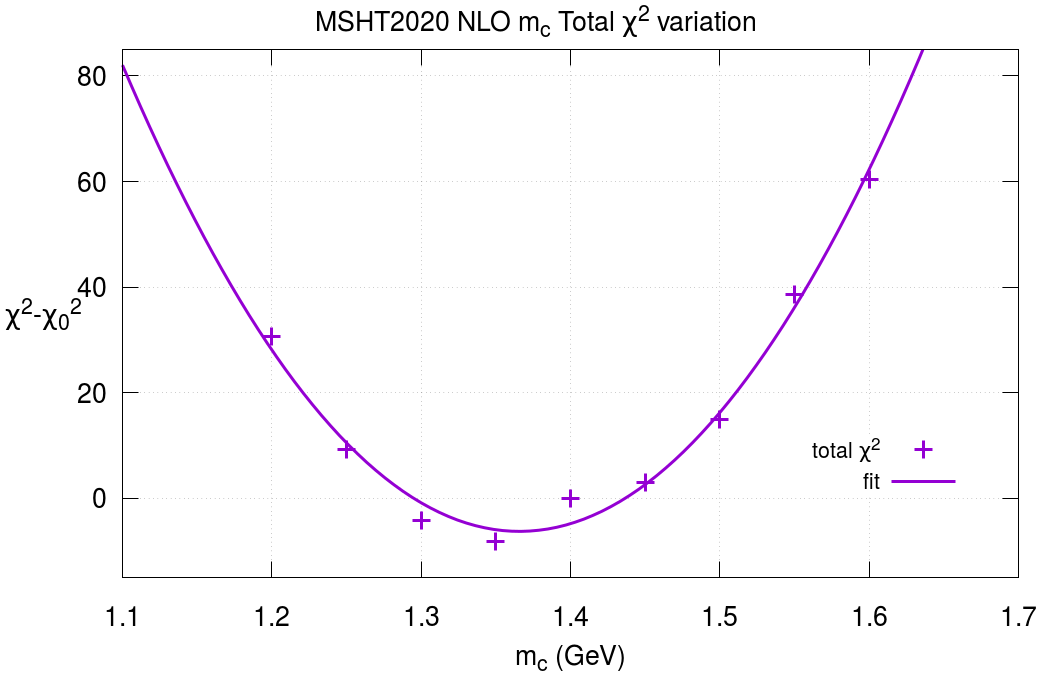}
\caption{\sf The fit quality versus the quark mass $m_c$ at NLO with $\alpha_S(M_Z^2)=0.118$ for (left) the reduced cross section for charm and bottom production
$\tilde{\sigma}^{\cc(\bb)}$ for the combined H1 and ZEUS data and (right) the full global fit.}
\label{fig:mcas118}
\end{center}
\end{figure}

\begin{figure} [h]
\begin{center}
\vspace*{-0.0cm}
\includegraphics[scale=0.22]{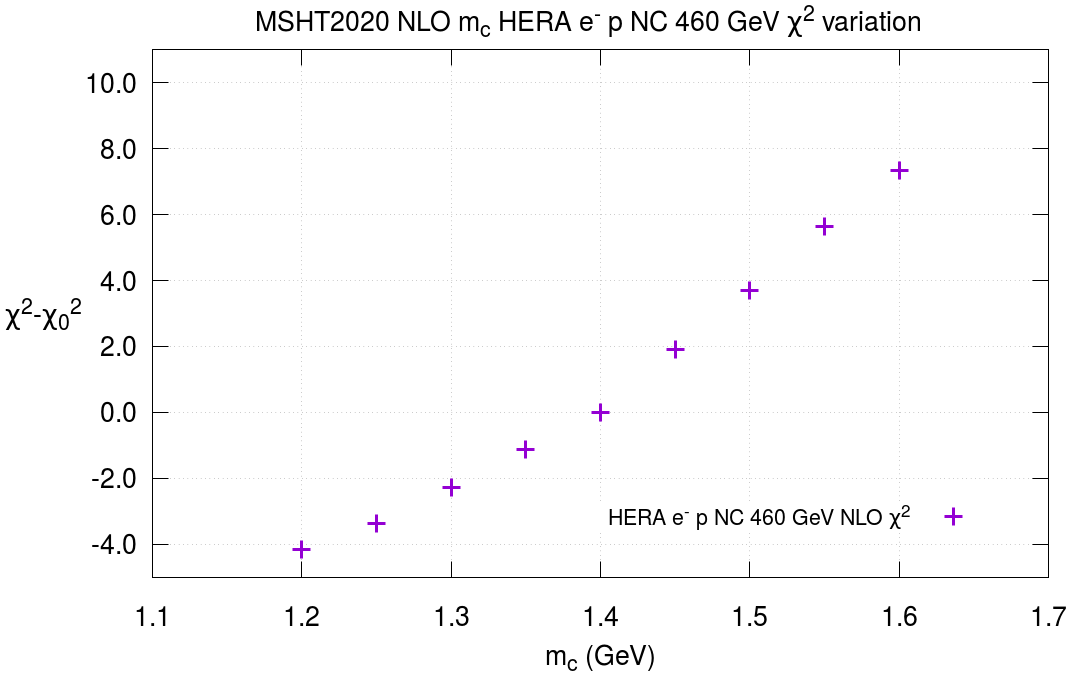}
\includegraphics[scale=0.22]{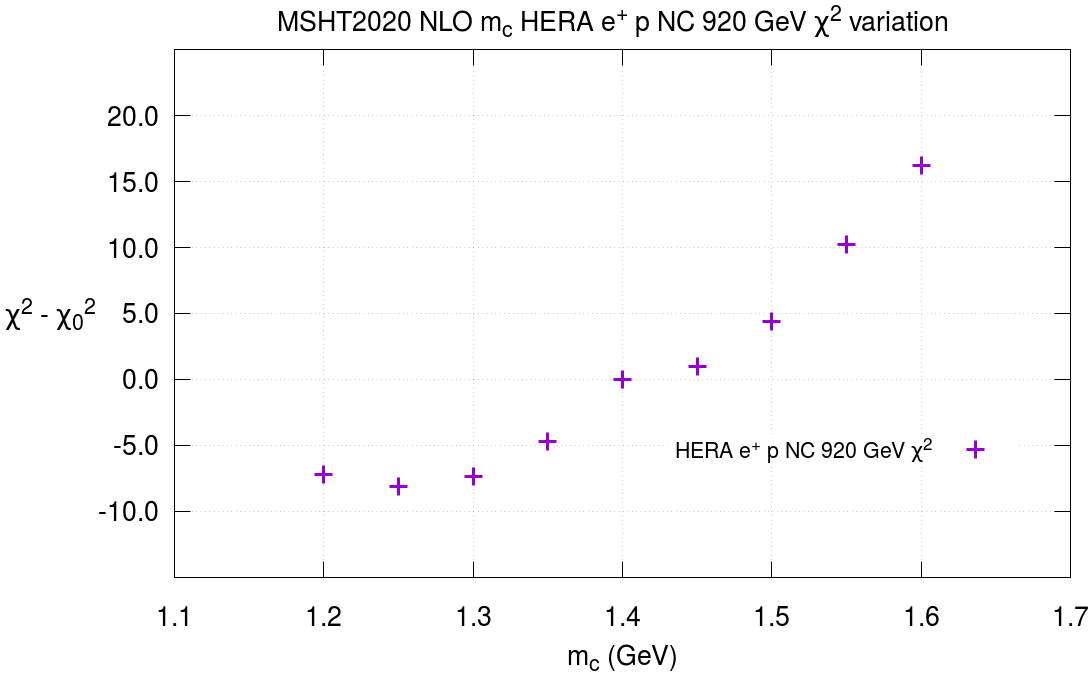}
\caption{\sf The fit quality versus the quark mass $m_c$ at NLO with $\alpha_S(M_Z^2)=0.118$ 
for (left) the total reduced cross section 
$\tilde{\sigma}$ for the combined H1 and ZEUS NC $e^-$ 460~GeV data and (right) NC $e^+$ 920~GeV data.}
\label{fig:mcas118HERA}
\end{center}
\end{figure}

\begin{figure} [h]
\begin{center}
\vspace*{-0.0cm}
\includegraphics[scale=0.22]{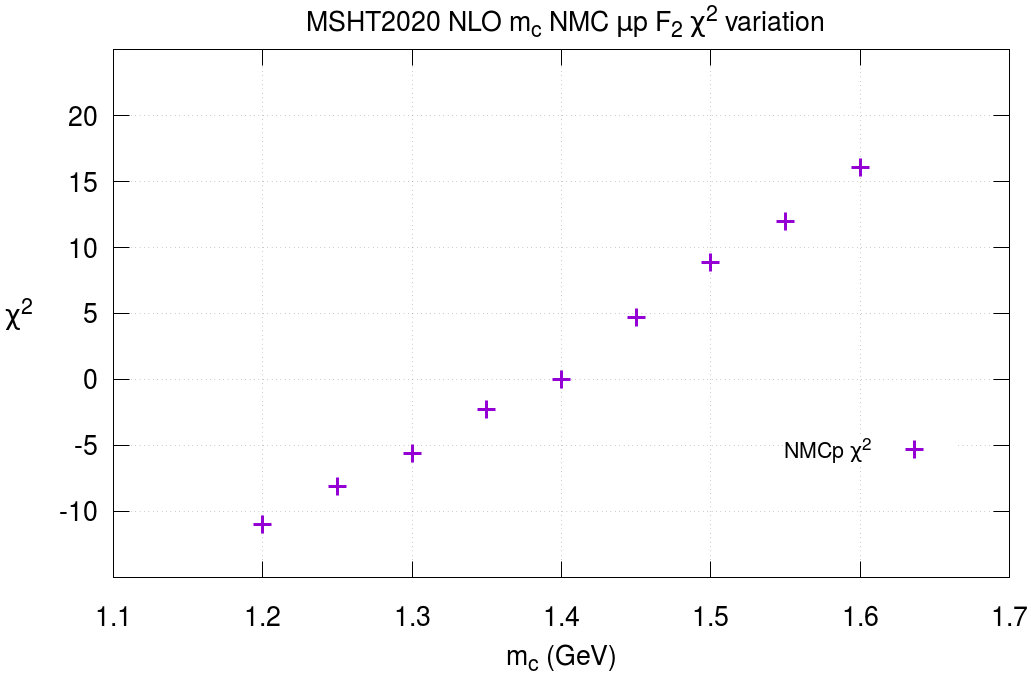}
\includegraphics[scale=0.22]{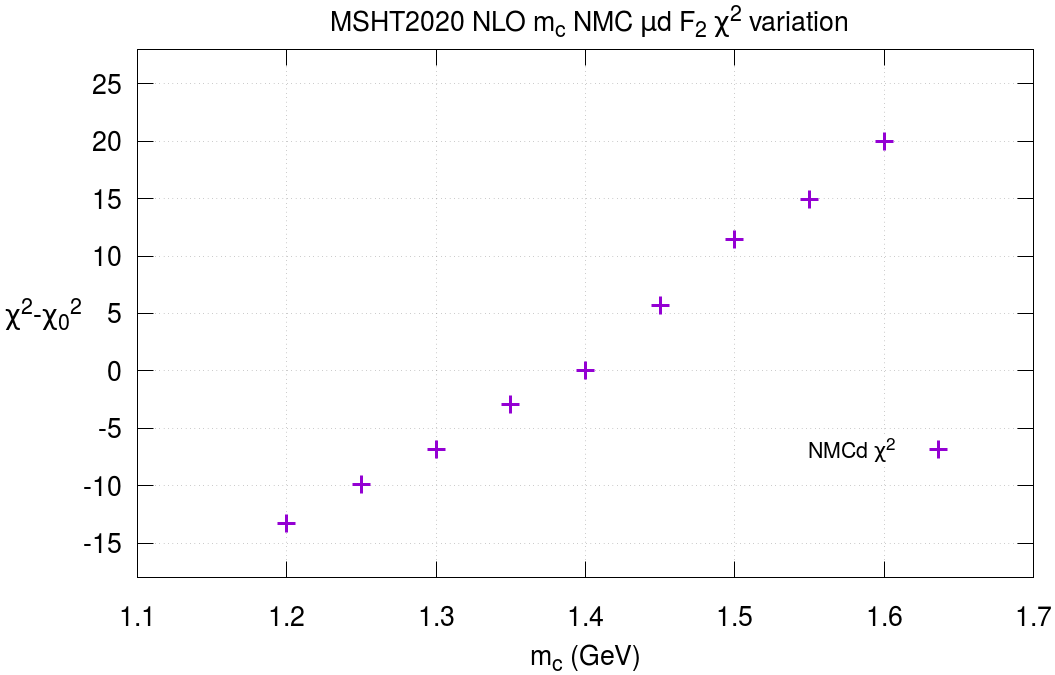}
\caption{\sf The quality of the fit versus the quark mass $m_c$ at NLO with $\alpha_S(M_Z^2)=0.118$ for (left) 
the NMC $F^p_2(x,Q^2)$ data and (right) the $F^d_2(x,Q^2)$ data.}
\label{fig:mcas118NMC}
\end{center}
\end{figure}

\begin{figure} [h]
\begin{center}
\vspace*{-0.0cm}
\includegraphics[scale=0.22]{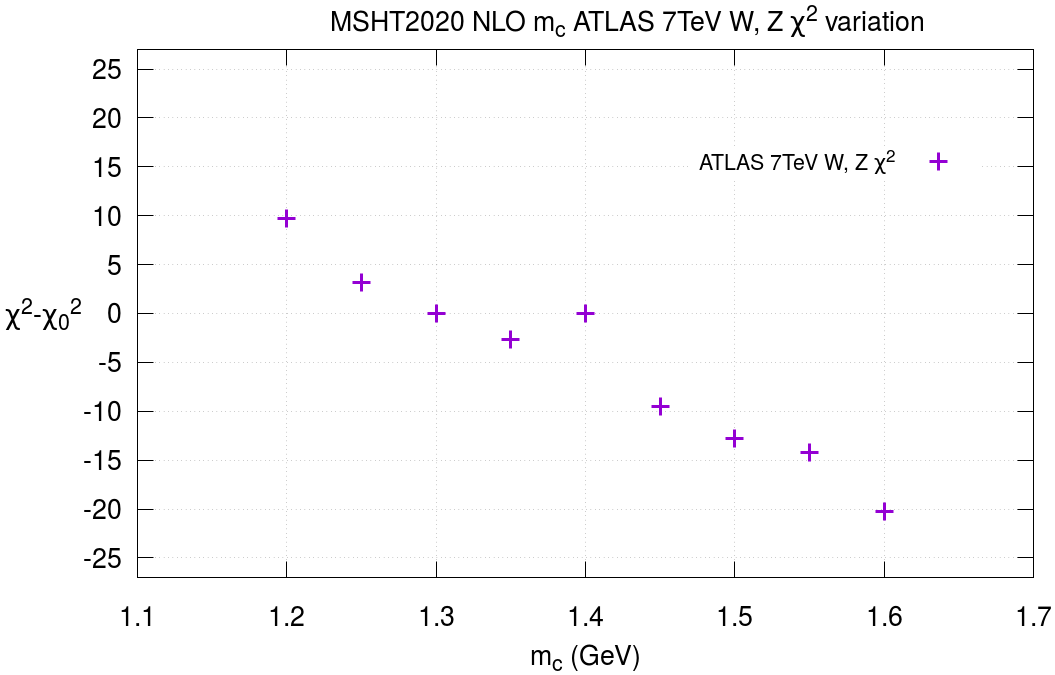}
\includegraphics[scale=0.22]{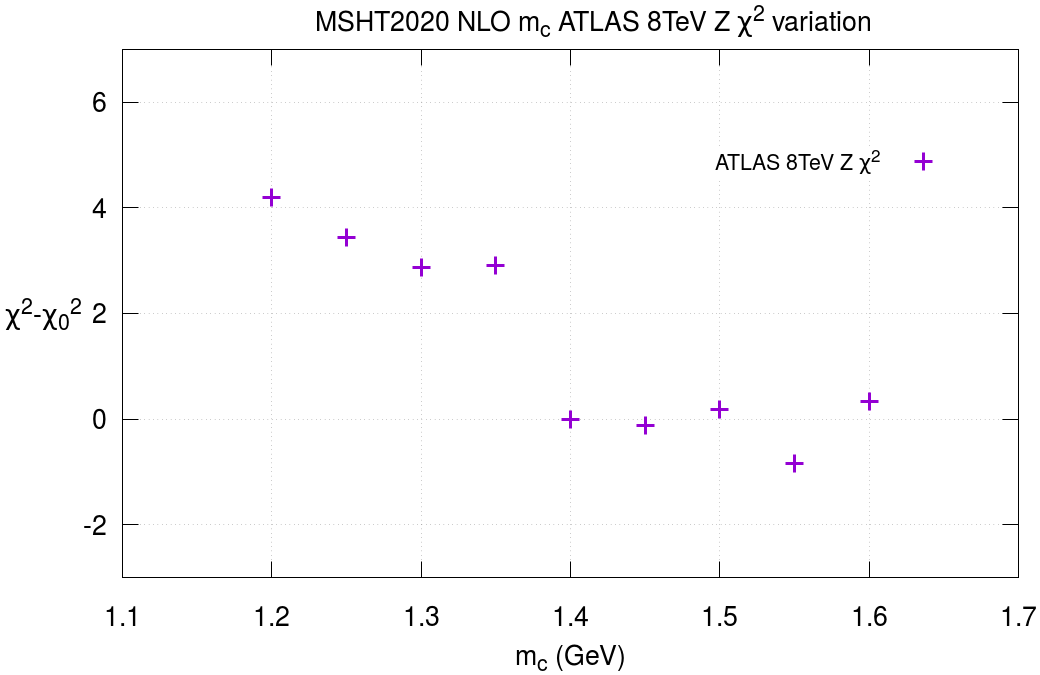}
\caption{\sf The quality of the fit versus the quark mass $m_c$ at NLO with $\alpha_S(M_Z^2)=0.118$ for (left) 
the ATLAS 7~TeV $W,Z$  data and (right) the ATLAS 8~TeV $Z$ data.}
\label{fig:mcas118ATLAS}
\end{center}
\end{figure}

\begin{figure} [h]
\begin{center}
\vspace*{-0.0cm}
\includegraphics[scale=0.22]{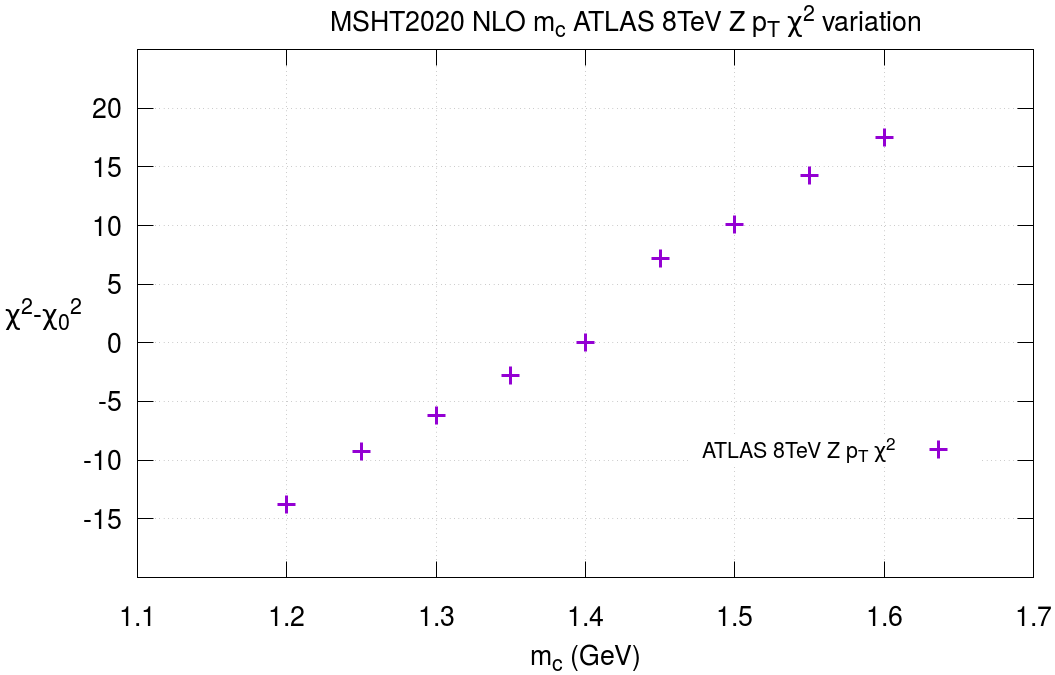}
\includegraphics[scale=0.22]{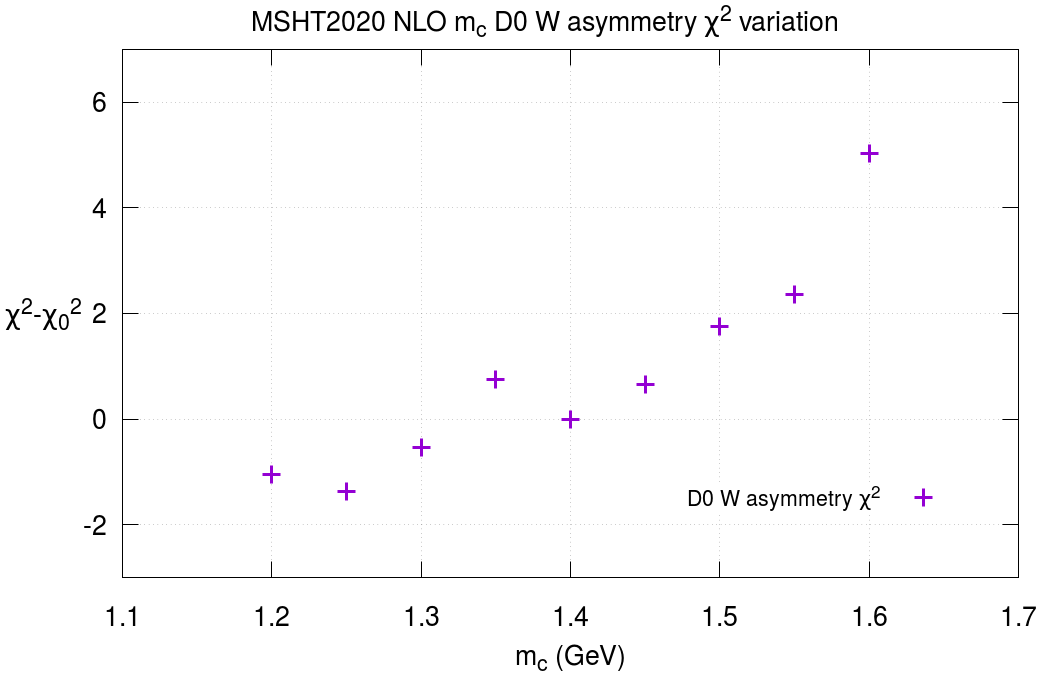}
\caption{\sf The quality of the fit versus the quark mass $m_c$ at NLO with $\alpha_S(M_Z^2)=0.118$ for (left) 
the ATLAS 8~TeV $Z$ $p_T$  data and (right) the D{\O} $W$ asymmetry data.}
\label{fig:mcas118ATLASD0}
\end{center}
\end{figure}

We repeat the global analysis in \cite{MSHT20} for values of $m_c=1.2-1.6~\GeV$
in steps of $0.05~\GeV$. As in \cite{MSHT20} we use the ``optimal'' version    
\cite{Thorne} of the TR' general mass variable flavour number scheme 
GM-VFNS \cite{TR1}. This uses the NNLO coefficient functions calculated in 
\cite{Laenen:1992zk}, and at NNLO
we require some degree of approximation in the vicinity of 
$Q^2 \sim m_h^2$ as the ${\cal O}(\alpha_S^3)$ heavy flavour coefficient
functions in the fixed flavour number scheme (FFNS) are still not known 
exactly, though the leading small-$x$ term~\cite{Catani:1990eg} and threshold 
logarithms~\cite{Laenen:1998kp,Kawamura:2012cr} have been calculated. 
We assume all heavy flavour is generated by evolution from the gluon
and light quarks, i.e. there is no intrinsic heavy flavour.  
We perform the analysis with $\alpha_S(M_Z^2)$ left as a free parameter in 
the fit at both NLO and NNLO, but also present our results at fixed values of the 
coupling of $\alpha_S(M_Z^2)=0.118$ at NLO at NNLO. We will concentrate on the 
results and PDFs with fixed coupling, as 
the standard MSHT PDFs were made available at these values.  At NLO PDFs are made available 
with $\a=0.118$ and our default value of $0.120$.     

We present results in terms of the $\chi^2$ for the total set of data
in the global fit and for just the data on the reduced cross section, $\tilde{\sigma}^{\cc(\bb)}$, for open 
charm production at HERA \cite{HERAhf}. For these variations, as well as the fit $\Delta\chi^2$ values shown by the points, we also provide a quadratic fit line as a guide to the behaviour. This is shown at NLO with 
$\alpha_S(M_Z^2)=0.118$ in Fig.~\ref{fig:mcas118}. The variation in the  
quality of the fit to the HERA combined charm and bottom cross section data is 
very significant. The heavy flavour data clearly prefer a value close to 
$m_c=1.5-1.55~\GeV$, above our default value of $m_c=1.4~\GeV$. 
The deterioration is clearly such as to 
make values of $m_c<1.3~\GeV$ strongly disfavoured.
However, there is a different variation in the fit quality to the global data set, 
with a clear preference
for values near to $m_c=1.35~\GeV$.  The main constraint comes 
from the $\chi^2$ for the inclusive HERA cross section data, shown in Fig.~\ref{fig:mcas118HERA}, but there is 
also a distinct preference 
for a low value of the mass from the $\chi^2$ for the NMC structure function data, shown in Fig.~\ref{fig:mcas118NMC}, where the 
data for $x\sim 0.01$ and $Q^2\sim 4~\GeV^2$ are sensitive to the turn-on of 
the charm contribution to the structure function. 
There is also clear sensitivity in the ATLAS 7~TeV $W,Z$ data, the ATLAS 8~TeV $Z$
data and $Z$ $p_T$ data and the D{\O} $W$ asymmetry, see Figs.~\ref{fig:mcas118ATLAS} and ~\ref{fig:mcas118ATLASD0}.
Overall, there is 
some element of tension between the preferred value from the global fit and 
the fit to charm data. We do not attempt to make a rigorous determination of 
the best value of the mass or its uncertainty as we believe there are more precise and 
better controlled methods for this. However, a rough indication of the 
uncertainty could be obtained from the $\chi^2$ profiles by treating $m_c$ 
in the same manner as the standard PDF eigenvectors and applying the dynamic 
tolerance procedure. 

\begin{table}
\begin{center}
\renewcommand\arraystretch{1.25}

\begin{tabular}{|l|l|l|l|}
\hline
   $m_c$ (GeV)       &  $\chi^2_{\rm global}$  &   $\chi^2_{\tilde{\sigma}^{\cc}}$ &  
$\alpha_S(M_Z^2)$        \\
          &  4363 pts &  79 pts &  \\
\hline
1.2   &  5823  & 198   & 0.1200     \\
1.25   &  5795  & 175   & 0.1201     \\
1.3  & 5793  &  157  &  0.1200   \\
1.35  & 5776  &  142  &  0.1202   \\
1.4  &  5772   &  130  & 0.1203     \\
1.45  &  5782   &  123  & 0.1204     \\
1.5  & 5799  &  118  &  0.1204   \\
1.55  & 5808  &  116  &  0.1204   \\
1.55  & 5838  &  118  &  0.1206   \\
\hline
    \end{tabular}
\end{center}

\caption{\sf The quality of the fit versus the quark mass $m_c$ at NLO with $\alpha_S(M_Z^2)$ 
left as a free parameter.}
\label{tab:mcasfree}   
\end{table}

\begin{figure} [h]
\begin{center}
\vspace*{-0.0cm}
\includegraphics[scale=0.22]{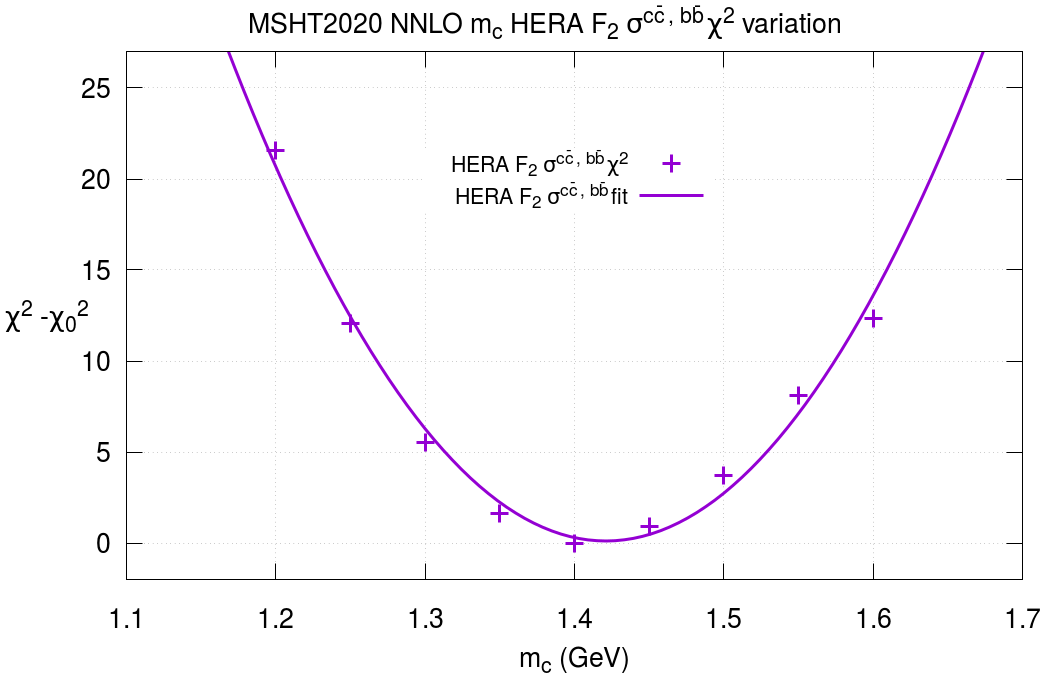}
\includegraphics[scale=0.22]{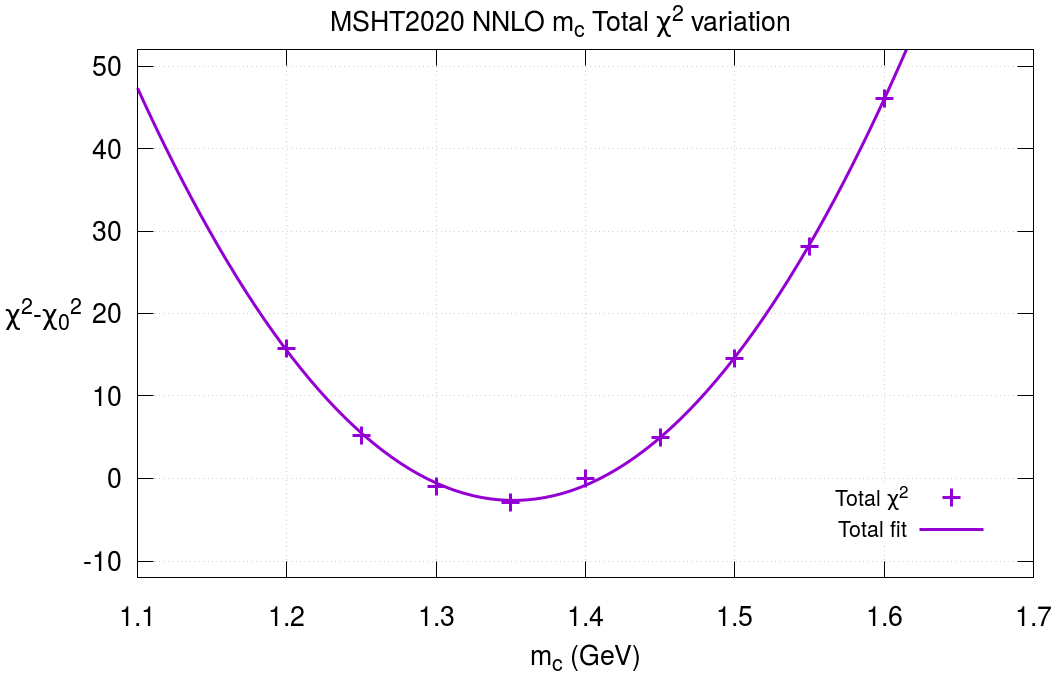}
\caption{\sf The quality of the fit versus the quark mass $m_c$ at NNLO with $\alpha_S(M_Z^2)=0.118$ for (left) the reduced cross section for charm and bottom production
$\tilde{\sigma}^{\cc(\bb)}$ for the combined H1 and ZEUS data and (right) the full global fit.}
\label{fig:mcNNLOas118}
\end{center}
\end{figure}

\begin{figure} [h]
\begin{center}
\vspace*{-0.0cm}
\includegraphics[scale=0.23]{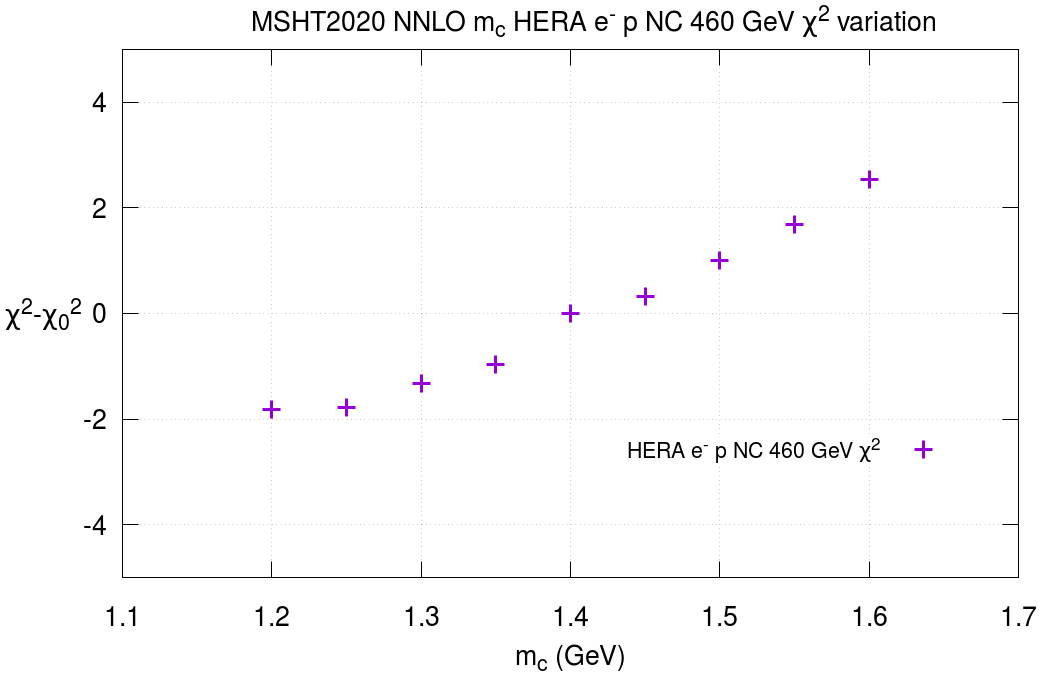}
\includegraphics[scale=0.23]{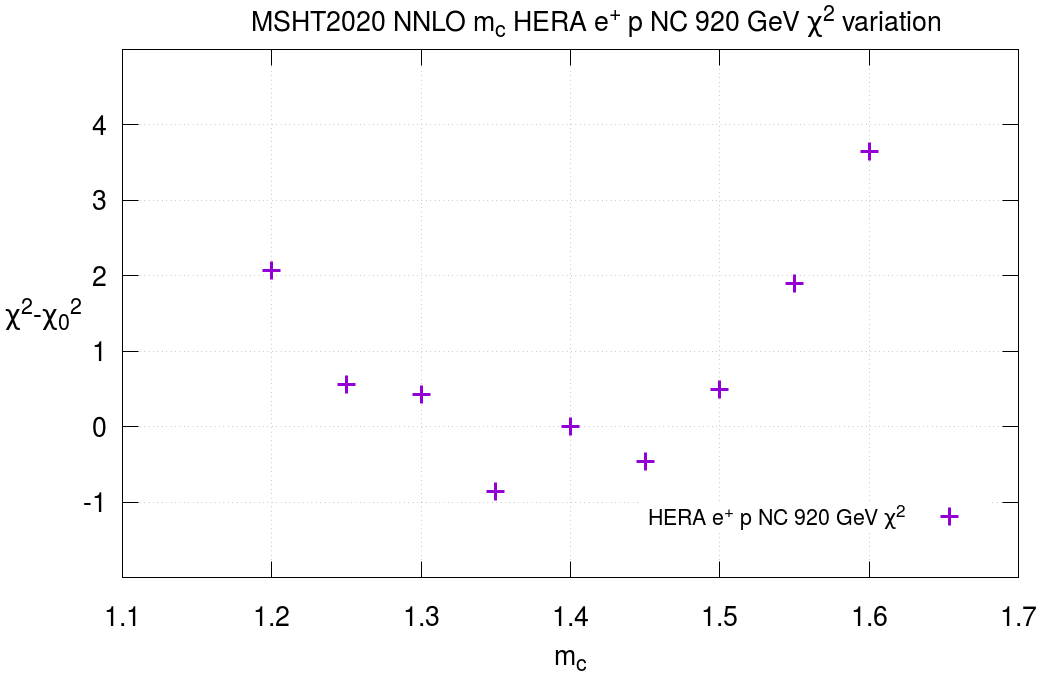}
\caption{\sf The quality of the fit versus the quark mass $m_c$ at NNLO with $\alpha_S(M_Z^2)=0.118$ 
for (left) the total reduced cross section 
$\tilde{\sigma}$ for the combined H1 and ZEUS NC $e^-$ 460~GeV data and (right) NC $e^+$ 920~GeV data.}
\label{fig:mcNNLOas118HERA}
\end{center}
\end{figure}

\begin{figure} [h]
\begin{center}
\vspace*{-0.0cm}
\includegraphics[scale=0.23]{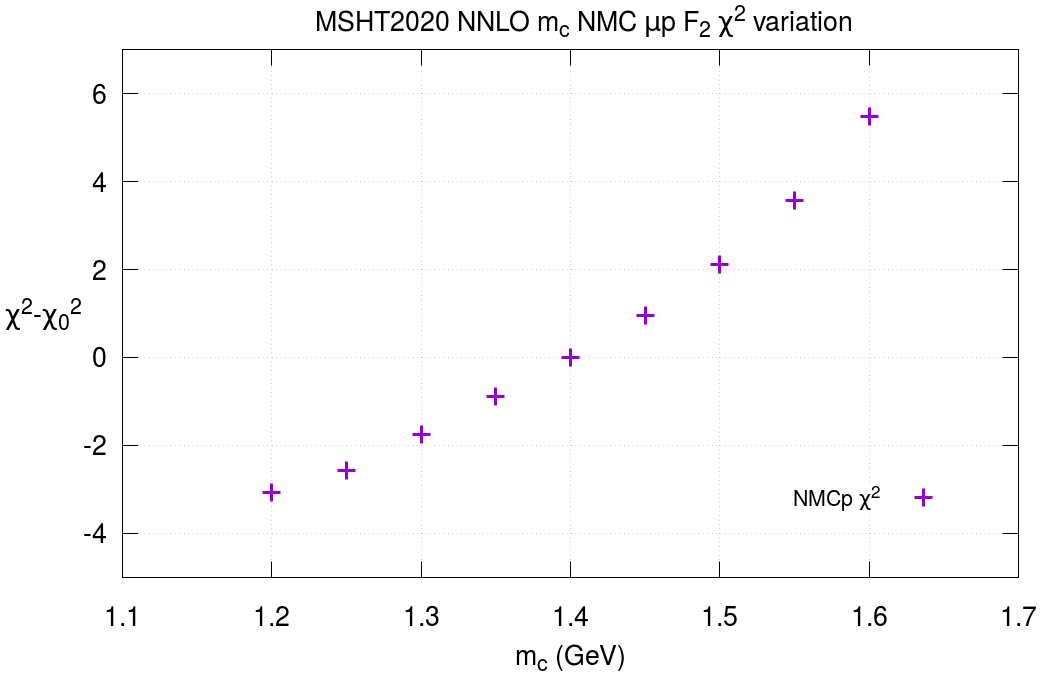}
\includegraphics[scale=0.23]{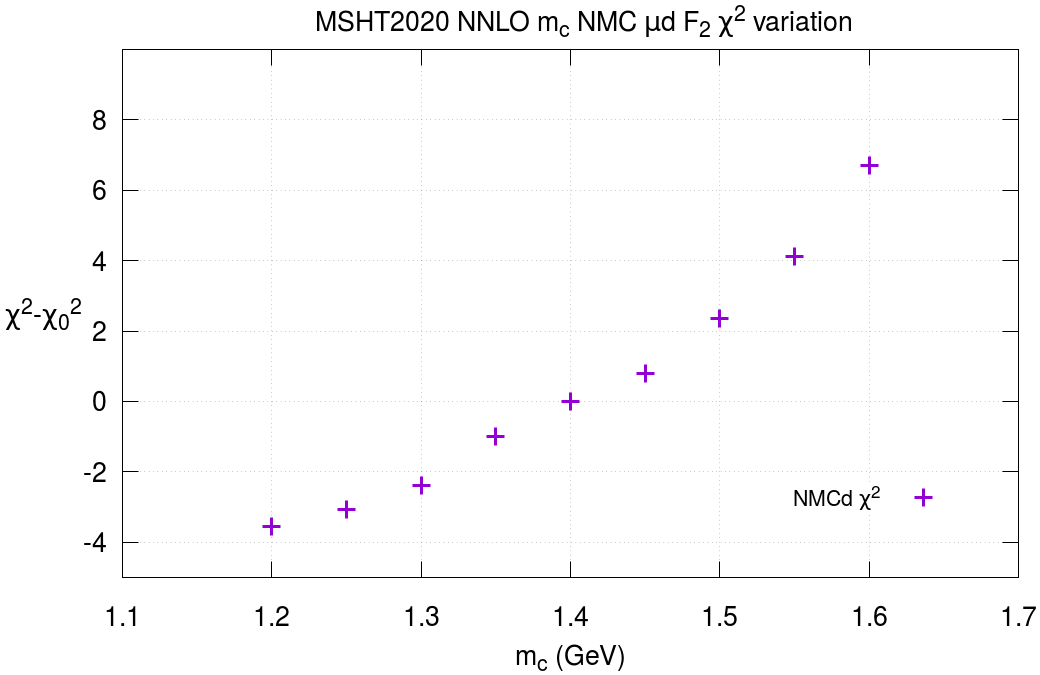}
\caption{\sf The quality of the fit versus the quark mass $m_c$ at NNLO with $\alpha_S(M_Z^2)=0.118$ for (left) 
the NMC $F^p_2(x,Q^2)$ data and (right) the $F^d_2(x,Q^2)$ data.}
\label{fig:mcNNLOas118NMC}
\end{center}
\end{figure}

\begin{figure} [h]
\begin{center}
\vspace*{-0.0cm}
\includegraphics[scale=0.23]{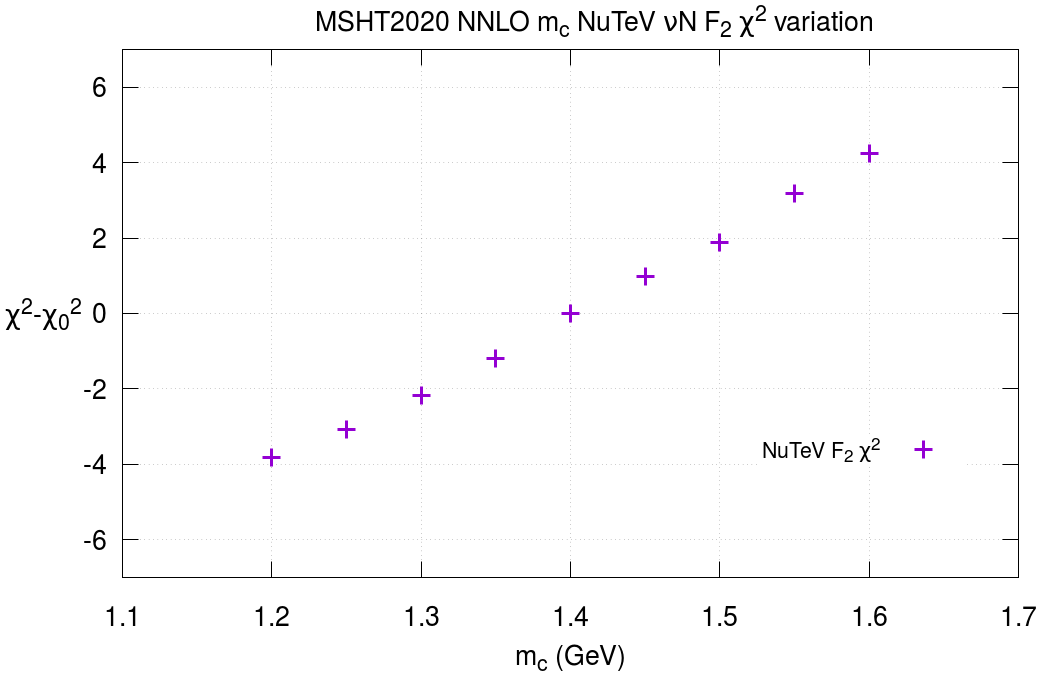}
\includegraphics[scale=0.23]{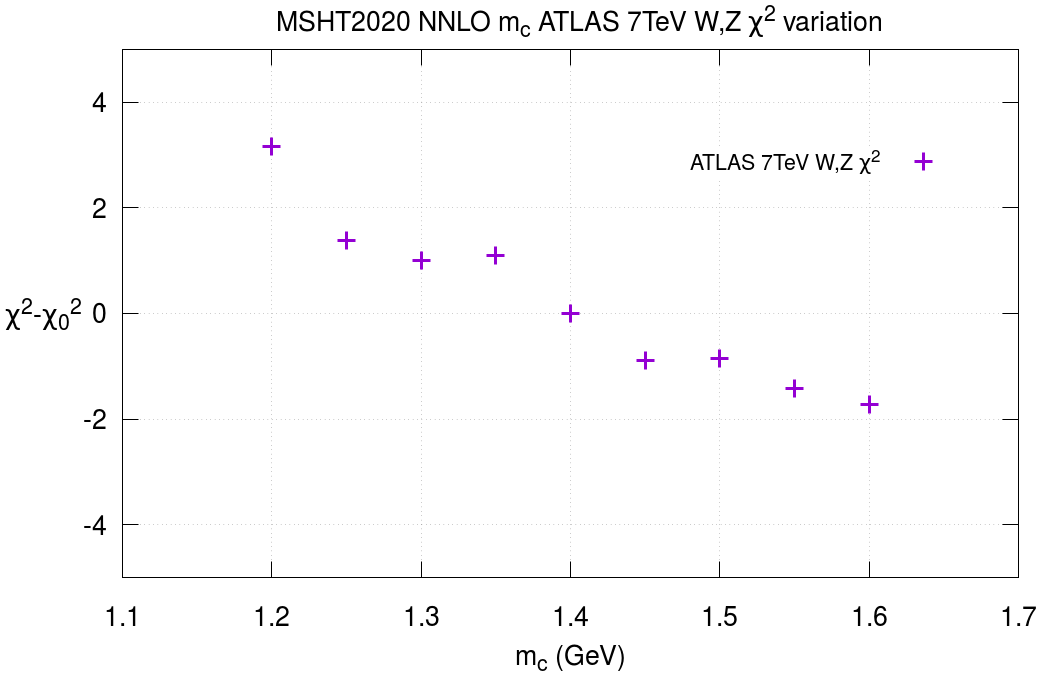}
\caption{\sf The quality of the fit versus the quark mass $m_c$ at NNLO with $\alpha_S(M_Z^2)=0.118$ for (left) the NuTeV $F_2(x,Q^2)$ data
 and (right) the ATLAS 7~TeV $W,Z$ data.}
\label{fig:mcNNLOas118ATLAS}
\end{center}
\end{figure}

\begin{figure} [h]
\begin{center}
\vspace*{-0.0cm}
\includegraphics[scale=0.22]{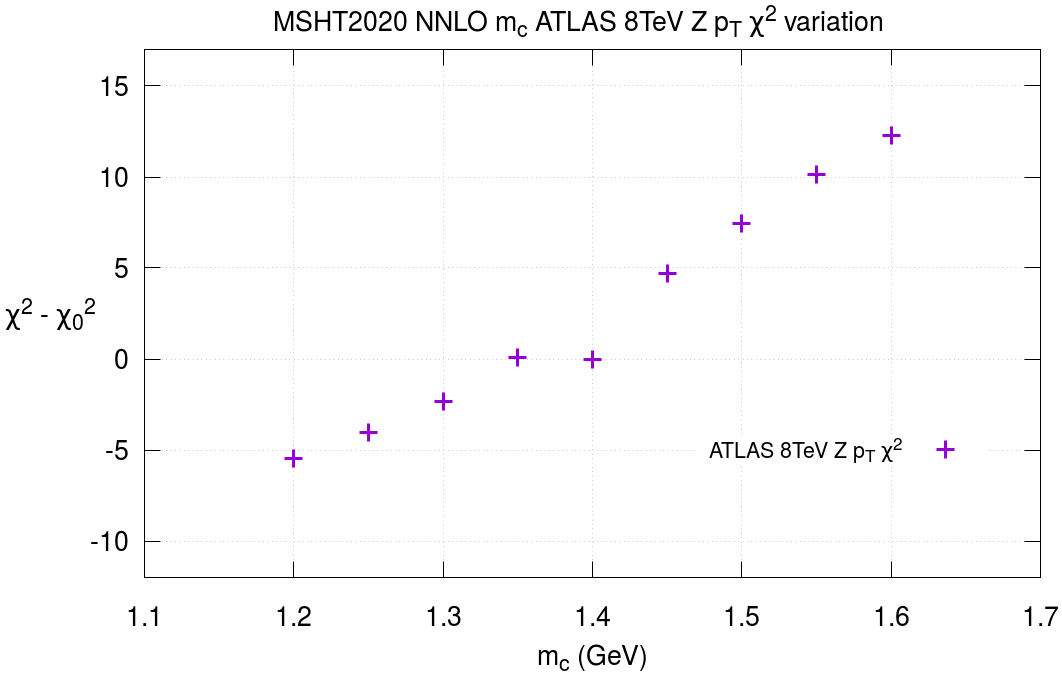}
\includegraphics[scale=0.22]{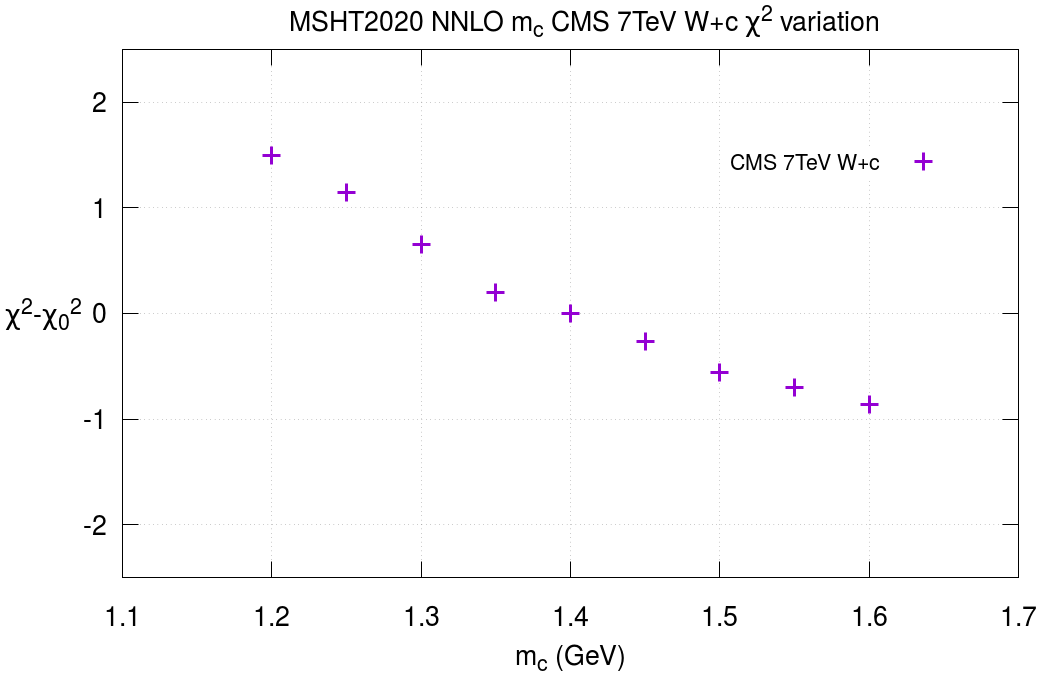}
\caption{\sf The quality of the fit versus the quark mass $m_c$ at NNLO with $\alpha_S(M_Z^2)=0.118$ for (left) 
the ATLAS 8~TeV $Z$ $p_T$  data and (right) the CMS 7~TeV $W+c$ jet data.}
\label{fig:mcNNLOas118ATLASCMS}
\end{center}
\end{figure}

\begin{table}
\begin{center}
\renewcommand\arraystretch{1.25}
\begin{tabular}{|l|l|l|l|}
\hline
   $m_c$ (GeV)       &  $\chi^2_{\rm global}$  &   $\chi^2_{\tilde{\sigma}^{\cc}}$ &  
$\alpha_S(M_Z^2)$        \\
          &  4363 pts &  79 pts &  \\
\hline
1.2   &  5134  & 153   & 0.1172     \\
1.25   &  5123  & 143   & 0.1172     \\
1.3  & 5118  &  137  &  0.1173   \\
1.35  & 5117  &  133  &  0.1173   \\
1.4  &  5119   &  132  & 0.1174     \\
1.45  &  5125   &  132  & 0.1175     \\
1.5  & 5136  &  135  &  0.1175   \\
1.55  & 5150  & 140  &  0.1176   \\
1.6  & 5168  & 144  &  0.1177   \\
\hline
    \end{tabular}
\end{center}

\caption{\sf The quality of the fit versus the quark mass $m_c$ at NNLO with $\alpha_S(M_Z^2)$
left free.}
\label{tab:mcNNLOasfree}   
\end{table}

The analogous results  for the global fit quality with $\alpha_S(M_Z^2)$ left 
free  are shown in Table \ref{tab:mcasfree}, where the corresponding $\alpha_S(M_Z^2)$ values 
are shown as well.  The results with free  $\alpha_S(M_Z^2)$ show
the preferred value of $\alpha_S(M_Z^2)$ falling slightly with 
lower values of $m_c$. However, at all masses the value of $\a$ remains close to the 
$\a=0.120$ used in the default fits so the variation of $\chi^2$ with $m_c$ is very similar
to the fixed $\a$ case and the minimum remains near $m_c=1.35-1.4~\GeV$. 

The results of the same analysis at NNLO are shown for $\alpha_S(M_Z^2)=0.118$
and $\alpha_S(M_Z^2)$ left free in Fig.~\ref{fig:mcNNLOas118} and Table
\ref{tab:mcNNLOasfree}, respectively, where again  in the latter case the 
corresponding $\alpha_S(M_Z^2)$ values are shown. 
The variation in the fit quality for the HERA combined charm and bottom cross section data is 
much reduced compared to NLO. The heavy flavour data clearly prefer a value at NNLO close to 
the default of $m_c=1.4~\GeV$. 
The deterioration is clearly such as to 
make very low values of $m_c$ strongly disfavoured, in contrast to MMHT14.
The variation in the fit quality to the global data set is quite similar this time,
with a preference
for values near to $m_c=1.35~\GeV$.  
Compared to NLO there is little constraint coming 
from the inclusive HERA cross section data, shown in Fig.~\ref{fig:mcNNLOas118HERA}. However, there is still a distinct preference 
for a low value of the mass from NMC structure function data shown in Fig.~\ref{fig:mcNNLOas118NMC}, where again the 
data for $x\sim 0.01$ and $Q^2\sim 4~\GeV^2$ are sensitive to the turn-on of 
the charm contribution to the structure function and prefer a lower value giving quicker
evolution. At NNLO there is also a more clear similar effect for NuTeV $F_2(x,Q^2)$ data in Fig.~\ref{fig:mcNNLOas118ATLAS} (left).  The ATLAS 7~TeV $W,Z$ data again
distinctly prefer a high value of $m_c$, see Fig.~\ref{fig:mcNNLOas118ATLAS} (right). At NLO the fit to these data is so poor that it is difficult
to attach as much importance to this result, but at NNLO it is more significant. The larger 
charm mass means suppression of the charm quark distribution. The charm quark (and antiquark) make 
more contribution to $W$ production at the LHC via charm-antistrange annihilation 
(or strange-anticharm) than to $Z$ production, due to charm-anticharm annihilation. Hence, a charm 
suppression lowers the $W^{+,-}$ cross section compared to the $Z$ cross section, particularly at 
lower rapidity where charm and strange quark distributions are more comparable to those of the up and down quark than at high $x$ where valence quarks 
dominate. This is what 
the data prefer, and charm suppression effectively acts in the same manner as strange quark enhancement,
which is a well-known feature of these data. Similar charm suppression is effectively seen in the 
NNPDF3.1 PDFs \cite{Ball:2017nwa}, but in that case due to direct suppression for $0.01<x<0.1$ in the input fitted 
charm rather than via the mass. There is also significant  sensitivity to the charm mass in the ATLAS 8~TeV
$Z p_T$ data and the CMS $W+c$ jet data, shown in Fig.~\ref{fig:mcNNLOas118ATLASCMS}. 

The analogous results  for the global fit quality with $\alpha_S(M_Z^2)$ left
free at NNLO are shown in Table \ref{tab:mcNNLOasfree}, where the corresponding $\alpha_S(M_Z^2)$ values
are shown.  As at NLO the results with free  $\alpha_S(M_Z^2)$ show
the preferred value of $\alpha_S(M_Z^2)$ falling slightly with
lower values of $m_c$. The value of $\a$ remains close to the best fit value of
$\a=0.1174$ and again the variation of $\chi^2$ with $m_c$ is very similar
to the fixed $\a$ case with the global minimum remaining at $m_c=1.35~\GeV$.

Broadly speaking, the results at NLO and NNLO are similar, but with 
greater $\chi^2$ variation at NLO. In both cases the preferred 
value of $m_c$ in the global fit is now only slightly below our default value, and much closer than
 for the MMHT14 study \cite{MMHThq}.

\subsection{Dependence on $m_b$}

\begin{figure} [h]
\begin{center}
\vspace*{-0.0cm}
\includegraphics[scale=0.22]{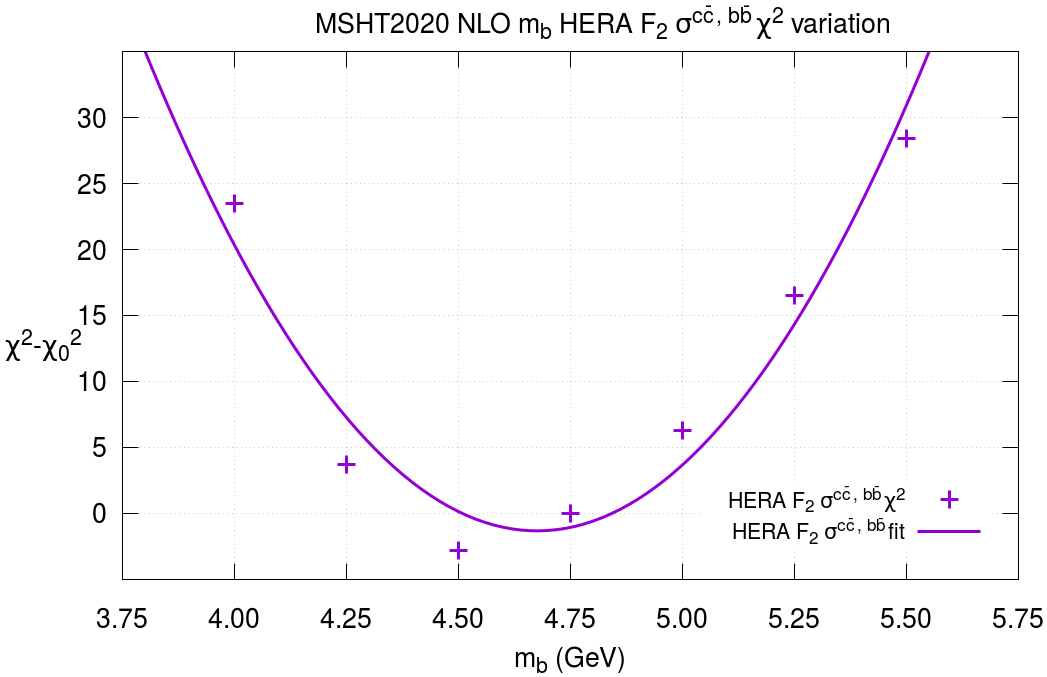}
\includegraphics[scale=0.22]{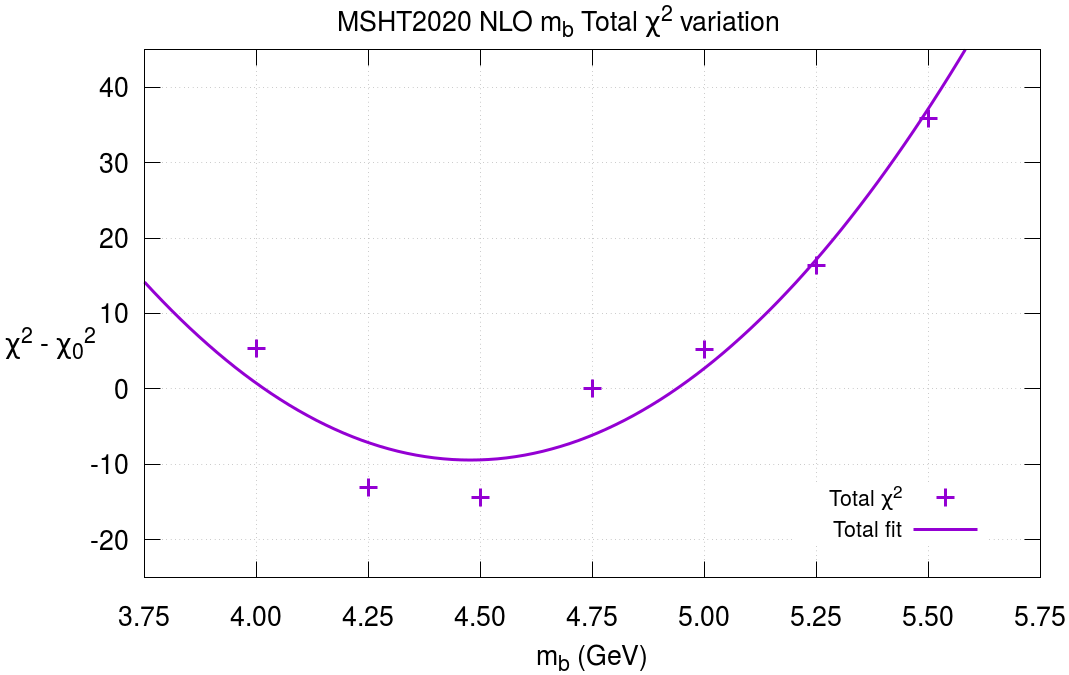}
\caption{\sf The quality of the fit versus the quark mass $m_b$ at NLO with $\alpha_S(M_Z^2)=0.118$ for (left) 
the reduced cross section for heavy flavour production
$\tilde{\sigma}^{\cc(\bb)}$ for the H1 and ZEUS data and (right) the global fit.}
\label{fig:mbas118}
\end{center}
\end{figure}

We repeat essentially the same procedure for the bottom quark mass, $m_b$. We now vary the values of 
$m_b$ in the range $4-5.5~\GeV$ in steps of $0.25~\GeV$. 
The results for the NLO PDFs with $\alpha_S(M_Z^2)=0.118$  
are shown in Fig.~\ref{fig:mbas118}. In this case we again provide a quadratic fit line to the ${\Delta \chi^2}$ values, in order to indicate the behaviour, though in this case the changes are sufficiently small that very small fluctuations in the fit quality make the quadratic behaviour less clear. There is a  
tendency to prefer slightly low values of $m_b$, similar to the results in 
\cite{MMHThq}. For the predictions to the heavy flavour
cross section data, the preference is for 
low values of $m_b \sim 4.5-4.75~\GeV$ but this is slightly lower, 
$m_b \sim 4.25-4.5~\GeV$ in the global fit. There is not a very clear pull 
from most data sets other than the heavy flavour data, so it is largely a cumulative effect, 
though at NLO the  ATLAS 7~TeV $W,Z$ data quite strongly disfavours $m_b>5~\GeV$ and the 
inclusive HERA combined data also prefer lower values. In the case of $m_b$ it is likely the 
fit quality of some data sets is affected as much, or even more, by the change in the details of the running of the coupling as by the change in the PDFs.

\begin{figure} [h]
\begin{center}
\vspace*{-0.0cm}
\includegraphics[scale=0.22]{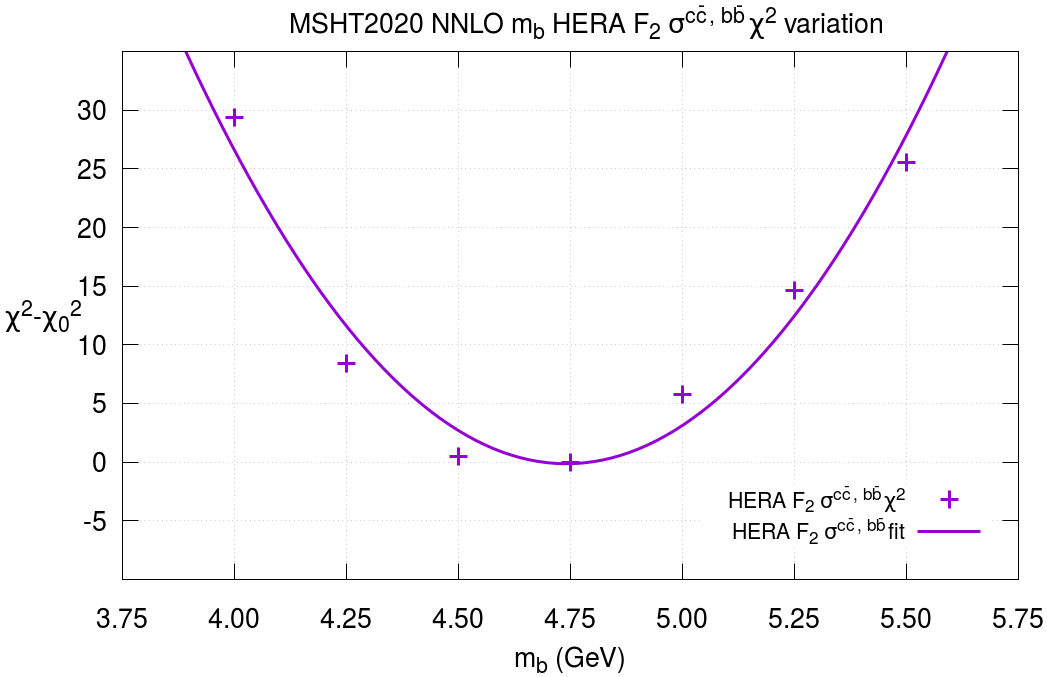}
\includegraphics[scale=0.22]{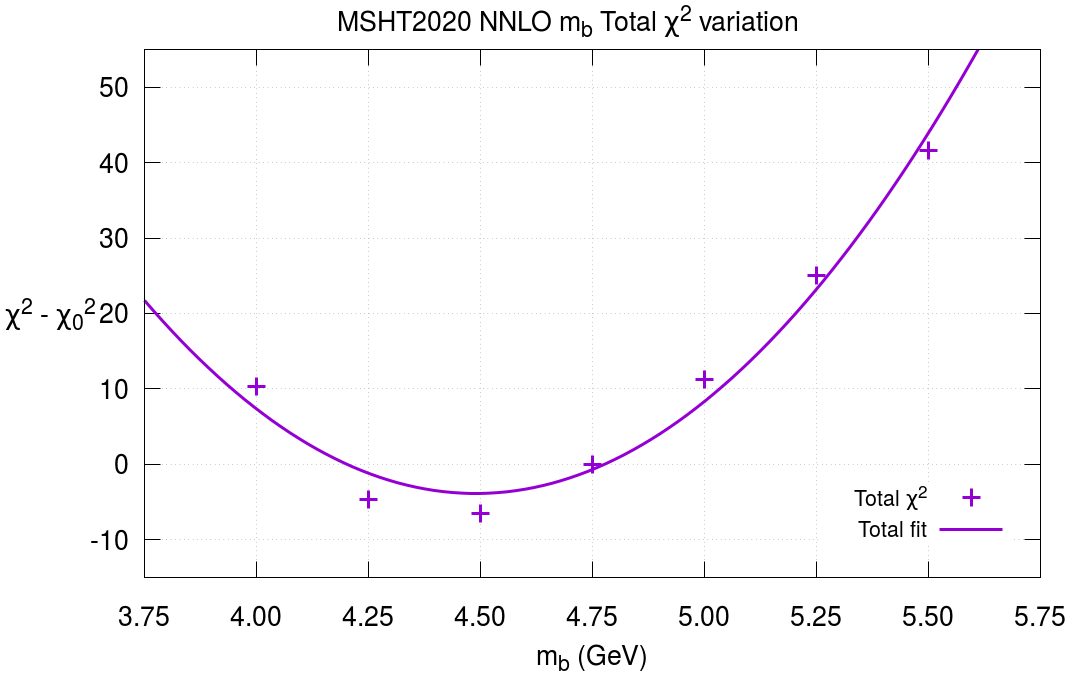}
\caption{\sf The quality of the fit versus the quark mass $m_b$ at NNLO with $\alpha_S(M_Z^2)=0.118$ for (left) 
the reduced cross section for heavy flavour production
$\tilde{\sigma}^{\cc(\bb)}$ for the H1 and ZEUS data and (right) the global fit.}
\label{fig:mbasNNLO118}
\end{center}
\end{figure}

The results for the NNLO fit with $\alpha_S(M_Z^2)=0.118$ are shown in Fig.~\ref{fig:mbasNNLO118}. The global fit is fairly weakly dependent 
on $m_b$,  
and prefers a value $m_b=4.25-4.75~\GeV$. As in the NLO case the $\chi^2$ for the 
prediction for $\tilde{\sigma}^{\cc(\bb)}$ is better for slightly higher values of $m_b$
and the $\chi^2$ minimises for  $m_b=4.75~\GeV$, which is our default value. The inclusive HERA combined data again prefers 
lower values of $m_b$, and in this case this is the dominant reason for the global fit having a 
minimum in $\chi^2$ a little below the fit to heavy flavour data. 

In summary, the constraints on $m_b$ are relatively weak, and at both NLO and NNLO there is general
compatibility with our default value of $m_b=4.75~\GeV$, particularly from the most direct constraint 
from HERA heavy flavour data, although the global fit prefers a slightly lower value. 

\subsection{Changes in the PDFs}

\begin{figure}
\begin{center}
\includegraphics[scale=0.205]{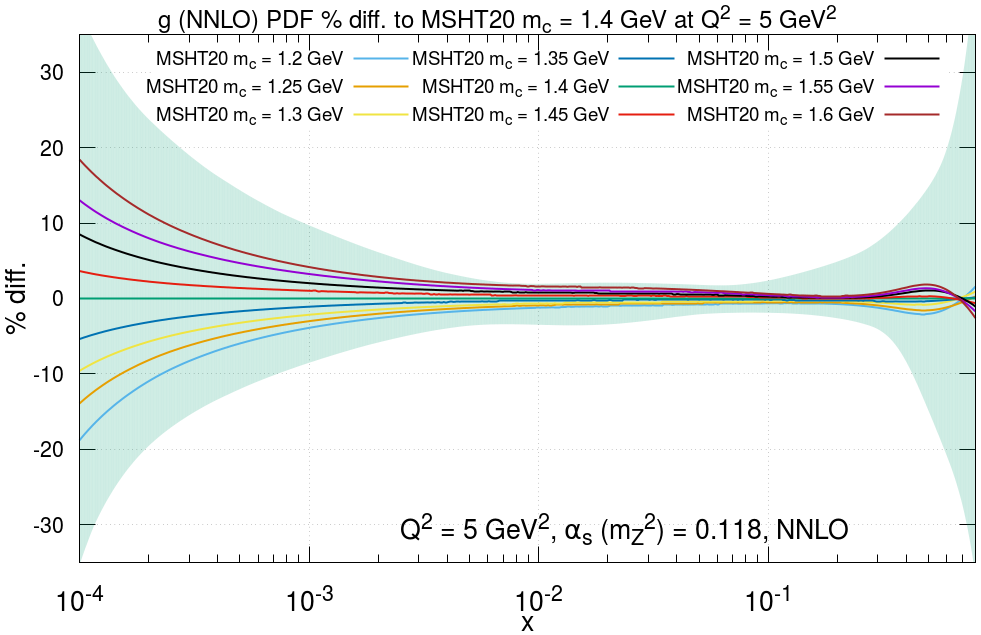}
\includegraphics[scale=0.205]{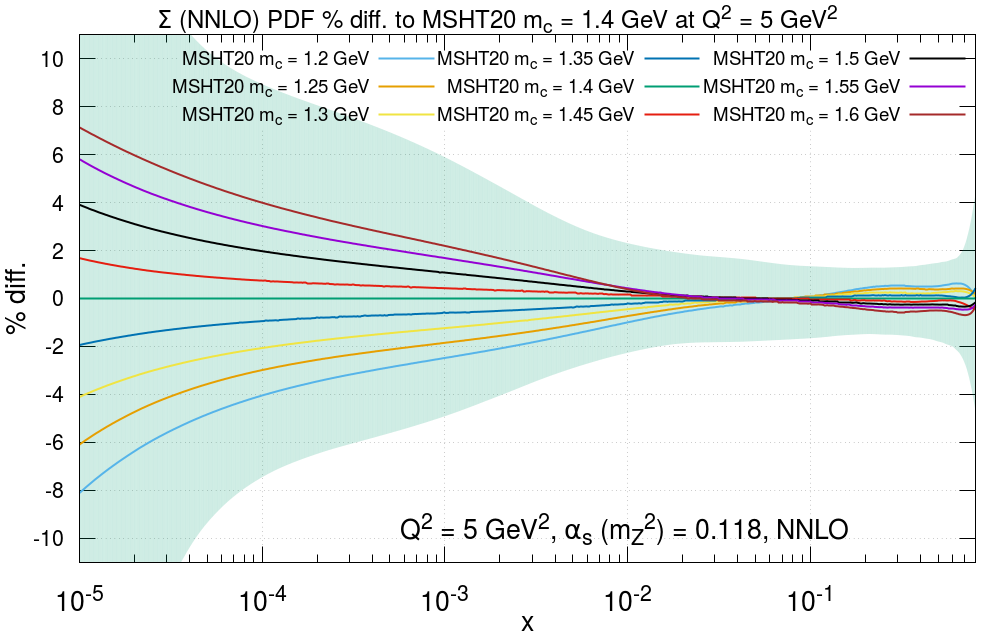}
\includegraphics[scale=0.205]{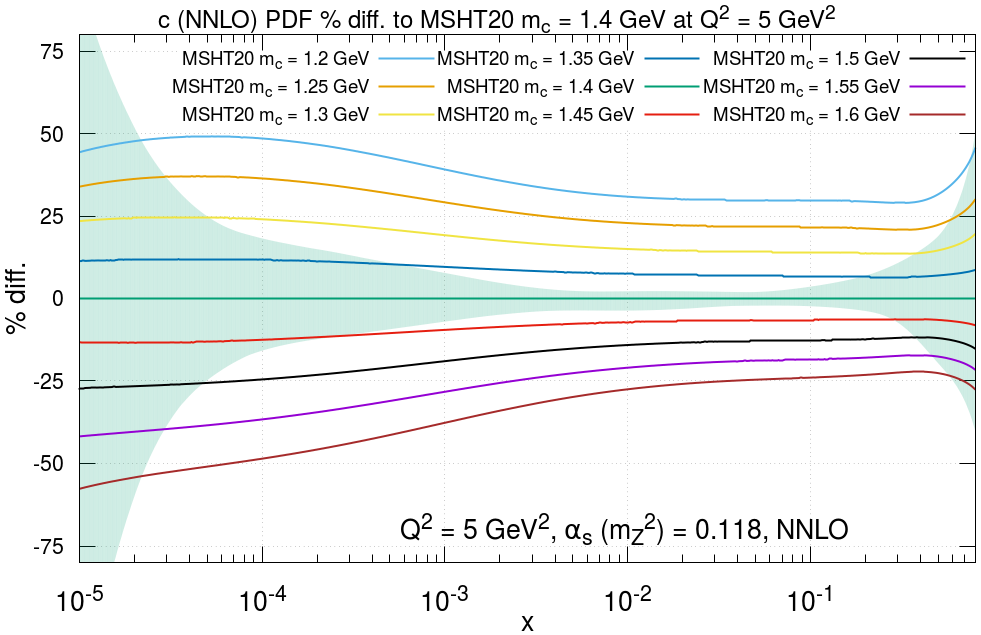}
\caption{\sf The $m_c$ dependence of the gluon, light-quark singlet and charm distributions
at NNLO for $Q^2=5~\GeV^2$, compared to the 
standard MSHT20 distributions with $m_c=1.4~\GeV$ and $m_b=4.75~\GeV$.}
\label{fig:PDFsmcq4}
\end{center}
\end{figure}

\begin{figure}
\begin{center}
\includegraphics[scale=0.205]{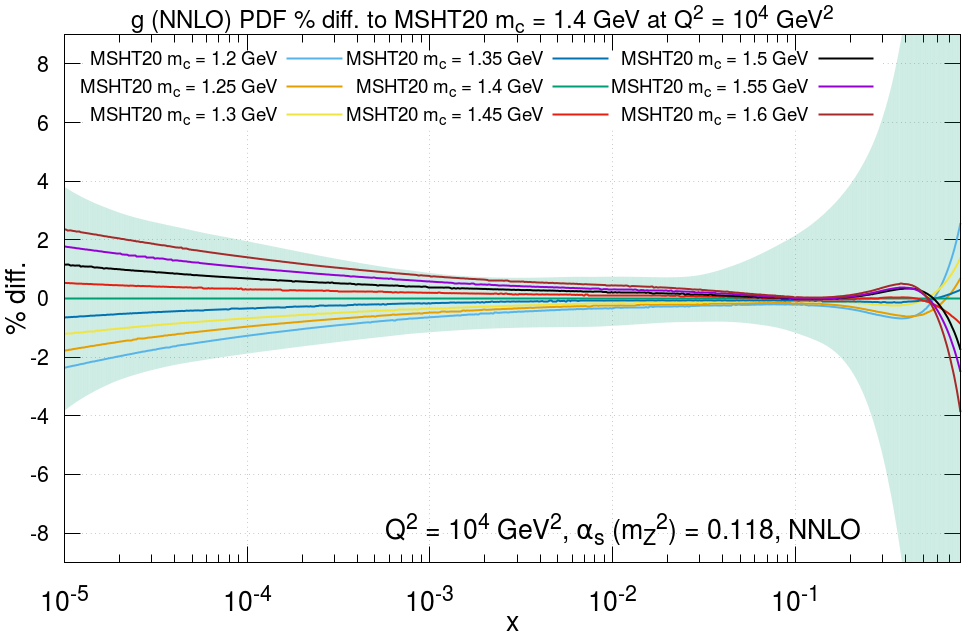}
\includegraphics[scale=0.205]{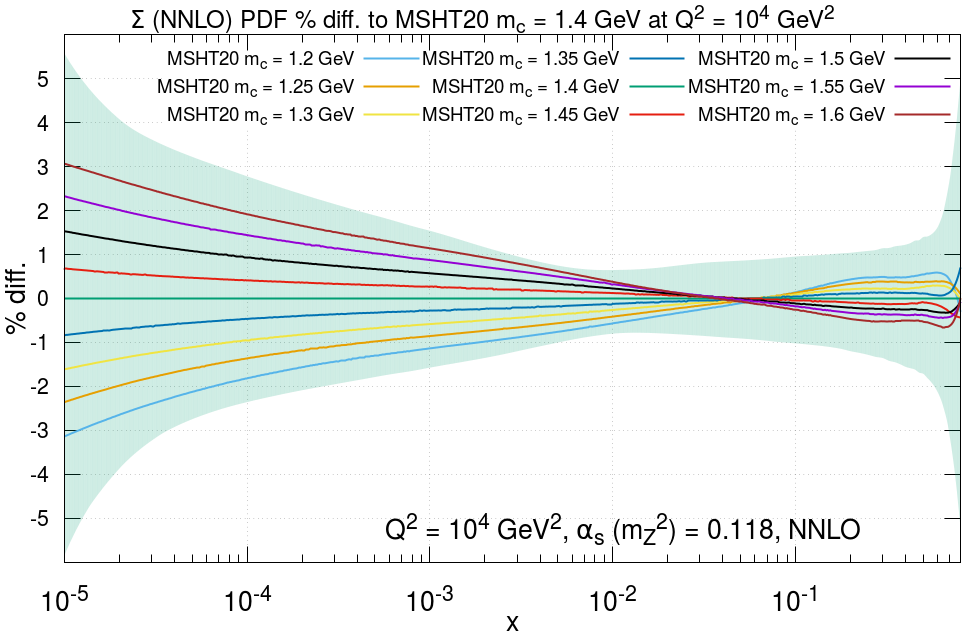}
\includegraphics[scale=0.205]{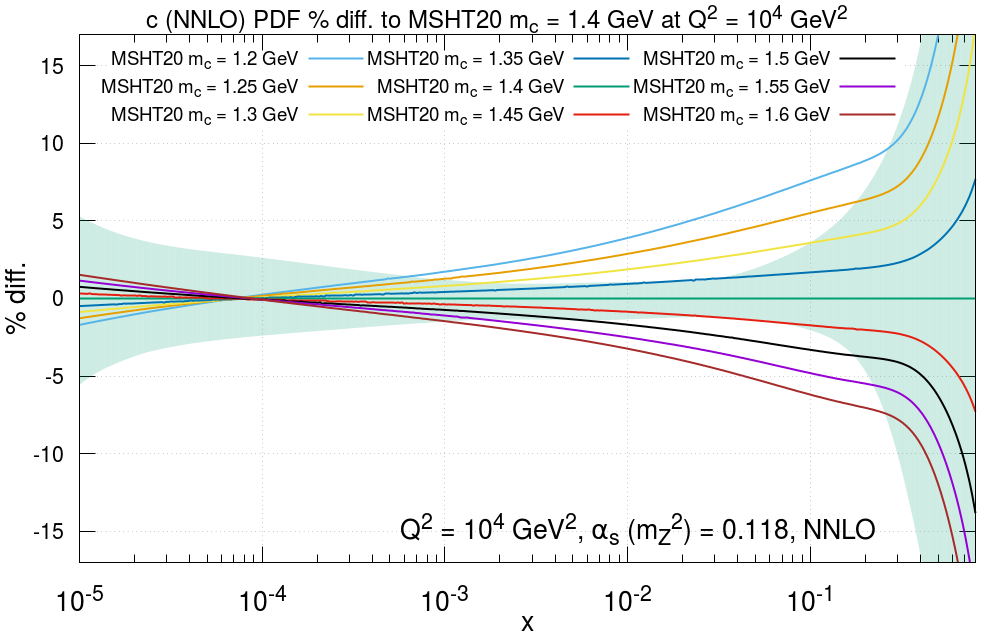}
\caption{\sf The $m_c$ dependence of the gluon, light-quark singlet and charm distributions
at NNLO for $Q^2=10^4~\GeV^2$, compared to the 
standard MSHT20 distributions with $m_c=1.4~\GeV$ and $m_b=4.75~\GeV$.}
\label{fig:PDFsmcq10000}
\end{center}
\end{figure}

We show how the NNLO PDFs change for the whole range of $m_c$ variations from $m_c=1.2-1.6~\GeV$, comparing them to 
the central PDFs with $m_c = 1.4~\GeV$ and their uncertainty bands in Figs.~\ref{fig:PDFsmcq4} and~\ref{fig:PDFsmcq10000}.  
We see at $Q^2=5$ GeV$^2$ (that is, quite close to the transition point $Q^2=m_c^2$) 
that the change in the gluon is well within its uncertainty band, though there is a slight increase 
in the gluon, mainly at smaller $x$ with higher $m_c$. The increased gluon with higher $m_c$ quickens the evolution of the 
structure function, which is suppressed by larger mass, so that these effects compensate each other. Similarly the light quark
singlet distribution increases slightly at low $Q^2$ and small $x$ for larger 
$m_c$ to make up for the smaller charm contribution to the small $x$ structure function, and decreases a little at 
higher $x$ in order
to maintain momentum conservation. This trend is maintained at larger scales (e.g. as shown at $Q^2=10^4 \GeV^2$ in Fig.~\ref{fig:PDFsmcq10000}), helped at small $x$ by the increased gluon, 
with a crossing point at $x=0.06$. For both the gluon and the singlet quark distributions, 
however, even at low $Q^2$ the changes are within uncertainties for these
variations in $m_c$. The charm distribution increases at low $Q^2$ for decreasing
$m_c$, and vice versa, simply due to the change in evolution 
length, $\ln(Q^2/m_c^2)$. This effect is  well outside the uncertainty band 
of the central fit, but this should be expected. We have identified the transition 
point at which heavy flavour evolution begins with the quark mass. This has 
the advantage that the boundary condition for evolution is zero up to NLO
(with our further assumption that there is no intrinsic charm), though there
is a finite ${\cal O}(\alpha_S^2)$ boundary condition at NNLO in the GM-VFNS, 
available in \cite{Buza:1996wv}. In principle the results on the charm 
distribution at relatively low scales, such as that in Fig.~\ref{fig:PDFsmcq4}
are sensitive to this choice of transition point at finite order, though as the order in QCD 
increases the correction for changes due to different choices of 
transition point arising from the corresponding changes 
in the boundary conditions become smaller and smaller, ambiguities 
always being of higher order than the calculation, see e.g.~\cite{quarkmatching}.
At scales more typical of LHC physics, however, the relative change in 
evolution length for the charm distribution is much reduced, as are the 
residual effects of choices relating to choice of transition point and 
intrinsic charm. This can be seen in both the significantly reduced variation in the charm PDF with varying $m_c$ in Fig.~\ref{fig:PDFsmcq10000}, and also partly in the notably reduced uncertainty bands on the central fit charm PDF. As a result, at these scales the change in the charm distribution 
is of the same general size as the PDF 
uncertainty for fixed $m_c$ for much of the $x$ range, with a crossing point at $x \sim 10^{-4}$, although the variation around $10^{-3}-0.2$ is larger,  as seen in Fig.~\ref{fig:PDFsmcq10000}. However, as seen in Section~\ref{Sec:Heavy_quark_mass_dependence}, the variations performed for $m_c$ here are wider than favoured by the $\chi^2$ variations or by the allowed variations in masses determined by other means.
We also note that the charm structure function at these high scales is 
reasonably well represented by the charm distribution, while at low scales,
which certainly includes $Q^2=5~\GeV^2$, this is not true. 
Indeed, at NNLO the boundary
condition for the charm distribution is negative at very low $x$ if 
the transition point is $m_c^2$, but this is more than compensated for 
by the gluon and light quark initiated cross section. 

\begin{figure}
\begin{center}
\includegraphics[scale=0.205]{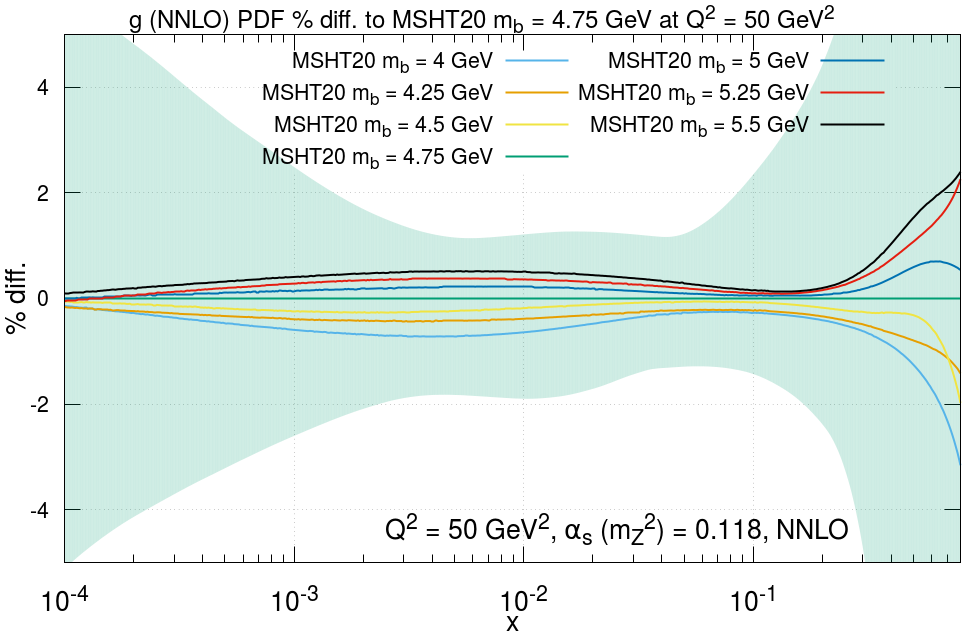}
\includegraphics[scale=0.205]{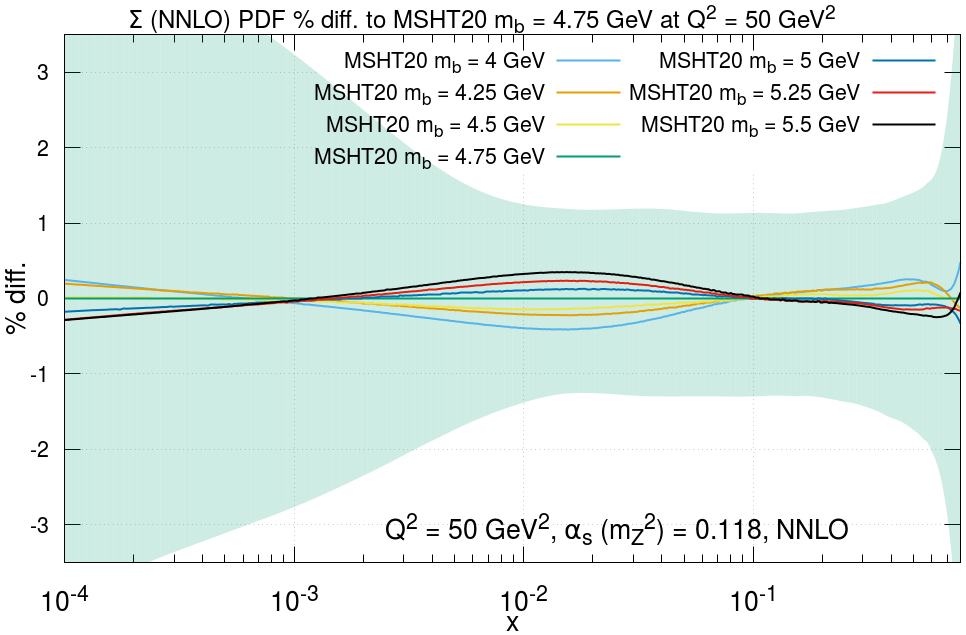}
\includegraphics[scale=0.205]{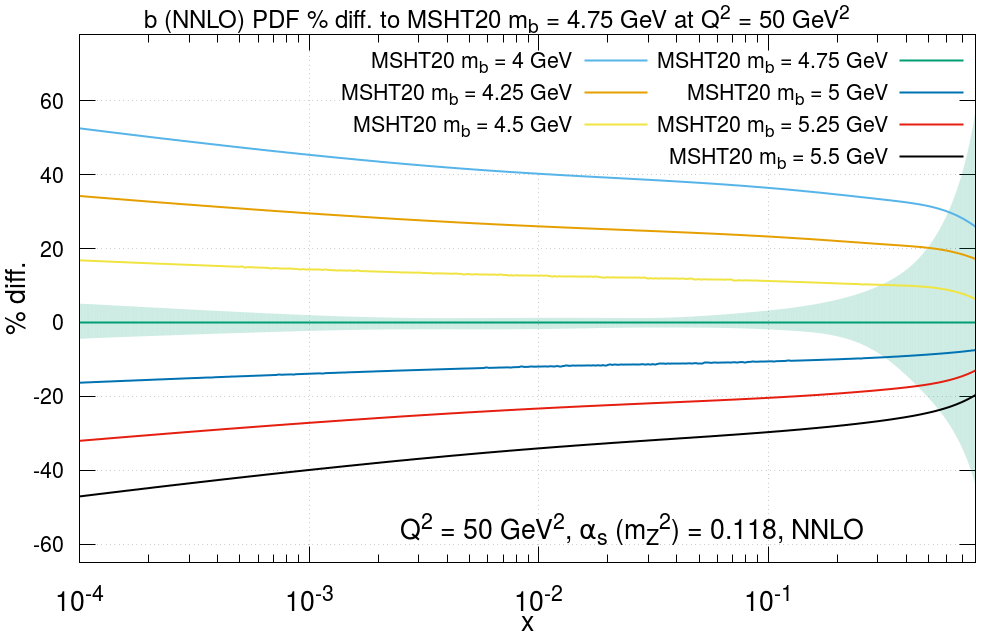}
\caption{\sf The $m_b$ dependence of the gluon, light-quark singlet and bottom distributions
at NNLO for $Q^2=50~\GeV^2$, compared to the 
standard MSHT20 distributions with $m_c=1.4~\GeV$ and $m_b=4.75~\GeV$.
}
\label{fig:PDFsmbq4}
\end{center}
\end{figure}

\begin{figure}
\begin{center}
\includegraphics[scale=0.205]{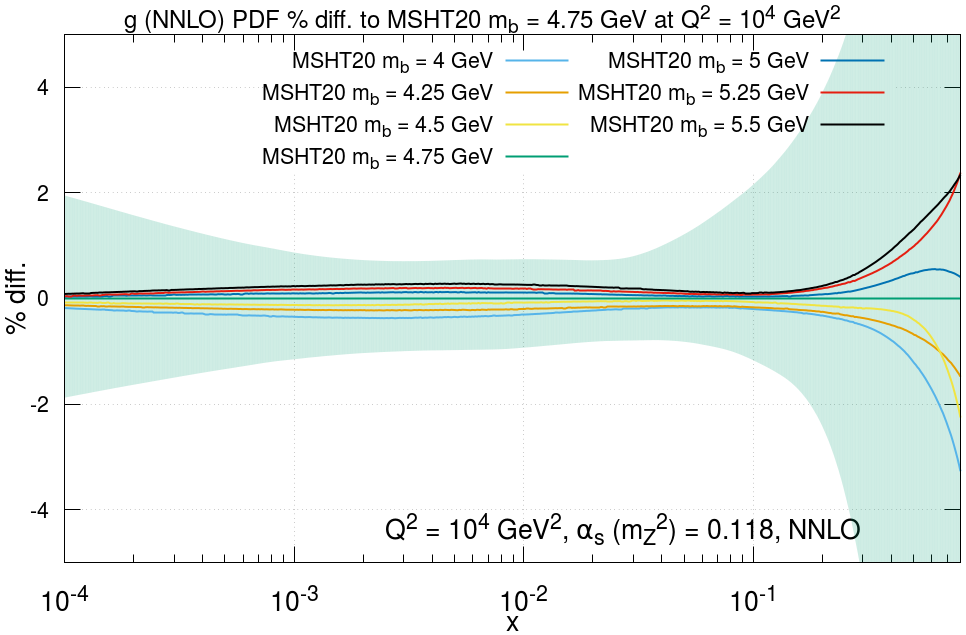}
\includegraphics[scale=0.205]{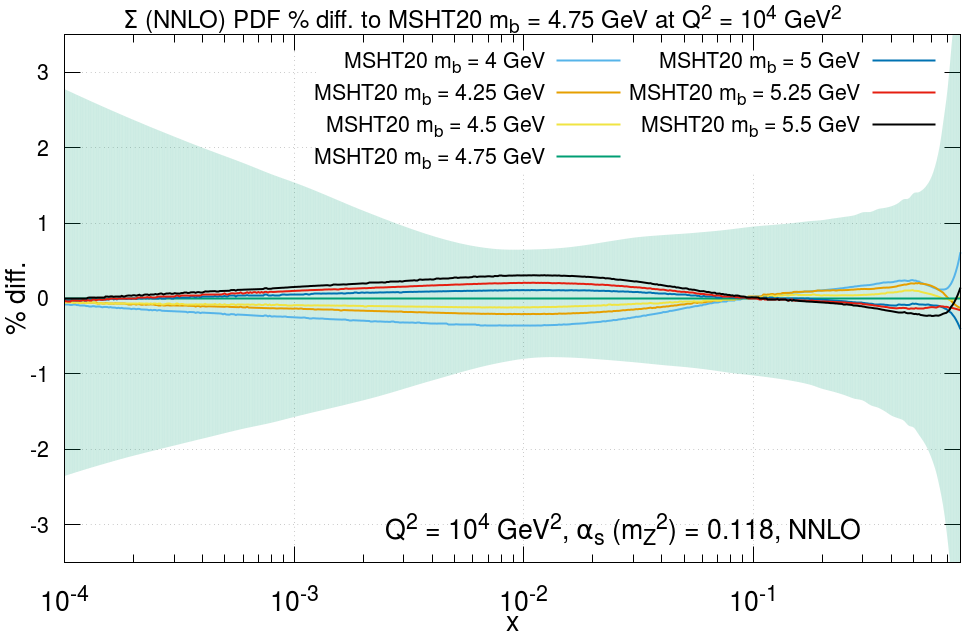}
\includegraphics[scale=0.205]{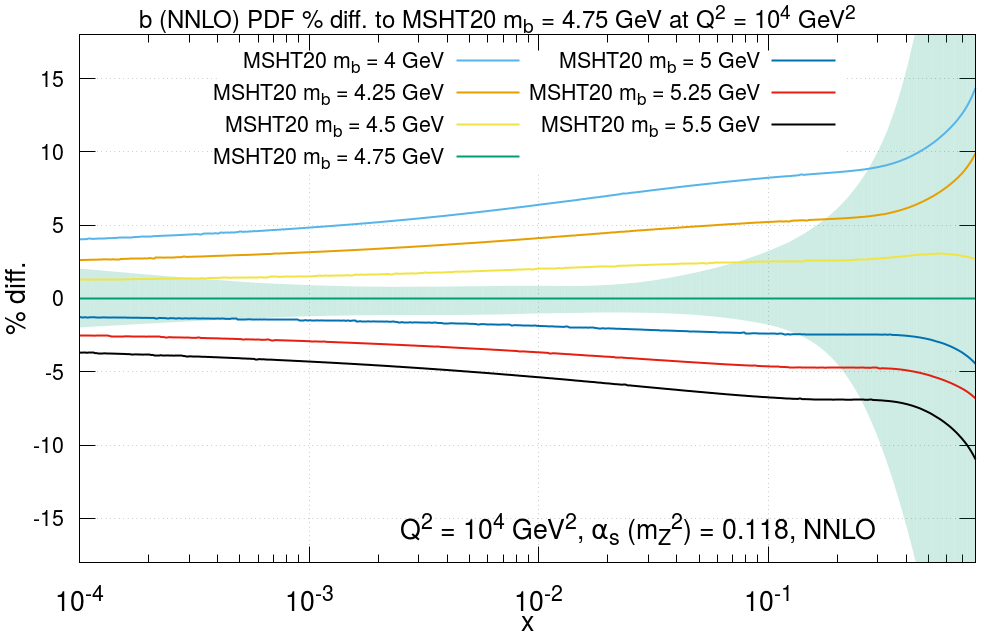}
\caption{\sf The $m_b$ dependence of the gluon, light-quark singlet and bottom distributions
at NNLO for $Q^2=10^4~\GeV^2$, compared to the 
standard MSHT20 distributions with $m_c=1.4~\GeV$ and $m_b=4.75~\GeV$.
}
\label{fig:PDFsmbq10000}
\end{center}
\end{figure}

The relative changes in the gluon and light quarks for variations in $m_b$ are 
significantly reduced, due to the much smaller impact of the bottom contribution 
to the structure functions from the charge-squared weighting, as can be seen 
in Figs.~\ref{fig:PDFsmbq4} and~\ref{fig:PDFsmbq10000}, where we show NNLO PDFs for $m_b=4~\GeV$ to
$m_b=5.5~\GeV$. At $Q^2=50\, {\GeV^2}$ the relative change in the 
bottom distribution for a $\sim 10\%$ change in the mass is similar to that 
for the same type of variation for $m_c$. However, the extent to which
this remains at $Q^2=10^4~\GeV^2$ is greater than the charm case due 
to the smaller evolution length.

\subsection{Effect on benchmark cross sections \label{sec:cxmcmb}}

In this section  we show the variation with $m_c$ and $m_b$ 
for cross sections at the 
Tevatron, for $8~\TeV$ and $13~\TeV$ at the LHC (variations for  
$7~\TeV$ and $14~\TeV$ will be very similar to those at $8~\TeV$ and $13~\TeV$
respectively) as well as for a $100~\TeV$ FCC (pp). We calculate the cross sections for $W$ and $Z$ 
boson, Higgs boson via gluon--gluon fusion and top-quark pair production.
Again our primary aim is not to present definitive predictions or to compare in detail to 
other PDF sets, but to illustrate the relative influence of varying $m_c$ and $m_b$ for these 
benchmark processes.

We show the predictions for the default MSHT20 PDFs, with PDF uncertainties, 
and the relative changes due to changing $m_c$ from $1.25~\GeV$ to $1.55~\GeV$, and 
$m_b$ from $4.25~\GeV$ to $5.25~\GeV$, i.e. changing the default values by 
approximately (but a little larger than) $10\%$ in each direction in each case. The dependence of the benchmark predictions on the value of $m_c$ 
in Tables \ref{tab:sigmaWZNNLO_mcmbvar}~-~\ref{tab:sigmahNNLO_mcmbvar} largely reflects the behaviour 
of the gluon with $x$.
The changes in cross section to good approximation scale linearly in variation 
of masses away from the default values.

\subsubsection{$W$ and $Z$ production}

We begin with the predictions for the $W$ and $Z$ production cross sections. 
The results at NNLO are shown in Table \ref{tab:sigmaWZNNLO_mcmbvar}.
The PDF uncertainties on the cross 
sections are $2\%$ at the Tevatron and slightly smaller at the LHC and larger at the FCC.  The $m_c$ variation 
is about $0.4\%$ at the Tevatron, mainly smaller at 
the LHC, and much  larger for the FCC, i.e. about $1.2-1.5\%$. In all cases increased $m_c$ leads to an increase in the cross
sections. This is due to increased light quarks at small $x$, though the decrease in the charm quark has the opposite effect, 
most significantly at the LHC. The larger effect at the FCC is a consequence of the smaller $x$ sampled, where light quarks 
changes are larger while decreases in the charm quark are smaller. Changes in the cross section with $m_b$ variation 
are $0.3\%$ at most.      

\begin{table}[h]
\begin{center}
\renewcommand\arraystretch{1.25}
\begin{tabular}{|l|c|c|c|c|}
\hline
& $\sigma$& PDF unc.& $m_c$ var.& $m_b$ var.  \\
\hline
$\!\! W\,\, {\rm Tevatron}\,\,(1.96~\TeV)$   & 2.705    & ${}^{+0.054}_{-0.057}$ $\left({}^{+2.0\%}_{-2.1\%}\right)$ & ${}^{+0.010}_{-0.013}$  $\left({}^{+0.37\%}_{-0.47\%}\right)$ & ${}^{-0.0079}_{+0.0029}$  $\left({}^{-0.29\%}_{+0.11\%}\right)$       \\   
$\!\! Z \,\,{\rm Tevatron}\,\,(1.96~\TeV)$   & 0.2506& ${}^{+0.0045}_{-0.0046}$  $\left({}^{+1.8\%}_{-1.8\%}\right)$ &${}^{+0.0009}_{-0.0012}$ $\left({}^{+0.37\%}_{-0.47\%}\right)$ & ${}^{-0.0006}_{+0.0003}$ $\left({}^{-0.26\%}_{+0.11\%}\right)$ \\     
\hline                                                                                                                                                                                                                                                   
$\!\! W^+ \,\,{\rm LHC}\,\, (8~\TeV)$        &7.075    & ${}^{+0.099}_{-0.110}$ $\left({}^{+1.4\%}_{-1.6\%}\right)$  & ${}^{+0.008}_{-0.014}$ $\left({}^{+0.12\%}_{-0.19\%}\right)$  &  ${}^{+0.013}_{-0.010}$ $\left({}^{+0.18\%}_{-0.14\%}\right)$    \\    
$\!\! W^- \,\,{\rm LHC}\,\, (8~\TeV)$        & 4.955  & ${}^{+0.071}_{-0.083}$ $\left({}^{+1.4\%}_{-1.7\%}\right)$  & ${}^{+0.005}_{-0.009}$ $\left({}^{+0.09\%}_{-0.19\%}\right)$    & ${}^{+0.009}_{-0.007}$ $\left({}^{+0.18\%}_{-0.15\%}\right)$ \\    
$\!\! Z \,\,{\rm LHC}\,\, (8~\TeV)$          & 1.122   & ${}^{+0.014}_{-0.017}$ $\left({}^{+1.3\%}_{-1.4\%}\right)$ & ${}^{+0.003}_{-0.004}$ $\left({}^{+0.24\%}_{-0.34\%}\right )$  & ${}^{+0.0006}_{-0.00004}$ $\left({}^{+0.05\%}_{-0.003\%}\right)$ \\ \hline    
$\!\! W^+ \,\,{\rm LHC}\,\, (13~\TeV)$       & 11.53      & ${}^{+0.16}_{-0.18}$ $\left({}^{+1.4\%}_{-1.6\%}\right)$  & ${}^{+0.024}_{-0.028}$ $\left({}^{+0.21\%}_{-0.24\%}\right)$     & ${}^{+0.025}_{-0.022}$ $\left({}^{+0.22\%}_{-0.19\%}\right)$  \\    
$\!\! W^- \,\,{\rm LHC}\,\, (13~\TeV)$       & 8.512     &${}^{+0.12}_{-0.14}$ $\left({}^{+1.4\%}_{-1.6\%}\right)$   & ${}^{+0.013}_{-0.019}$ $\left({}^{+0.15\%}_{-0.23\%}\right)$     & ${}^{+0.018}_{-0.017}$ $\left({}^{+0.21\%}_{-0.19\%}\right)$       \\    
$\!\! Z \,\,{\rm LHC}\,\, (13~\TeV)$        & 1.914  & ${}^{+0.024}_{-0.029}$ $\left({}^{+1.3\%}_{-1.5\%}\right)$  & ${}^{+0.006}_{-0.008}$ $\left({}^{+0.33\%}_{-0.40\%}\right)$      & ${}^{+0.0006}_{+0.0004}$ $\left({}^{+0.03\%}_{+0.02\%}\right)$   \\ 
\hline
$\!\! W^+ \,\,{\rm FCC}\,\, (100~\TeV)$       & 70.82      & ${}^{+2.46}_{-3.08}$ $\left({}^{+3.6\%}_{-4.4\%}\right)$  & ${}^{+0.93}_{-0.92}$ $\left({}^{+1.3\%}_{-1.3\%}\right)$ & ${}^{+0.12}_{-0.12}$ $\left({}^{+0.17\%}_{-0.17\%}\right)$  \\    
$\!\! W^- \,\,{\rm FCC}\,\, (100~\TeV)$       & 60.39     &${}^{+1.65}_{-2.04}$ $\left({}^{+2.9\%}_{-3.3\%}\right)$   & ${}^{+0.70}_{-0.71}$ $\left({}^{+1.2\%}_{-1.2\%}\right)$  & ${}^{+0.10}_{-0.10}$ $\left({}^{+0.17\%}_{-0.16\%}\right)$   \\    
$\!\! Z \,\,{\rm FCC}\,\, (100~\TeV)$         & 13.50  & ${}^{+0.40}_{-0.47}$ $\left({}^{+3.1\%}_{-3.4\%}\right)$  & ${}^{+0.20}_{-0.19}$ $\left({}^{+1.5\%}_{-1.4\%}\right)$ & ${}^{-0.03}_{+0.04}$ $\left({}^{-0.25\%}_{+0.33\%}\right)$   \\ 
\hline
    \end{tabular}
\end{center}
\caption{\sf Predictions for $W^\pm$ and $Z$ cross sections (in nb), including leptonic branching, obtained with the NNLO MSHT20 parton sets. The PDF uncertainties and $m_c$ and 
$m_b$  variations are  shown, where the $m_c$ variation corresponds to $\pm 0.15~\GeV$ and the $m_b$ variation corresponds to $\pm 0.5~\GeV$ , i.e. about
$10\%$ in each case.}
\label{tab:sigmaWZNNLO_mcmbvar}   
\end{table}

\subsubsection{Top-quark pair production}

In Table~\ref{tab:sigmatNNLO_mcmbvar} we show the analogous results for the top-quark pair  production 
cross section. The PDF uncertainties on the cross 
sections are around $2\%$ at the Tevatron and at the LHC, but a little smaller at 100~TeV as there is less sensitivity 
to the high-$x$ gluon. The $m_c$ variation is about $0.5\%$ at the Tevatron, between approximately $0.3\%$ and $0.4\%$ at the LHC 
and around $0.6\%$ at the FCC. At the Tevatron the cross section decreases with increasing $m_c$ due to the decrease in high $x$ light quarks
seen in Fig.~\ref{fig:PDFsmcq10000}, and the dominance of the quark channel at this collider. At the LHC and FCC, where
gluon gluon fusion is the dominant production mechanism, the 
cross section is positively correlated with $m_c$ due to the increase in the gluon distribution.  
Again, changes with $m_b$ are smaller but follow the same pattern as for $m_c$. 

\begin{table}[h]
\begin{center}
\renewcommand\arraystretch{1.25}
\vspace{0.5cm}
\begin{tabular}{|l|c|c|c|c|}
\hline
& $\sigma$& PDF unc.& $m_c$ var.& $m_b$ var. \\
\hline
$t\overline{t}$ $ {\rm Tevatron}\,\,(1.96~\TeV)$ & 7.24    & ${}^{+0.13}_{-0.12}$ $\left({}^{+1.8\%}_{-1.7\%}\right)$  & ${}_{+0.035}^{-0.035}$  $\left({}_{+0.48\%}^{-0.48\%}\right)$  & ${}^{-0.009}_{+0.013}$  $\left({}^{-0.12\%}_{+0.19\%}\right)$   \\   
$t\overline{t}$  ${\rm LHC}\,\, (8~\TeV)$              &243.1   & ${}^{+6.4}_{-3.9}$ $\left({}^{+2.6\%}_{-1.6\%}\right)$  &  ${}^{+0.8}_{-1.0}$ $\left({}^{+0.32\%}_{-0.42\%}\right)$   &  ${}_{-0.58}^{+0.54}$ $\left({}_{-0.24\%}^{+0.23\%}\right)$        \\    
$t\overline{t}$ ${\rm LHC}\,\, (13~\TeV)$             & 796.8   &${}^{+16.0}_{-10.6}$ $\left({}^{+2.0\%}_{-1.3\%}\right)$  & ${}_{-2.6}^{+2.9}$ $\left({}_{-0.33\%}^{+0.36\%}\right)$       & ${}_{-2.2}^{+2.0}$ $\left({}_{-0.27\%}^{+0.25\%}\right)$      \\
$t\overline{t}$ ${\rm FCC}\,\, (100~\TeV)$             & 34600   &${}^{+300}_{-400}$ $\left({}^{+0.9\%}_{-1.2\%}\right)$  & ${}_{-200}^{+200}$ $\left({}_{-0.58\%}^{+0.58\%}\right)$       & ${}_{-120}^{+90}$ $\left({}_{-0.34\%}^{+0.27\%}\right)$      \\
\hline
    \end{tabular}
\end{center}
\caption{\sf Predictions for $t\overline{t}$ cross sections (in pb), obtained with the NNLO MSHT20 parton sets. The PDF uncertainties and $m_c$ and 
$m_b$  variations are  shown, where the $m_c$ variation corresponds to $\pm 0.15~\GeV$ and the $m_b$ variation corresponds to $\pm 0.5~\GeV$.}
\label{tab:sigmatNNLO_mcmbvar}   
\end{table}

\subsubsection{Higgs boson production}

\begin{table}[h]
\begin{center}
\renewcommand\arraystretch{1.25}
\begin{tabular}{|l|c|c|c|c|}
\hline
& $\sigma$& PDF unc.& $m_c$ var.& $m_b$ var.\\
\hline
Higgs $ {\rm Tevatron}\,\,(1.96~\TeV)$   & 0.867    & ${}^{+0.030}_{-0.019}$ $\left({}^{+3.5\%}_{-2.2\%}\right)$ &  ${}^{+0.0028}_{-0.0034}$  $\left({}^{+0.32\%}_{-0.39\%}\right)$  & ${}^{+0.0028}_{-0.0030}$  $\left({}^{+0.32\%}_{-0.35\%}\right)$  \\       
Higgs  ${\rm LHC}\,\, (8~\TeV)$              &18.44    & ${}^{+0.24}_{-0.24}$ $\left({}^{+1.3\%}_{-1.3\%}\right)$  & ${}^{+0.10}_{-0.090}$ $\left({}^{+0.54\%}_{-0.49\%}\right)$   &  ${}^{+0.060}_{-0.070}$ $\left({}^{+0.33\%}_{-0.38\%}\right)$           \\    
Higgs ${\rm LHC}\,\, (13~\TeV)$            & 42.13   &${}^{+0.47}_{-0.51}$ $\left({}^{+1.1\%}_{-1.2\%}\right)$  & ${}^{+0.27}_{-0.23}$ $\left({}^{+0.64\%}_{-0.57\%}\right)$       & ${}^{+0.14}_{-0.16}$ $\left({}^{+0.32\%}_{-0.38\%}\right)$            \\    
Higgs ${\rm FCC}\,\, (100~\TeV)$       & 708.2   &${}^{+9.5}_{-12}$ $\left({}^{+1.3\%}_{-1.7\%}\right)$  & ${}^{+8.3}_{-7.3}$ $\left({}^{+1.2\%}_{-1.0\%}\right)$  & ${}^{+1.8}_{-2.4}$ $\left({}^{+0.25\%}_{-0.34\%}\right)$  \\   
\hline
    \end{tabular}
\end{center}
\caption{\sf Predictions for the Higgs boson cross sections (in pb), obtained with the NNLO MSHT20 parton sets. The PDF uncertainties and $m_c$ and 
$m_b$  variations are  shown, where the $m_c$ variation corresponds to $\pm 0.15~\GeV$ and the $m_b$ variation corresponds to $\pm 0.5~\GeV$.}
\label{tab:sigmahNNLO_mcmbvar}   
\end{table}

In Table \ref{tab:sigmahNNLO_mcmbvar} we show the uncertainties in the rate of Higgs boson production 
from gluon-gluon fusion. For this process the cross section always increases with increasing $m_c$, due to positive correlation between the gluon and $m_c$. 
At the Tevatron the
resultant uncertainty is $\sim 0.3\%$. At the LHC it is a little larger at $\sim 0.5-0.6\%$, whereas at 
the FCC it has increased to about $1\%$ due to the greater change in the small-$x$ gluon. Again changes with $m_b$ are smaller, 
but follow the same trend as for $m_c$.  

We recommend that in order to estimate the total uncertainty due to PDFs and the quark masses
it is best to add the uncertainty due to the variation in quark mass in quadrature with the PDF 
uncertainty, or the PDF+$\alpha_S$ uncertainty, if 
the $\alpha_S$ uncertainty is also used.

\section{PDFs in three- and four-flavour-number-schemes}

In our default studies we work in a general-mass variable-flavour-number-scheme (GM-VFNS) 
with a maximum of 5 active flavours. This means that we start evolution
at our input scale of $Q_0^2=1~\GeV^2$ with three active light flavours. 
At the transition point $m_c^2$ the charm quark starts evolution, from a non-zero value at 
NNLO and beyond, and then 
at $m_b^2$ the bottom quark also starts evolution. The evolution is in terms 
of massless splitting functions, and at high $Q^2$ the contribution from 
charm and bottom quarks lose all mass dependence other than that 
input via the perturbative  
boundary conditions at the chosen transition point. The explicit mass 
dependence is included at lower scales, but falls away like inverse powers 
as $Q^2/m^2_{c,b} \to \infty$. We do not currently ever consider the top quark as 
a parton, though this would probably need to change for detailed studies at $100~\TeV$.

We could alternatively keep the information about the heavy quarks only
in the coefficient functions, i.e. the heavy quarks would only be 
generated in the final state. This is called a fixed-flavour-number-scheme 
(FFNS). For example, we could decide that neither charm and bottom exist as partons,
and this would be a 3-flavour FFNS. Alternatively we could let charm 
evolution turn on, but never allow the bottom quark to be treated as a parton -- 
a 4-flavour FFNS. We will use this notation for PDFs where the bottom quark is absent, 
but strictly speaking it 
is a GM-VFNS with a maximum of 4 active flavours as the charm quark will not exist 
for scales below $m_c^2$.

One might produce the partons for the 3- and 4-flavour FFNS by performing 
global fits in these schemes. However, as discussed in  
\cite{Thorne:2008xf}, the fit to structure function data is not optimal 
in these schemes. Indeed, definite evidence for this has been provided in 
\cite{Thorne,ThorneFFNS,NNPDFgmvfns}. Moreover, 
much of the data (for example, on inclusive jets and $W,Z$ production
at hadron colliders) is not known to NNLO in these schemes, and is 
very largely at scales where $m_{c,b}$ are very small compared to the scales, and 
effectively act as if massless. So it seems clear that the 
GM-VFNS are more appropriate. Hence, in \cite{Martin:2006qz} it was decided
to make available PDFs in the 3- and 4-flavour schemes simply by using the input 
PDFs obtained in the GM-VFNS, but with evolution of the bottom quark, or both 
the bottom and charm quark, turned off. This procedure was continued in \cite{MSTWhq} and
\cite{MMHThq}, and is 
the common choice for PDF groups who fit using a GM-VFNS but make 
PDFs available with a maximum of 3- or 4- active flavours. Hence, here, we continue 
to make this choice for the MSHT20 PDFs. 

We make PDFs available with a maximum of 3 or 4 active flavours for 
the NLO central PDFs and their uncertainty eigenvectors for both 
the standard choices of $\alpha^{n_{f,\max}=5}_S(M_Z^2)$ of 0.118 and 0.120, and for the 
NNLO central PDF and the uncertainty eigenvectors for  
the standard choice of $\alpha^{n_{f,\max}=5}_S(M_Z^2)$ of 0.118.\footnote{In doing 
so we notice a typo in Table 10 of \cite{MSHT20}, where for eigenvector 8 in the negative 
direction we should have $t=1.68, T=1.92$ and the most constraining data set is the D{\O} $W$ asymmetry.}  
We also provide PDF 
sets with $\a$ displaced by 0.001 from these default values, 
in order to facilitate the calculation of $\alpha_S$ uncertainties in the 
different flavour schemes. Finally, we make available PDF sets with different 
values of $m_c$ and $m_b$ in the different fixed-flavour schemes. 

By default, when the charm or bottom quark evolution is turned off, we
also turn off the contribution of the same quark to the running coupling. 
This is because most calculations use this convention when these quarks are treated entirely 
as final state particles. This results in the coupling 
running more quickly above $m_c$ and $m_b$. In particular, if the coupling at $Q_0^2$ is chosen so that 
$\alpha^{n_{f,\max}=5}_S(M_Z^2)\approx 0.118$, then we find that $\alpha^{n_{f,\max}=3}_S(M_Z^2)\approx 0.105$
and $\alpha^{n_{f,\max}=4}_S(M_Z^2)\approx 0.113$. 

   
\begin{figure} [h]
\begin{center}
\includegraphics[scale=0.23]{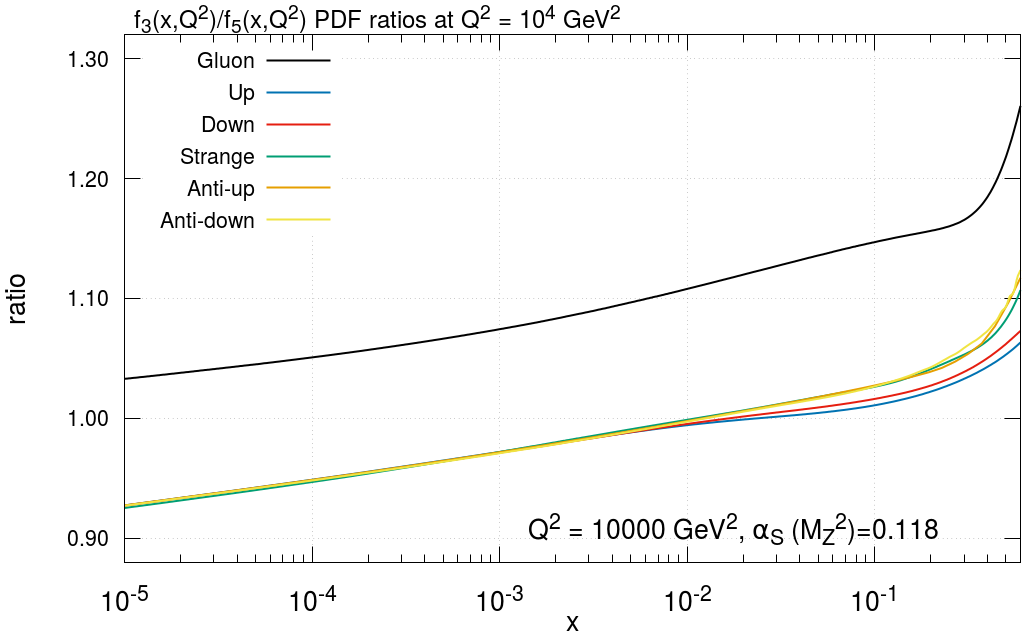}
\includegraphics[scale=0.23]{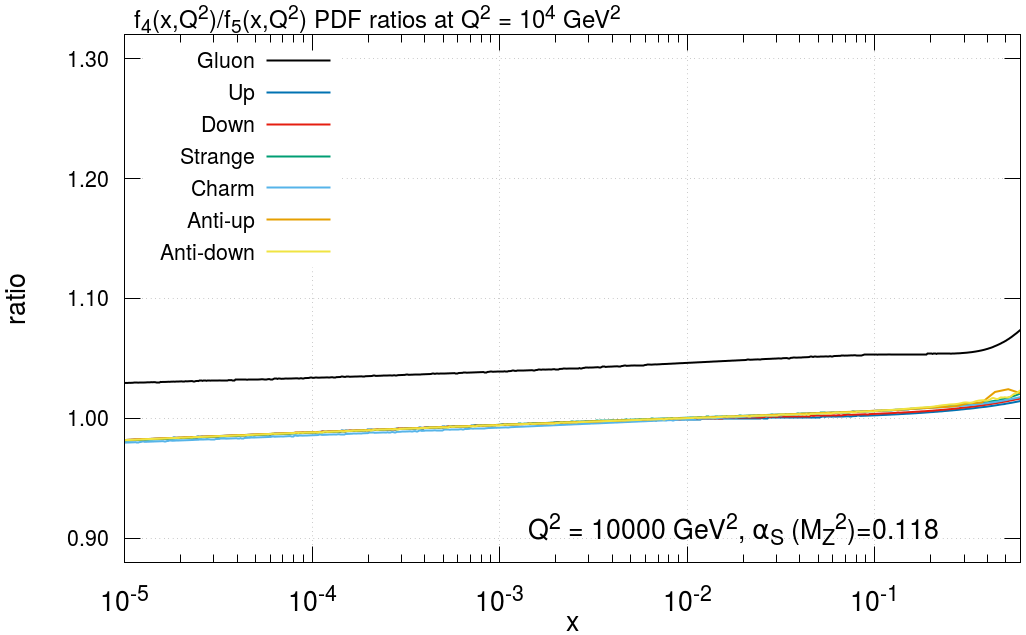}
\caption{\sf The ratio of the different fixed flavour PDFs to the standard 5 flavour PDFs at NNLO and 
at $Q^2=10^4~\GeV^2$. The 3 and 4 flavour schemes are show in the left and right plots respectively.
}
\label{fig:PDFs3and4}
\end{center}
\end{figure}  

The variation of the PDFs defined with a maximum number of 3 and 4 flavours, 
compared to our default of 5 flavours, is shown at $Q^2=10^4~\GeV^2$ in 
Fig. \ref{fig:PDFs3and4} for NNLO PDFs. The the differences 
are due to two different effects. For fewer active quarks there is less gluon branching, so the 
gluon is by definition larger if the flavour number is smaller. However, it is also the case that as $Q^2$ increases the 
coupling gets smaller as there are fewer active quarks. Hence, evolution is generally 
slower, which means that partons decrease less quickly for large $x$ and 
grow less quickly at small $x$. The latter effect dominates for the evolution of the light quarks, so these are smaller at small $x$ where evolution acts to increase the distribution and 
vice versa at high $x$. 
However, for the gluon the two effects compete at small $x$. Overall, the reduction in quark branching 
is the greater effect, mainly due to the fact that this sets in suddenly at flavour transition points, and hence directly 
affects the evolution immediately. The effect on the evolution of the coupling sets in directly at
transitions points, but the resultant effect on the value of the coupling, and hence on PDF 
evolution, is cumulative, and hence less direct. At high $x$ for the gluon the two effects 
act in the same direction, leading to a large enhancement of the gluon for smaller active 
flavour number in this region. As expected, the effects are the same in the 3 and 4 flavour schemes, but larger in the former.

\section{PDF set availability   \label{sec:access}}

We provide the MSHT20 PDFs in the \texttt{LHAPDF} format~\cite{Buckley:2014ana}, available at:
\\
\\
\href{http://lhapdf.hepforge.org/}{\texttt{http://lhapdf.hepforge.org/}}
\\
\\
as well as on the repository:
\\
\\ 
\href{http://www.hep.ucl.ac.uk/msht/}{\texttt{http://www.hep.ucl.ac.uk/msht/}}
\\
\\
We present NLO and NNLO central sets of PDFs for a range of $\alpha_S(M_Z^2)$ values, from 0.108 to 0.130, in increments of 0.001, the central (i.e. 0) member of the NLO set is $\alpha_S(M_Z^2)=0.120$ whilst for the NNLO set the central member is $\alpha_S(M_Z^2)=0.118$:
\\
\\
\href{http://www.hep.ucl.ac.uk/msht/Grids/MSHT20nnlo_as_largerange.tar.gz}{\texttt{MSHT20nnlo\_as\_largerange}}\\
\href{http://www.hep.ucl.ac.uk/msht/Grids/MSHT20nlo_as_largerange.tar.gz}{\texttt{MSHT20nlo\_as\_largerange}}
\\
\\
We provide central sets for 3 and 4 active flavours, over a smaller range of  $\alpha_S(M_Z^2)$ values, from 0.117 to 0.121 for NLO (so as to include both our NLO sets at $\alpha_S(M_Z^2)=0.118, 0.120$) and from 0.117 to 0.119 for NNLO,  in increments of 0.001:
\\
\\
\href{http://www.hep.ucl.ac.uk/msht/Grids/MSHT20nnlo_as_smallrange_nf3.tar.gz}{\texttt{MSHT20nnlo\_as\_smallrange\_nf3}}, \href{http://www.hep.ucl.ac.uk/msht/Grids/MSHT20nnlo_as_smallrange_nf4.tar.gz}{\texttt{MSHT20nnlo\_as\_smallrange\_nf4}}\\
\href{http://www.hep.ucl.ac.uk/msht/Grids/MSHT20nlo_as_smallrange_nf3.tar.gz}{\texttt{MSHT20nlo\_as\_smallrange\_nf3}}, \href{http://www.hep.ucl.ac.uk/msht/Grids/MSHT20nlo_as_smallrange_nf4.tar.gz}{\texttt{MSHT20nlo\_as\_smallrange\_nf4}}.  
\\
\\
We provide central sets for different values of the charm mass, $m_c$, from 1.2 to 1.6 GeV, in increments of 0.05 GeV, and the bottom mass, $m_b$, from 4 to 5.5 GeV, in increments of 0.25 GeV. These are given at NNLO ($\alpha_S(M_Z^2)$=0.118) and NLO (at both $\alpha_S(M_Z^2)$=0.118, 0.120), for 3, 4 and 5 flavours:
\\
\\
\href{http://www.hep.ucl.ac.uk/msht/Grids/MSHT20nnlo_mcrange_nf3.tar.gz}{\texttt{MSHT20nnlo\_mcrange\_nf3}}, \href{http://www.hep.ucl.ac.uk/msht/Grids/MSHT20nnlo_mcrange_nf4.tar.gz}{\texttt{MSHT20nnlo\_mcrange\_nf4}}, \href{http://www.hep.ucl.ac.uk/msht/Grids/MSHT20nnlo_mcrange_nf5.tar.gz}{\texttt{MSHT20nnlo\_mcrange\_nf5}}\\
\href{http://www.hep.ucl.ac.uk/msht/Grids/MSHT20nnlo_mbrange_nf3.tar.gz}{\texttt{MSHT20nnlo\_mbrange\_nf3}}, \href{http://www.hep.ucl.ac.uk/msht/Grids/MSHT20nnlo_mbrange_nf4.tar.gz}{\texttt{MSHT20nnlo\_mbrange\_nf4}}, \href{http://www.hep.ucl.ac.uk/msht/Grids/MSHT20nnlo_mbrange_nf5.tar.gz}{\texttt{MSHT20nnlo\_mbrange\_nf5}}\\
\href{http://www.hep.ucl.ac.uk/msht/Grids/MSHT20nlo_mcrange_nf3.tar.gz}{\texttt{MSHT20nlo\_mcrange\_nf3}}, \href{http://www.hep.ucl.ac.uk/msht/Grids/MSHT20nlo_mcrange_nf4.tar.gz}{\texttt{MSHT20nlo\_mcrange\_nf4}}, \href{http://www.hep.ucl.ac.uk/msht/Grids/MSHT20nlo_mcrange_nf5.tar.gz}{\texttt{MSHT20nlo\_mcrange\_nf5}}\\
\href{http://www.hep.ucl.ac.uk/msht/Grids/MSHT20nlo_mbrange_nf3.tar.gz}{\texttt{MSHT20nlo\_mbrange\_nf3}}, \href{http://www.hep.ucl.ac.uk/msht/Grids/MSHT20nlo_mbrange_nf4.tar.gz}{\texttt{MSHT20nlo\_mbrange\_nf4}}, \href{http://www.hep.ucl.ac.uk/msht/Grids/MSHT20nlo_mbrange_nf5.tar.gz}{\texttt{MSHT20nlo\_mbrange\_nf5}}\\
\href{http://www.hep.ucl.ac.uk/msht/Grids/MSHT20nlo_as120_mcrange_nf3.tar.gz}{\texttt{MSHT20nlo\_as120\_mcrange\_nf3}}, \href{http://www.hep.ucl.ac.uk/msht/Grids/MSHT20nlo_as120_mcrange_nf4.tar.gz}{\texttt{MSHT20nlo\_as120\_mcrange\_nf4}}, \href{http://www.hep.ucl.ac.uk/msht/Grids/MSHT20nlo_as120_mcrange_nf5.tar.gz}{\texttt{MSHT20nlo\_as120\_mcrange\_nf5}}\\
\href{http://www.hep.ucl.ac.uk/msht/Grids/MSHT20nlo_as120_mbrange_nf3.tar.gz}{\texttt{MSHT20nlo\_as120\_mbrange\_nf3}}, \href{http://www.hep.ucl.ac.uk/msht/Grids/MSHT20nlo_as120_mbrange_nf4.tar.gz}{\texttt{MSHT20nlo\_as120\_mbrange\_nf4}},\href{http://www.hep.ucl.ac.uk/msht/Grids/MSHT20nlo_as120_mbrange_nf5.tar.gz}{\texttt{MSHT20nlo\_as120\_mbrange\_nf5}}.
\\
\\
Note the NLO sets without the $\alpha_S(M_Z^2)$ value indicated in the name (e.g. {\texttt{MSHT20nlo\_mcrange\_nf3} and {\texttt{MSHT20nlo\_mbrange\_nf3}) are at $\alpha_S(M_Z^2)=0.118$, matching the naming of the NNLO sets which are also at $\alpha_S(M_Z^2)=0.118$. 
\\
\\
Finally, we provide eigenvector sets for 3 and 4 flavours:
\\
\\
\href{http://www.hep.ucl.ac.uk/msht/Grids/MSHT20nnlo_nf3.tar.gz}{\texttt{MSHT20nnlo\_nf3}}, \href{http://www.hep.ucl.ac.uk/msht/Grids/MSHT20nnlo_nf4.tar.gz}{\texttt{MSHT20nnlo\_nf4}}\\
\href{http://www.hep.ucl.ac.uk/msht/Grids/MSHT20nlo_as120_nf3.tar.gz}{\texttt{MSHT20nlo\_as120\_nf3}}, \href{http://www.hep.ucl.ac.uk/msht/Grids/MSHT20nlo_as120_nf4.tar.gz}{\texttt{MSHT20nlo\_as120\_nf4}}\\
\href{http://www.hep.ucl.ac.uk/msht/Grids/MSHT20nlo_nf3.tar.gz}{\texttt{MSHT20nlo\_nf3}}, \href{http://www.hep.ucl.ac.uk/msht/Grids/MSHT20nlo_nf4.tar.gz}{\texttt{MSHT20nlo\_nf4}}.
\\
\\
We note again for clarity that for sets with 3 and 4 flavours, the quoted $\alpha_S(M_Z^2)$ values correspond to evolving the coupling from $Q_0=1$ GeV at a value that for 5 active flavours corresponds to this $\alpha_S(M_Z^2)$, but with 3 or 4 active flavours, i.e. the values specified here and in the info files for $\alpha_S(M_Z^2)$ are 5-flavour scheme values which are then translated internally in generating the 3 and 4 flavour number scheme sets.

\section{Conclusions \label{sec:x}}

The PDFs  determined from global fits to deep--inelastic and  hard--scattering data 
are highly correlated to the value of $\a$ used, and any changes in 
the values of $\a$ must be accompanied by changes in the PDFs such that 
the optimum fit to data is still obtained. In~\cite{MSHT20} we produced updated PDFs and 
uncertainty eigenvector sets for specific values of $\a$, close to the best fit values. 
In this article we additionally present PDF sets
and the global fit quality 
at NLO and NNLO for a wide variety of $\a$ values, i.e. $\a=0.108$ to 
$\a=0.130$ in steps of $\Delta \a =0.001$. Hence, we illustrate in more 
detail the origin of our best fit $\a$ values of  
\begin{align}
  \text{NLO:}\qquad\alpha_S(M_Z^2) &= 0.1203\quad \pm 0.0015 \text{ (68\% C.L.)},\\
  \text{NNLO:}\qquad\alpha_S(M_Z^2) &= 0.1174\quad \pm 0.0013 \text{ (68\% C.L.)},
\end{align}
already presented in \cite{MSHT20}, but here also determine the uncertainties. 
We show the variation of the fit quality with $\a$ of numerous data sets, 
within the context of the global fit, and see which are the more and less 
constraining sets, and which prefer higher and lower values. We see that,
in common with the global fit, 
most data sets show a systematic trend of preferring a slightly lower 
$\a$ value at NNLO than at NLO, but note that whilst there are some trends, no particular type of 
data consistently prefers a high or low value of $\a$. 
There are examples of 
fixed target DIS data which prefer either high or low values and 
similarly for the collider data sets. Indeed, our best values of 
$\a$ are almost unchanged from the previous values~\cite{MMHTas} of 
$\a=0.1201$ (NLO) and $\a=0.1172$ (NNLO). 

They are also similar to the values obtained by NNPDF of 
$\a=0.1185$ (NNLO) and $\a=0.1207$ 
(NLO)~\cite{Ball:2018iqk}, and by CT18 of $\a=0.1164$ (NNLO)~\cite{Hou:2019efy}. However, our extraction disagrees with the recent 
value $\a=0.1147$ (NNLO) obtained by ABMP in \cite{Alekhin:2017kpj}, but agrees better with their value
$\a=0.1191$ at NLO in \cite{Alekhin:2018pai}. 
We find agreement of our NNLO value at the level of less than one sigma
with the world average value of $\a=0.1179 \pm 0.001$. Hence, our NNLO best fit value of $\a$ is in excellent agreement with the default value, $\a=0.118$, for
which eigenvector sets are made available. 
The PDF sets obtained at the 23 different values of $\a$ at NLO and NNLO are made publicly available. These will be useful in studies of 
$\a$ by other groups, but we warn against the naive extraction of the best-fit value of 
$\a$ using the fit quality to a particular data set, as discussed in \cite{Forte:2020pyp}.   
Indeed, the explicit presentation of the $\chi^2$ deterioration away from the best fit values 
of $\a$ in Table \ref{tab:deltachisq_global} provides additional information which should 
impact on studies of $\a$ variation while using our global PDFs.

In order to calculate the PDF$+\a$ uncertainty we recommend the 
approach introduced in \cite{CTEQalphas} of treating PDFs with 
$\a\pm \Delta \a$ as an extra eigenvector set. As shown in \cite{CTEQalphas},
provided certain conditions are met, the 
$\a$ uncertainty is correctly added to the PDF uncertainty by simply adding 
in quadrature the variation of any quantity under a change in coupling 
$\Delta \a$ as long as the
change in $\a$ is accompanied by the appropriate change in PDFs required by the 
global fit. As examples, we have calculated the total cross sections for the production 
of $W$, $Z$, top quark pairs and Higgs bosons at the Tevatron, LHC and FCC.  
For $W$ and $Z$ production, where the LO sub-process is 
${\cal O}(\alpha_S^0)$ the combined ``PDF+$\alpha_S$'' 
uncertainty is not much larger than the PDF-only uncertainty with a fixed 
$\alpha_S$.  The additional uncertainty due to $\alpha_S$ is more 
important for top quark pair production and Higgs boson production via 
gluon--gluon fusion, since the LO sub-process is now ${\cal O}(\alpha_S^2)$,
though the details depend significantly on the correlation between $\a$ and the contributing 
PDFs. For any particular process the details of the 
uncertainty can be explicitly calculated in a straightforward way using 
the PDFs we have provided in this paper, together with the procedure 
for combining PDF and $\a$ uncertainty discussed in Section~\ref{sec:PDFalpha}.

The additional purpose of this article is to present and make available PDF sets in the framework 
used to produce the MSHT20 PDFs, but with 
differing values of the charm, and of the bottom, quark masses. We do not strictly make a 
determination of the optimum values of these masses, but we do investigate and
note the effect the mass variation has on the quality of the fits to the data, 
concentrating on the final combined HERA heavy flavour cross section data~\cite{HERAhf} in particular. 
We note that for both the charm and bottom quarks 
our default values of 
\be
m_c=1.4~{\rm GeV},\qquad m_b=4.75~\GeV 
\label{eq:masses}
\ee
are close to the 
values preferred by the fit, more so than in the previous global PDFs~\cite{MMHThq},
and that these are consistent with the values of pole masses one would expect
by conversion from the values measured in the $\MS$ scheme. 
For instance, our NNLO studies indicate that $m_c \simeq 1.3 -1.4~\GeV$, $m_b \simeq 4.25 -  4.75~\GeV$.

We also make PDFs available with a maximum of 
3 or 4 active quark flavours. 
All publicly available sets can be found at~\cite{UCLsite} 
and will also be available from the LHAPDF library~\cite{LHAPDF}.

Finally,  we investigate the variation of the PDFs and the predicted cross sections for standard
processes corresponding to these variations in 
heavy-quark mass. For reasonable variations of $m_c$ the effects  are small, but not
insignificant, compared to PDF uncertainties. For variations in $m_b$ the
effect is smaller. Changes in PDFs with the value of $m_b$ are much smaller than PDF uncertainties, 
except for the bottom 
distribution itself, which can vary more than its uncertainty at a fixed
value of $m_b$. In summary, while currently the uncertainties 
on PDFs due to quark masses
are subleading in comparison to the PDF uncertainties, these are not negligible and we can expect them to become more important in the future, as precision requirements increase.

\section*{Acknowledgements}

We would like to thank numerous members of the PDF4LHC committee and working 
group for useful conversations. 
T. C. and R. S. T. thank the Science and Technology Facilities Council (STFC) for support via grant awards ST/P000274/1 and ST/T000856/1. L. H. L. thanks STFC for support via grant award ST/L000377/1.

\newpage

\bibliography{referencesashq.bib}

\begin{thebibliography}{10}

\bibitem{MSHT20}
S.~Bailey, T.~Cridge, L.~A. Harland-Lang, A.~D. Martin, and R.~S. Thorne,
\newblock Eur. Phys. J. C {\bf 81}, 341 (2021), 2012.04684.

\bibitem{MMHT14}
L.~A. Harland-Lang, A.~D. Martin, P.~Motylinski, and R.~S. Thorne,
\newblock Eur. Phys. J. C {\bf 75}, 204 (2015), 1412.3989.

\bibitem{PDG2020}
Particle Data Group, P.~Zyla {\em et~al.},
\newblock Prog. Theor. Exp. Phys. {\bf 8}, 083C01 (2020).

\bibitem{MMHTas}
L.~A. Harland-Lang, A.~D. Martin, P.~Motylinski, and R.~S. Thorne,
\newblock Eur. Phys. J. C {\bf 75}, 435 (2015), 1506.05682.

\bibitem{PDF4LHC1}
S.~Alekhin {\em et~al.},
\newblock (2011), 1101.0536.

\bibitem{PDF4LHC2}
M.~Botje {\em et~al.},
\newblock (2011), 1101.0538.

\bibitem{bench1}
G.~Watt,
\newblock JHEP {\bf 1109}, 069 (2011), 1106.5788.

\bibitem{bench2}
R.~D. Ball {\em et~al.},
\newblock JHEP {\bf 1304}, 125 (2013), 1211.5142.

\bibitem{PDF4LHC15}
J.~Butterworth {\em et~al.},
\newblock J. Phys. G {\bf 43}, 023001 (2016), 1510.03865.

\bibitem{Moch:2004pa}
S.~Moch, J.~Vermaseren, and A.~Vogt,
\newblock Nucl. Phys. B {\bf 688}, 101 (2004), hep-ph/0403192.

\bibitem{Vogt:2004mw}
A.~Vogt, S.~Moch, and J.~Vermaseren,
\newblock Nucl. Phys. B {\bf 691}, 129 (2004), hep-ph/0404111.

\bibitem{vanNeerven:1991nn}
W.~van Neerven and E.~Zijlstra,
\newblock Phys. Lett. B {\bf 272}, 127 (1991).

\bibitem{Zijlstra:1991qc}
E.~Zijlstra and W.~van Neerven,
\newblock Phys. Lett. B {\bf 273}, 476 (1991).

\bibitem{Zijlstra:1992kj}
E.~Zijlstra and W.~van Neerven,
\newblock Phys. Lett. B {\bf 297}, 377 (1992).

\bibitem{Zijlstra:1992qd}
E.~Zijlstra and W.~van Neerven,
\newblock Nucl. Phys. B {\bf 383}, 525 (1992).

\bibitem{Moch:2004xu}
S.~Moch, J.~Vermaseren, and A.~Vogt,
\newblock Phys. Lett. B {\bf 606}, 123 (2005), hep-ph/0411112.

\bibitem{Vermaseren:2005qc}
J.~Vermaseren, A.~Vogt, and S.~Moch,
\newblock Nucl. Phys. B {\bf 724}, 3 (2005), hep-ph/0504242.

\bibitem{Forte:2020pyp}
S.~Forte and Z.~Kassabov,
\newblock Eur. Phys. J. C {\bf 80}, 182 (2020), 2001.04986.

\bibitem{MSTW}
A.~D. Martin, W.~J. Stirling, R.~S. Thorne, and G.~Watt,
\newblock Eur.Phys.J. {\bf C63}, 189 (2009), 0901.0002.

\bibitem{Cooper-Sarkar:2020twv}
A.~M. Cooper-Sarkar, M.~Czakon, M.~A. Lim, A.~Mitov, and A.~S. Papanastasiou,
\newblock (2020), 2010.04171.

\bibitem{CTEQalphas}
H.-L. Lai {\em et~al.},
\newblock Phys.Rev. {\bf D82}, 054021 (2010), 1004.4624.

\bibitem{NNLOtop}
M.~Czakon, P.~Fiedler, and A.~Mitov,
\newblock Phys. Rev. Lett. {\bf 110}, 252004 (2013), 1303.6254.

\bibitem{MMHThq}
L.~A. Harland-Lang, A.~D. Martin, P.~Motylinski, and R.~S. Thorne,
\newblock Eur. Phys. J. C {\bf 76}, 10 (2016), 1510.02332.

\bibitem{Bauer:2004ve}
C.~W. Bauer, Z.~Ligeti, M.~Luke, A.~V. Manohar, and M.~Trott,
\newblock Phys. Rev. {\bf D70}, 094017 (2004), hep-ph/0408002.

\bibitem{Hoang:2005zw}
A.~H. Hoang and A.~V. Manohar,
\newblock Phys. Lett. {\bf B633}, 526 (2006), hep-ph/0509195.

\bibitem{Thorne}
R.~S. Thorne,
\newblock Phys.Rev. {\bf D86}, 074017 (2012), 1201.6180.

\bibitem{TR1}
R.~S. Thorne,
\newblock Phys.Rev. {\bf D73}, 054019 (2006), hep-ph/0601245.

\bibitem{Laenen:1992zk}
E.~Laenen, S.~Riemersma, J.~Smith, and W.~L. van Neerven,
\newblock Nucl. Phys. B {\bf 392}, 162 (1993).

\bibitem{Catani:1990eg}
S.~Catani, M.~Ciafaloni, and F.~Hautmann,
\newblock Nucl. Phys. B {\bf 366}, 135 (1991).

\bibitem{Laenen:1998kp}
E.~Laenen and S.-O. Moch,
\newblock Phys. Rev. D {\bf 59}, 034027 (1999), hep-ph/9809550.

\bibitem{Kawamura:2012cr}
H.~Kawamura, N.~Lo~Presti, S.~Moch, and A.~Vogt,
\newblock Nucl. Phys. B {\bf 864}, 399 (2012), 1205.5727.

\bibitem{HERAhf}
H1, ZEUS, H.~Abramowicz {\em et~al.},
\newblock Eur. Phys. J. C {\bf 78}, 473 (2018), 1804.01019.

\bibitem{Ball:2017nwa}
NNPDF, R.~D. Ball {\em et~al.},
\newblock Eur. Phys. J. C {\bf 77}, 663 (2017), 1706.00428.

\bibitem{Buza:1996wv}
M.~Buza, Y.~Matiounine, J.~Smith, and W.~L. van Neerven,
\newblock Eur. Phys. J. {\bf C1}, 301 (1998), hep-ph/9612398.

\bibitem{quarkmatching}
xFitter Developers Team, V.~Bertone {\em et~al.},
\newblock Eur. Phys. J. C {\bf 77}, 837 (2017), 1707.05343.

\bibitem{Thorne:2008xf}
R.~S. Thorne and W.~K. Tung,
\newblock {PQCD Formulations with Heavy Quark Masses and Global Analysis},
\newblock in {\em {Proceedings, workshop: HERA and the LHC workshop series on
  the implications of HERA for LHC physics}}, 2008, 0809.0714.

\bibitem{ThorneFFNS}
R.~Thorne,
\newblock Eur.Phys.J. {\bf C74}, 2958 (2014), 1402.3536.

\bibitem{NNPDFgmvfns}
The NNPDF Collaboration, R.~D. Ball {\em et~al.},
\newblock Phys.Lett. {\bf B723}, 330 (2013), 1303.1189.

\bibitem{Martin:2006qz}
A.~D. Martin, W.~J. Stirling, and R.~S. Thorne,
\newblock Phys. Lett. {\bf B636}, 259 (2006), hep-ph/0603143.

\bibitem{MSTWhq}
A.~Martin, W.~Stirling, R.~Thorne, and G.~Watt,
\newblock Eur.Phys.J. {\bf C70}, 51 (2010), 1007.2624.

\bibitem{Buckley:2014ana}
A.~Buckley {\em et~al.},
\newblock Eur. Phys. J. C {\bf 75}, 132 (2015), 1412.7420.

\bibitem{Ball:2018iqk}
NNPDF, R.~D. Ball {\em et~al.},
\newblock Eur. Phys. J. C {\bf 78}, 408 (2018), 1802.03398.

\bibitem{Hou:2019efy}
T.-J. Hou {\em et~al.},
\newblock Phys. Rev. D {\bf 103}, 014013 (2021), 1912.10053.

\bibitem{Alekhin:2017kpj}
S.~Alekhin, J.~Bl\"umlein, S.~Moch, and R.~Placakyte,
\newblock Phys. Rev. D {\bf 96}, 014011 (2017), 1701.05838.

\bibitem{Alekhin:2018pai}
S.~Alekhin, J.~Bl\"umlein, and S.~Moch,
\newblock Eur. Phys. J. C {\bf 78}, 477 (2018), 1803.07537.

\bibitem{UCLsite}
\texttt{http://www.hep.ucl.ac.uk/msht/}.

\bibitem{LHAPDF}
\texttt{http://lhapdf.hepforge.org}.

\end{thebibliography}

\bibliographystyle{h-physrev}

\end{document}